\documentclass[twocolumn]{aastex631}

\usepackage{comment}

\begin{document}

\title{The Same But Different: SiO$_2$ Clouds in Two L-Dwarfs with JWST}

\author[0000-0002-0786-7307]{Joshua D. Lothringer}
\affiliation{Space Telescope Science Institute, Baltimore, MD, USA}
\email{jlothringer@uvu.edu}

\author[0000-0003-3305-6281]{Jeff A. Valenti}
\affiliation{Space Telescope Science Institute, Baltimore, MD, USA}

\author[0000-0002-6721-3284]{Sarah E. Moran}
\altaffiliation{NHFP Sagan Fellow}
\affiliation{Space Telescope Science Institute, Baltimore, MD, USA}
\affiliation{University of Maryland, College Park, MD, USA}

\author[0009-0001-0008-0464]{Rhys Collins}
\affiliation{Department of Physics, University of Warwick, Coventry CV4 7AL, UK}

\author[0000-0001-9552-3709]{Siddharth Gandhi}
\affiliation{Department of Physics, University of Warwick, Coventry CV4 7AL, UK}
\affiliation{Centre for Exoplanets and Habitability, University of Warwick, Gibbet Hill Road, Coventry
CV4 7AL, UK}

\author{Seti Norris}
\affiliation{Space Telescope Science Institute, Baltimore, MD, USA}

\author[0000-0001-6396-8439]{William O. Balmer}
\email{balmer@northwestern.edu}
\altaffiliation{51 Pegasi b Fellow}
\affiliation{Center for Interdisciplinary Exploration and Research in Astrophysics (CIERA), Northwestern University, 1800 Sherman Avenue, Evanston, IL 60201, USA}

\author[0000-0002-8518-9601]{Peter Gao}
\affiliation{Earth and Planets Laboratory, Carnegie Institute of Washington, Washington DC, USA}

\author{Brayden Roberts}
\affiliation{Utah Valley University, Orem, UT}

\author[0000-0002-5251-2943]{Mark S. Marley}
\affiliation{Lunar and Planetary Laboratory, University of Arizona, Tucson, AZ, USA}

\author[0000-0003-4328-3867]{Hannah R. Wakeford}
\affiliation{School of Physics, University of Bristol, H.H. Wills Physics Laboratory, Bristol, UK, BS8 1TL}

\author[0000-0001-6050-7645]{David K. Sing}
\affiliation{Department of Earth and Planetary Sciences, Johns Hopkins University, Baltimore, MD}
\affiliation{William H.\ Miller III Department of Physics \& Astronomy,
Johns Hopkins University, Baltimore, MD 21218, USA}

\begin{abstract}

The precision of brown-dwarf spectra from JWST enables unprecedented insight into the atmospheric and cloud properties of sub-stellar atmospheres. Here, we compare both self-consistent radiative-convective equilibrium (RCE) and retrieval atmosphere models to high-SNR 1--18~$\mu$m spectra of two field L-dwarfs from JWST, J0624 (L5) and J2148 (L6--L6.5). Our observations include spectral features from H$_2$O, CO, CO$_2$, CH$_4$, Na, and K and cover approximately 98\% of the outgoing thermal emission from the two objects. We first compare the data to three independent grids of 1D-RCE models: BT-Settl, exoREM, and Sonora Diamondback. While exoREM provides the best fit of the grid models, no 1D-RCE model provides a satisfactory fit to the data, highlighting the complexity of fitting cloudy L-dwarf spectra. Nonetheless, the temperature and cloud properties can be constrained with the 1D-RCE models, while metallicity, C/O ratio, and surface gravity are not well-measured. Next, we use petitRADTRANS to retrieve atmosphere and cloud properties, finding strong evidence for the presence of SiO$_2$ clouds in both objects. The SiO$_2$ is found to be amorphous in structure for J0624 but crystalline in J2148, highlighting important differences in objects of similar spectral type. 
We also used a non-Bayesian optimization framework to explore a wider variety of cloud scenarios, finding evidence for the inclusion of iron-rich magnesium silicates (Mg$_{0.5}$Fe$_{0.5}$SiO$_3$, MgFeSiO$_4$) in addition to SiO$_2$ in both objects. These results show both the complexity and diversity of brown dwarf cloud properties, while also highlighting the constraining power of broad-wavelength JWST observations of cloudy L-dwarfs. 

\end{abstract}


\section{Introduction} \label{sec:intro}


Condensate clouds are expected to dominate the thermal emission spectra of L-type brown dwarfs. Atmosphere models have long predicted the formation of solid high-temperature minerals at the infrared photospheres of substellar objects with effective temperatures between approximately 1300 and 1600\,K \citep[e.g.,][]{allard:2001,tsuji:2002,ackerman:2001,visscher:2010}. Such particles tend to have significant opacity throughout the infrared, suppressing flux emerging from deeper atmospheric layers and yielding characteristically red near-infrared colors \citep{cushing:2005,cushing:2006}. 

As an L-type brown dwarf continues to cool over time, it will transition to a T-type dwarf, marked by a sudden shift towards bluer colors \citep{tsuji:2003}. This has long been interpreted as a gradual ``clearing" of the clouds as they become patchy and begin to sink below the photosphere \citep{saumon:2008,marley:2010}. This hypothesis is supported by the population of highly-variable objects along the L-T transition thought to be showing cloud ``patchiness" \citep[e.g.,][]{apai:2013,vos:2017,biller:2024,oliverosgomez:2025,nasedkin:2025} and by the depth of the 10 $\mu$m silicate feature, maximized for late L-dwarfs and gradually disappearing in T-dwarfs \citep{cushing:2006,suarez:2022, barber:2026}.

Nonetheless, other effects may also contribute to the L-T transition, including fingering convection mimicking clouds in the outgoing spectrum by making the temperature structure more isothermal \citep{tremblin:2015}, though this has also been contested \citep{leconte:2018}. Observations of both the cloud features themselves in the MIR and their effect on the emergent spectrum across the IR can chart the formation and evolution of these clouds through parameter space. 

The composition, crystallinity, particle size distribution, vertical extent, and patchiness of clouds in brown dwarfs and exoplanets can all be measured with high-SNR infrared spectra \citep{cushing:2006,luna:2021,moran:2024}. The shape of the spectral energy distribution, combined with multiple molecular spectral features, can help constrain these properties, but the most information-rich measurements come from mid-IR spectra of the 10~$\mu$m silicate feature. Differences in composition (e.g., SiO$_2$ vs. MgSiO$_3$ vs. Mg$_2$SiO$_4$), crystallinity (amorphous versus crystalline), and particle size all influence the level of solid-state absorption at mid-IR wavelengths \citep{cushing:2006}. For example, observations of the cloudy L-dwarf J2224 with IRTF/SpeX, Gemini/GNIRS, and \textit{Spitzer} were fit with atmosphere retrievals, identifying sub-micron particles of SiO$_2$ and MgSiO$_3$ above a layer of cloud deck of Fe clouds \citep{burningham:2021}. Such detailed constraints can then inform microphysical models of cloud formation.

In this work, we fit JWST observations from 1--28~$\mu$m with atmosphere model grids and retrievals to measure the atmospheric and cloud properties in two L-dwarfs, J21481633+4003594 (hereafter J2148) \citep{looper:2008} and J06244595-4521548 (hereafter J0624). These two objects were chosen due to their relative proximity to Earth ($\sim10$pc) and their classification as somewhat dissimilar cloudy objects. While J2148 appears to have a very strong 10\,$\mu$m silicate absorption feature \citep{looper:2008} and very red colors ($J-K = 2.38$), J0624 has a somewhat more muted MIR absorption \citep{suarez:2022} and bluer colors ($J-K = 1.9$), indicating a potential difference in cloud properties. Table~\ref{table:properties} lists fundamental properties of our two targets from past literature.

Section~\ref{sec:obs} describes our observations, followed by a comparison to self-consistent model grids in Section~\ref{sec:rce}. Section~\ref{sec:retrievals} describes our retrieval analysis methods and Section~\ref{sec:retrieval_results} details the results for each object. A discussion of these results, comparing the two objects to themselves and the broader substellar population, is presented in Section~\ref{sec:disc}. We conclude with a brief summary in Section~\ref{sec:conc}.

\section{Observations}\label{sec:obs}

JWST observed J0624 on 2023-Apr-04 and J2148 on 2023-Sep-20 in program 2288 \citep{jwstprop:2288}. We obtained NIRSpec \citep{jakobsen:2022} spectra with all three $R\sim2700$ gratings: G140H/F100LP (0.97--1.82\,$\mu$m), G235H (1.66--3.05\,$\mu$m), G395H (2.87--5.14\,$\mu$m). For each NIRSpec grating, we obtained spectra through two NIRSpec slits: S200A1 and S200A2. For each NIRSpec slit, we obtained spectra at two dither locations. Thus, for each target we obtained four dithered spectra with each NIRSpec grating and 12 spectra in total. For all the NIRSpec spectra, we used the NRSIRS2RAPID readout pattern, which reads the entire detector and mitigates 1/f noise to some extent.

We also obtained $R\sim3000$ spectra with the MIRI Medium Resolution Spectrometer (MRS; \citealp{wells:2015}) using three settings (short, medium, long). Each setting produces four separate spectra (channels) that cover disjoint wavelength ranges. Taken together, the three settings cover 4.9--28\,$\mu$m in 12 segments, but we ignore wavelengths longer than 18\,$\mu$m where SNR is below 20. For each setting, we obtained spectra at four dither positions. Thus, for each target, we obtained a total of 48 MIRI spectra. We also obtained simultaneous images of an adjacent field with the F1800W, F2100W, and F2550W filters, but do not report on those images here.

We sequenced exposures in ascending (J0624) or descending (J2148) order, so that consecutive exposures had some spectral overlap. The order was dictated by observatory constraints. Flux in the spectral overlap regions agrees to within the expected accuracy of flux calibration, including the order overlap between the long-wavelength end of NIRSpec and the short wavelength end of MIRI.

We used version 1.16.0 \citep{jwstpipe1160} of the JWST pipeline \citep{bushouse:2020} with calibration reference files in CRDS context jwst\_1296.pmap to produce extracted and calibrated 1D spectra. Prior to running stage 3 of the pipeline, we masked out pixel (set DQ=1) with zero flux and the last 20 pixels of NRS2 spectra, which did not seem to be well calibrated.

\begin{figure*}
    \centering
    \includegraphics[width=0.9\linewidth]{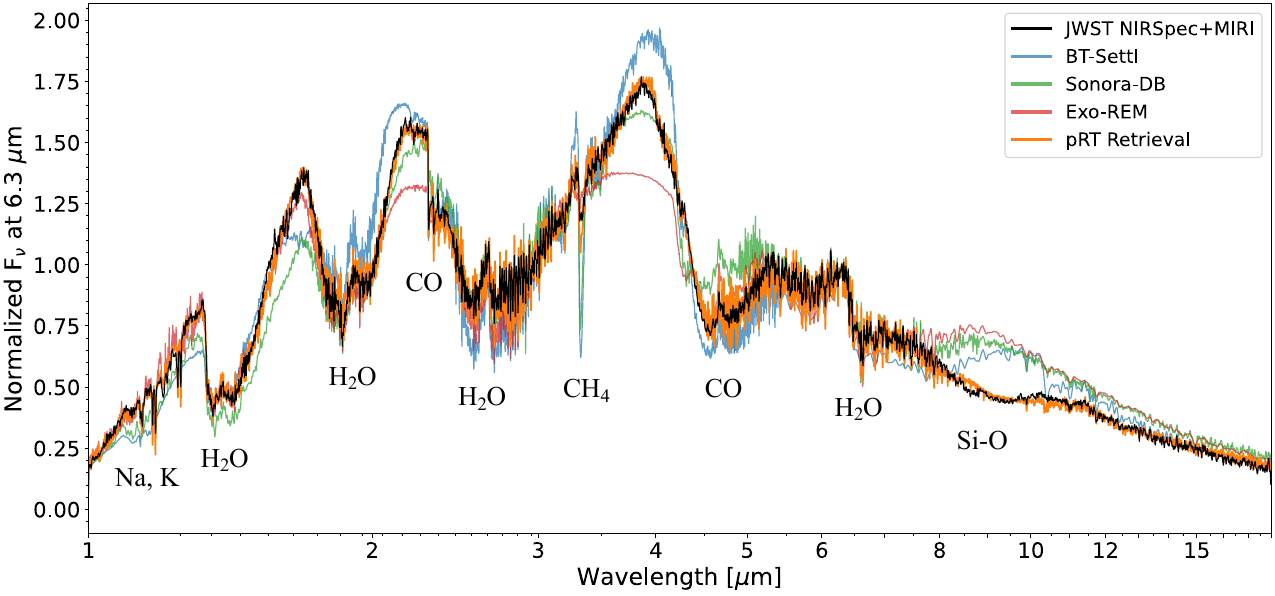}
    \caption{Best-fit 1D-RCE and retrieval models for J0624 compared to our observations. All spectra are shown at R$\sim$1,000, though the retrieval (orange) was fit at R$\sim$100 for computational efficiency. All fitting routines were done in F$_\lambda$-space, but the flux shown here is scaled as F$_\nu$ to better visualize the dynamic range across the entire wavelength range.}
    \label{fig:J0624_spec}
\end{figure*}

\begin{figure*}
    \centering
    \includegraphics[width=0.9\linewidth]{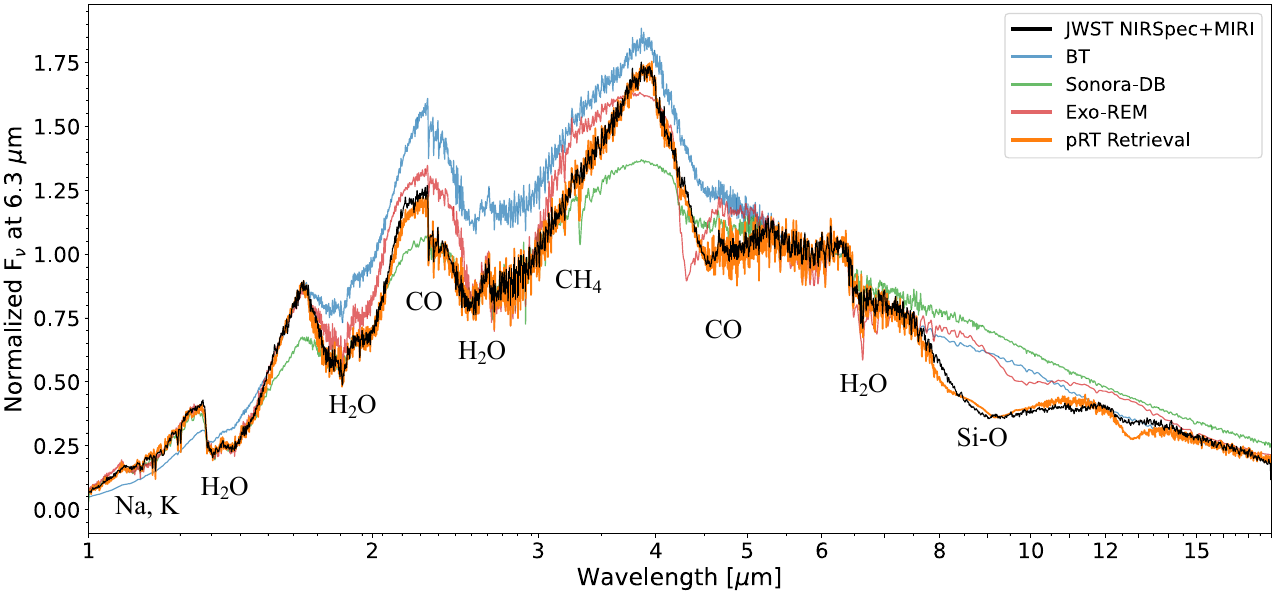}
    \caption{Same as Figure~\ref{fig:J0624_spec}, but for J2148.}
    \label{fig:J2148_spec}
\end{figure*}

After running the JWST science calibration pipeline, we combined the three pipeline-produced NIRSpec spectra (one per grating) and nine pipeline-produced MIRI spectra (one per segment, ignoring three segments beyond 18\,$\mu$m) for each target into a single spectrum with a log-uniform wavelength scale $\Delta\lambda/\lambda = 3000$. For each input spectrum, we found a cubic b-spline that exactly reproduced the input spectrum after integrating the b-spline analytically over the wavelength domain of each input wavelength bin. We then integrated that same b-spline over the wavelength domain of each output wavelength bin. To estimate uncertainty in each output bin, we calculated the standard deviation of trial fluxes in each output bin from 100 Monte Carlo trials with input fluxes perturbed according to the input uncertainty. Figure~\ref{fig:J0624_spec} and \ref{fig:J2148_spec} show the spectra. 

\begin{deluxetable*}{c|cc|cc}
\tablecaption{Previous Literature Target Properties \label{table:properties}}
\tablehead{Name     &J0624               &Ref   &J2148                   &Ref}
\startdata
Right Ascension (HH MM SS.sss)& 06 24 45.9449285271 $\pm$ 0.3122 mas     &  1   & 21 48 16.2986583891  $\pm$ 0.2342 mas  & 1 \\
Declination         & -45 21 54.758017706 $\pm$ 0.3951 mas     &  1   & +40 03 59.331855665  $\pm$ 0.2807 mas  & 1 \\
Parallax (mas)      & 82.03 $\pm$ 0.36        &  1   & 123.67 $\pm$ 0.36      & 1 \\
Distance (pc)       & 12.191 $\pm$ 0.054      &  1   & 8.086  $\pm$ 0.023     & 1 \\
\hline
Opt SpT             & L5                      &  2   & L6                     & 3 \\
IR SpT              & L5                      &  4   & L6.5p                  & 5 \\
J$_{2MASS}$         &  14.480 $\pm$ 0.029     &  6   & 14.147 $\pm$ 0.029     & 6 \\
H$_{2MASS}$         &  13.335 $\pm$ 0.028     &  6   & 12.78 $\pm$ 0.03       & 6 \\
K$_{2MASS}$         &  12.595 $\pm$ 0.026     &  6   & 11.7655 $\pm$ 0.023    & 6 \\
log(L/L${sun}$)     & -4.33 $\pm$ 0.05        &  7   & -4.394 $\pm$ 0.004     & 7 \\
Age (Gyr)           & 0.500–10.00             &  7   & 0.500–10.00            & 7 \\
\hline
log($g$)            & 5.14 $\pm$ 0.27         &  7   & 5.13 $\pm$ 0.28        & 7 \\
Radius (R$_J$)      & 0.99 $\pm$ 0.10         &  7   & 0.99 $\pm$ 0.10        & 7 \\
T$_{\rm{eff}}$ (K)       & 1501 $\pm$ 85           &  7   & 1446 $\pm$ 72          & 7 \\
Mass (M$_J$)        & 56.04 $\pm$ 18.52       &  7   & 55.03 $\pm$ 18.55      & 7 \\
Inclination ($^{\circ}$)& Not Available         &  --  & $88^{+8}_{-2}$          & 8 \\
\enddata
\tablerefs{(1) \citet{Gaia:2023}; (2) \citet{reid:2008}; (3) \citet{looper:2008}; (4) \citet{manjavacas:2018}; (5) \citet{kirkpatrick:2010}; (6) \citet{cutri:2003}; (7) \citet{filippazzo:2015}; (8) \citet{vos:2017}.}
\end{deluxetable*}

\begin{deluxetable*}{c|ccccccccc}
\tablecaption{Parameters of the Forward Model Grids \label{tab:model_grids}}
\tablehead{
\colhead{Grid} & 
\colhead{$T_{\rm eff}$ (K)} & 
\colhead{$\Delta T_{\rm eff}$ (K)} &
\colhead{$\log{(g_{\rm cgs})}$} & 
\colhead{$\Delta \log{g}$} & 
\colhead{[M/H]} & 
\colhead{$\Delta$[M/H]} & 
\colhead{$f_{\rm sed}$} & 
\colhead{C/O} &
\colhead{$\Delta$C/O}  
}
\startdata
BT-Settl & 400--2600 & 100 & 3.5--5.5 & 0.5 & 0.0\tablenotemark{a} & --- & --- & 0.55 (Solar\tablenotemark{a}) & ---  \\
Sonora-DB& 900--2400 & 100 & 3.5--5.5 & 0.5 & $-0.5-0.5$\tablenotemark{b}  & 0.5 & 1,2,3,4,8& 0.458 (Solar\tablenotemark{b}) & ---  \\
Exo-REM& 1000--2000 & 50 & 3.5--5.0 & 0.5 &  $-0.5-1.0$\tablenotemark{c} & 0.5 & ---  & 0.1--0.8  & 0.05 \\
\enddata
\tablenotetext{a}{BT-Settl metallicity and C/O defined by \cite{asplund:2009}. Model grid ranges from \citet{allard:2012}.}
\tablenotetext{b}{Sonora-DB metallicity and C/O defined by \cite{lodders:2010}.}
\tablenotetext{c}{Exo-REM metallicity and C/O defined by \cite{asplund:2021}.}

\end{deluxetable*}

\begin{deluxetable*}{c|cc|cc|cc}
\tablecaption{Best-fit grid values for J0624 \label{table:J0624:fits_v2}}
\tablehead{\colhead{} & \colhead{BT Settl}   &\colhead{BT Settl}  & \colhead{Sonora Diamondback} & \colhead{Sonora Diamondback} & \colhead{Exo-REM} & \colhead{Exo-REM} \\
Property & Normalized 6.3$\mu$m & Absolute flux& Normalized 6.3$\mu$m & Absolute flux& Normalized 6.3$\mu$m & Absolute flux }
\startdata
$T_\mathrm{eff}$ (K) & 1500.0 & 1500.0 & 1400.0 & 1500.0 & 1500.0 & 1450.0 \\
log($g$) & 4.5 & 4.5 & 4.5 & 4.0 & 5.0 & 4.5 \\
$[M/H]$ & 0.0 & 0.0 & 0.5 & 0.5 & 3.16 & 1.0 \\
C/O & - & - & 0.5 & 0.5 & 0.8 & 0.65 \\
$f_\mathrm{sed}$ & - & - & 3.0 & 3.0 & - & - \\
\tableline
$\chi^2_{\nu}$\tablenotemark{a}  & 1094.31 & 1311.68 & 881.36 & 623.06 & 567.02 & 554.65 \\
\enddata
\tablenotetext{a}{$N$ = 8727.}
\end{deluxetable*}

\begin{deluxetable*}{c|cc|cc|cc}
\tablecaption{Best-fit grid values for 2148 \label{table:J2148:fits_v2}}
\tablehead{\colhead{} & \colhead{BT Settl}   &\colhead{BT Settl}  & \colhead{Sonora Diamondback} & \colhead{Sonora Diamondback} & \colhead{Exo-REM} & \colhead{Exo-REM} \\
Property & Normalized 6.3$\mu$m & Absolute flux& Normalized 6.3$\mu$m & Absolute flux& Normalized 6.3$\mu$m & Absolute flux }
\startdata
$T_\mathrm{eff}$ (K) & 1500.0 & 1500.0 & 1200.0 & 1500.0 & 1350.0 & 1550.0 \\
log($g$) & 4.0 & 3.5 & 3.5 & 5.5 & 4.0 & 5.0 \\
$[M/H]$ & 0.0 & 0.0 & 0.0 & 0.0 & 10.0 & 1.0 \\
C/O & - & - & 0.5 & 0.5 & 0.3 & 0.8 \\
$f_\mathrm{sed}$ & - & - & 1.0 & 1.0 & - & - \\
\tableline
$\chi^2_{\nu}$\tablenotemark{a} & 4910.56 & 4440.46 & 1550.8 & 4082.93 & 912.97 & 1907.14 \\
\enddata
\tablenotetext{a}{$N$ = 8727.}
\end{deluxetable*}

\section{Radiative-Convective Equilibrium Models}\label{sec:rce}

We first compare the observed spectra of J0624 and J2148 to three different grids of self-consistent 1-D models: BT-Settl \citep{allard:2012,Allard2012_SVO}, Sonora-Diamondback \citep[Version 2][]{morley:2024,morley_2024_12735103}, and 2021-era Exo-REM \citep{baudino:2024, exoremYGP2021}. Each grid uses a different set of opacity sources, model resolution, radiative transfer algorithm, cloud parameterization, and grid dimensions. We show the range, dimensions, and sampling of the grids in Table \ref{tab:model_grids}.

We compare each model to the observations by first convolving the model to the approximate instrumental resolution and then summing the flux at the wavelengths associated with each pixel or wavelength bin. We compare to the absolute flux from the observations, in which we scale the observed fluxes to the distance of the objects and assume a radius of 0.95 $R_{\rm Jup}$ for each object. We also compare the grids to the normalized fluxes at 6.3~$\mu$m to account for uncertainties in the object radii and distances. After either treatment, a $\chi^2$ is calculated.

\subsection{1D-RCE Grid Comparison Results}

Figures~\ref{fig:J0624_spec} and \ref{fig:J2148_spec} show the best-fit models from each grid to the observations of J0624 and J2148, respectively. The best-fit parameters from each grid are listed in Table~\ref{table:J0624:fits_v2} and \ref{table:J2148:fits_v2}, along with the $\chi^2$ goodness-of-fit.

In general, the 1D-RCE model grids do not replicate the observed fluxes well. This, combined with the high precision of the data, result in $\chi^2$ values in the hundreds to thousands. The best-fitting model grid is Exo-REM with $\chi^2$ values in the 500s for J0624 and between 900 and 2000 for J2148. The Sonora Diamondback models fit the second-best while BT-Settl had the highest $\chi^2$ values. These results were consistent between J0624 and J2148, as well as whether the spectral comparison was between absolute or normalized fluxes. In general, the grids with the more sampling of parameter space (i.e., closer space grid points) faired best. For example, Exo-REM's flexibility with composition via both [M/H] and C/O free parameters, likely helped fit the observations better.

In each fit to the observations, the grids often fit a portion of the spectrum well \citep[e.g.,][]{petrus:2024}, but miss large features of the observed spectra with no model adequately fitting the whole wavelength range. The shape of the NIR continuum regions (J-, H-, K-bands) are particularly poorly fit. Historically, the 4\,$\mu$m continuum region has been especially hard to fit \citep{stephens:2009}, but these objects seem to give poor fits in the other continuum ranges (e.g., 1.7, 2.2\,$\mu$m) to the grids, perhaps because our targets were chosen for their unique cloudiness.

Furthermore, no 1D-RCE model from any grid fit the MIR absorption seen at around 8--10\,$\mu$m. A spectral feature of this strength is likely caused by abundant small-particle silicates, which forward models general have difficulty producing self-consistently \citep{cushing:2006, luna:2021}. For example, the Sonora Diamondback models calculate the particle size from the \cite{ackerman:2001} model with a given fixed $f_{\rm sed}$ and a fixed log-normal particle size distribution of 2 \citep{morley:2024}. BT-Settl and Exo-REM both use different simplified versions of the \cite{rossow:1978} microphysical cloud model to self-consistently model the particle sizes \citep{allard:2012,charnay:2018}. In all cases, these forward models either predict too large of particles or wide size distributions, which may wash out the silicate feature, supporting the use of more complex or flexible particle size parameterizations \citep{powell:2018}.

Despite the relatively poor fits to the data, the best-fitting models in each case often agree across grids. For both objects, when comparing to the absolute fluxes, each model grid gives a best-fit $T_{\rm eff}$ within 50~K of 1500~K, which agrees well with the L5 and L6-L6.5 spectral type of J0624 and J2148, respectively \citep{stephens:2009}. This is not too surprising, since the observations cover the vast majority of the outgoing flux from each object, and thus the mean flux level will be highly sensitive to $T_{\rm eff}$ given an assumed radius, which the grids ought to agree on.

For BT-Settl and Exo-REM, the best-fit $T_{\rm eff}$ agreed to within 50~K between normalized and absolute flux fits for J0624 and to within 200~K for J2148. The Sonora Diamondback grid consistently preferred a lower $T_{\rm eff}$ in normalized flux fits relative to absolute flux fits. We hypothesize that this is an attempt by the grid to produce a redder spectrum, more in agreement with these cloudy L-dwarfs. 

\begin{figure*}[h]
\gridline{\fig{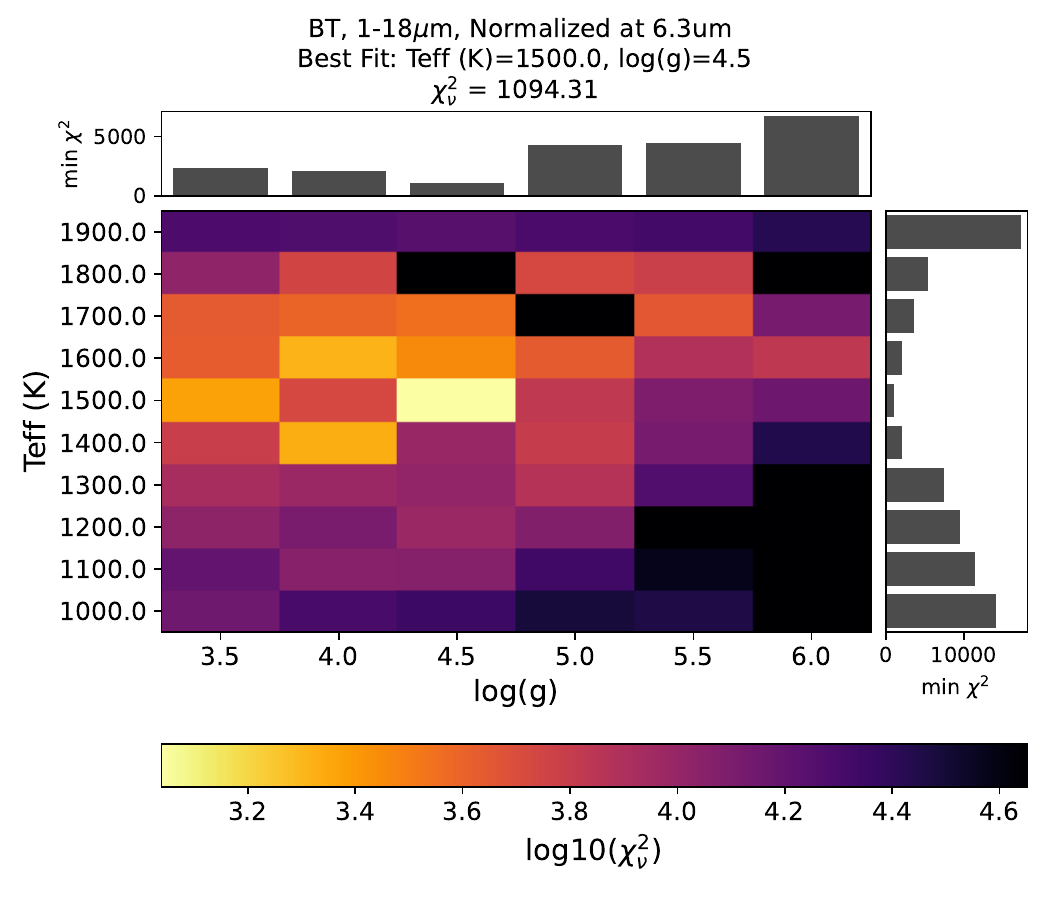}{0.45\textwidth}{(J0624)}
          \fig{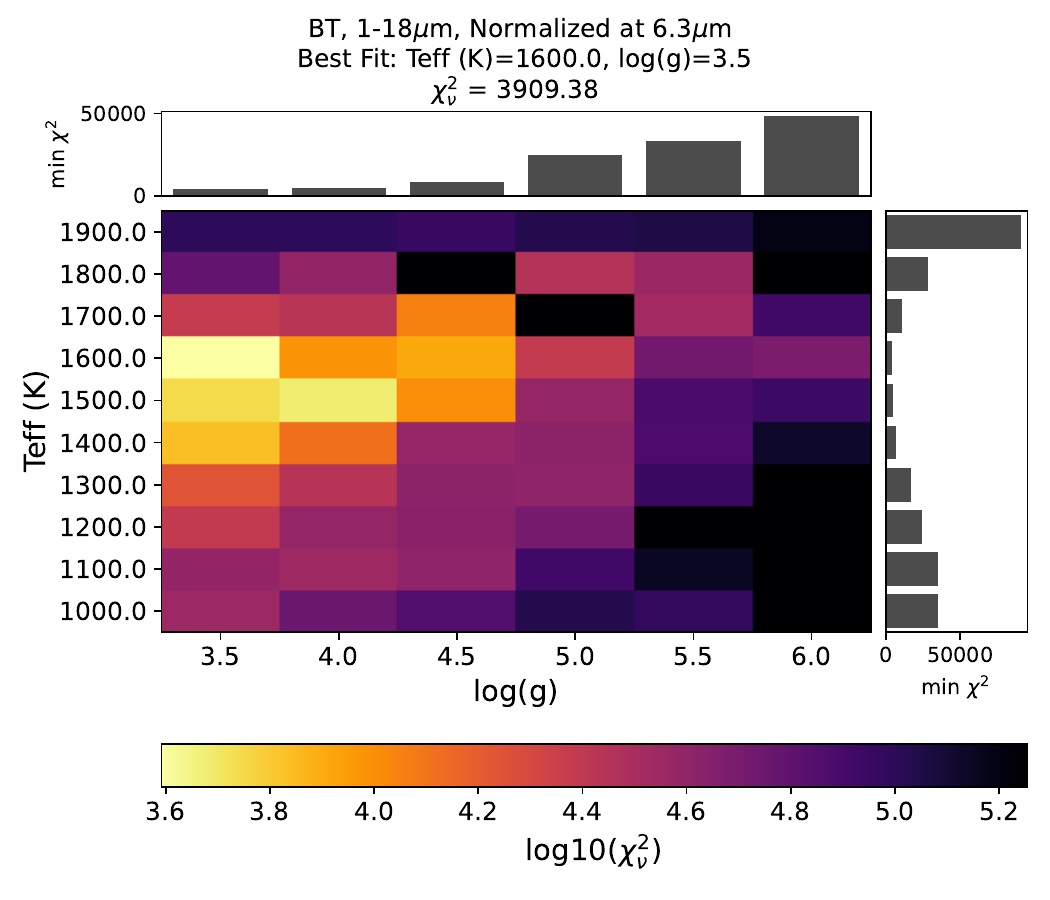}{0.45\textwidth}{(J2148)}
          }
\gridline{\fig{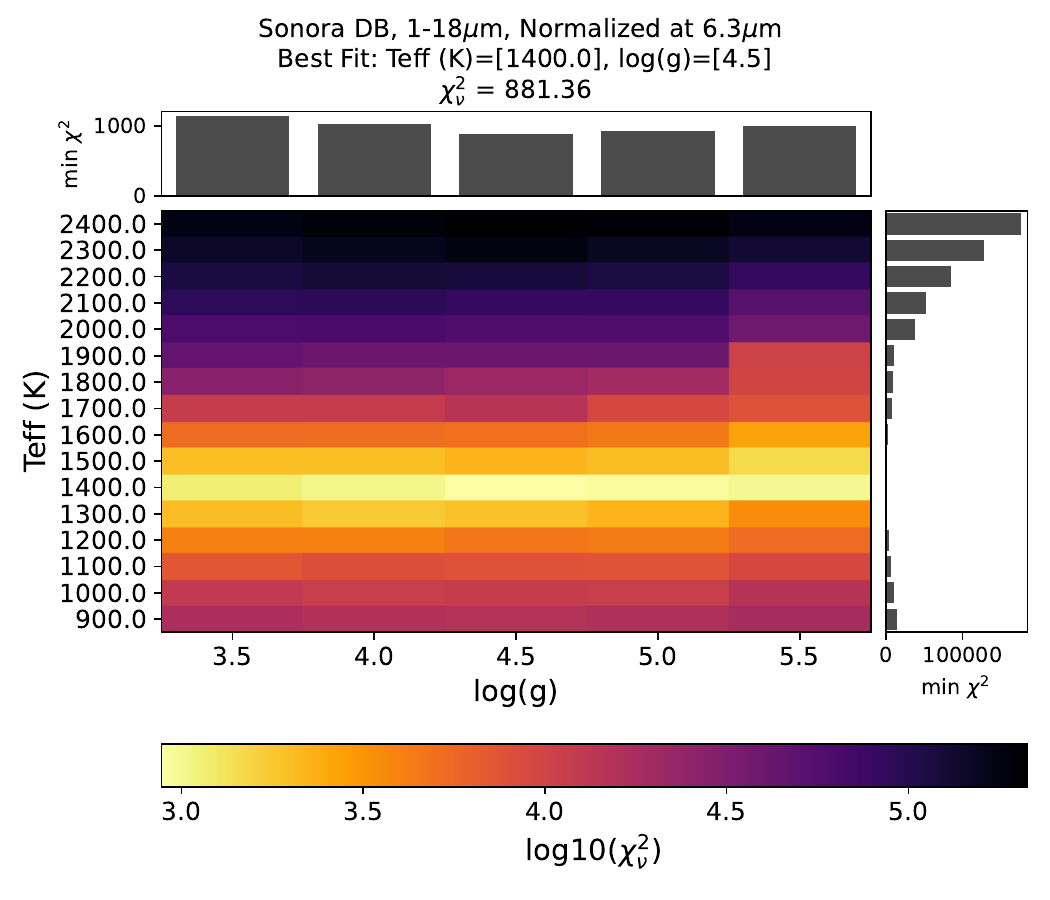}{0.45\textwidth}{(J0624)} 
          \fig{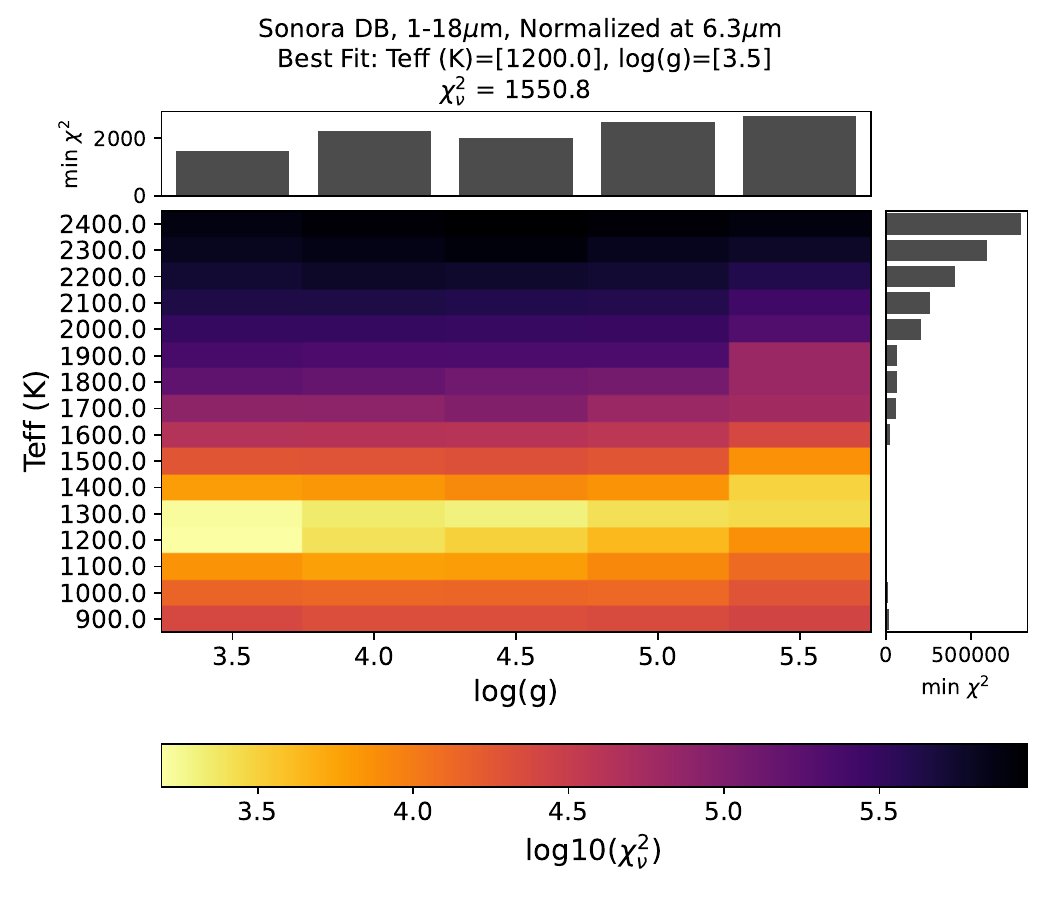}{0.45\textwidth}{(J2148)}
          }
\gridline{\fig{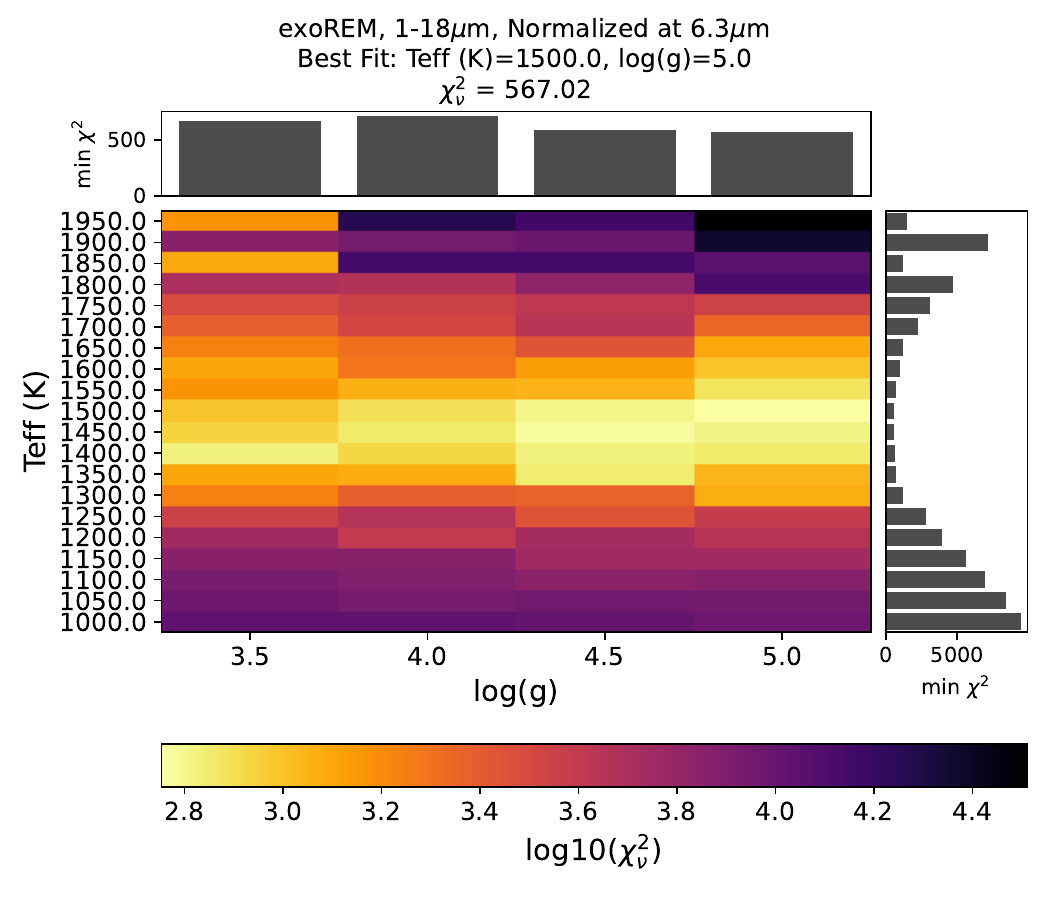}{0.45\textwidth}{(J0624)}
          \fig{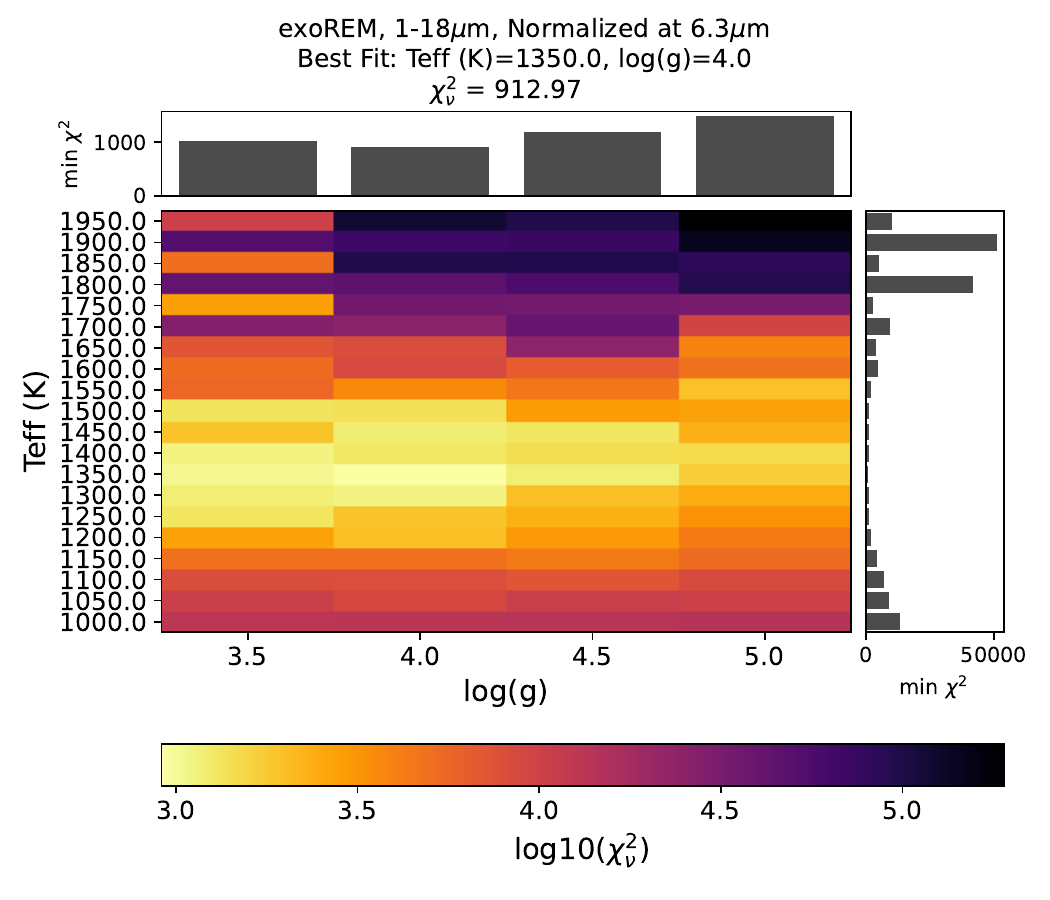}{0.45\textwidth}{(J2148)}
          }

\caption{$\chi^2$ maps for the T$_{\rm{eff}}$ and log10(g$_{cgs}$) for each grid in our comparison to the J0624 (left) and J2148 (right) 1-18$\mu$m, R$\sim$3,000 data. Bar charts above and to the right of the $\chi^2$ maps show the minimum $\chi^2$ found in that that row or column. \label{fig:chi_teff_logg}}
\end{figure*}

\begin{figure*}[h]
\gridline{\fig{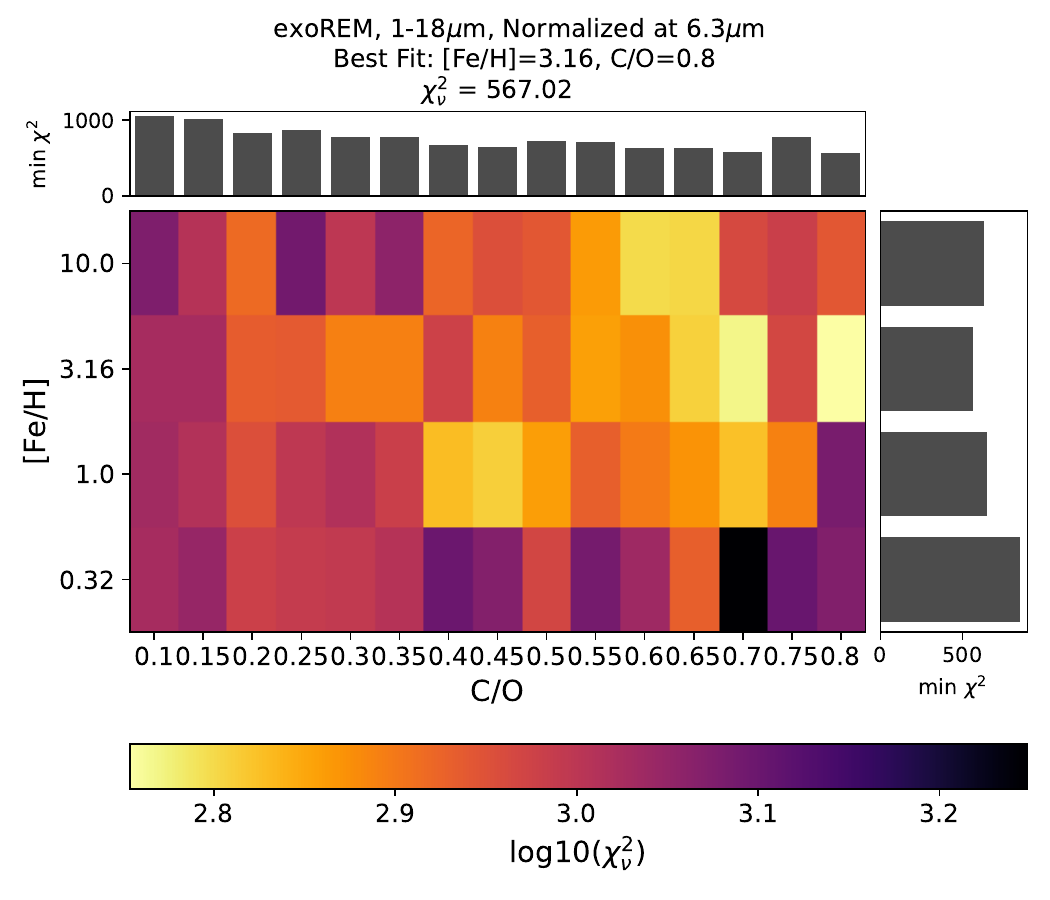}{0.5\textwidth}{(J0624)}
          \fig{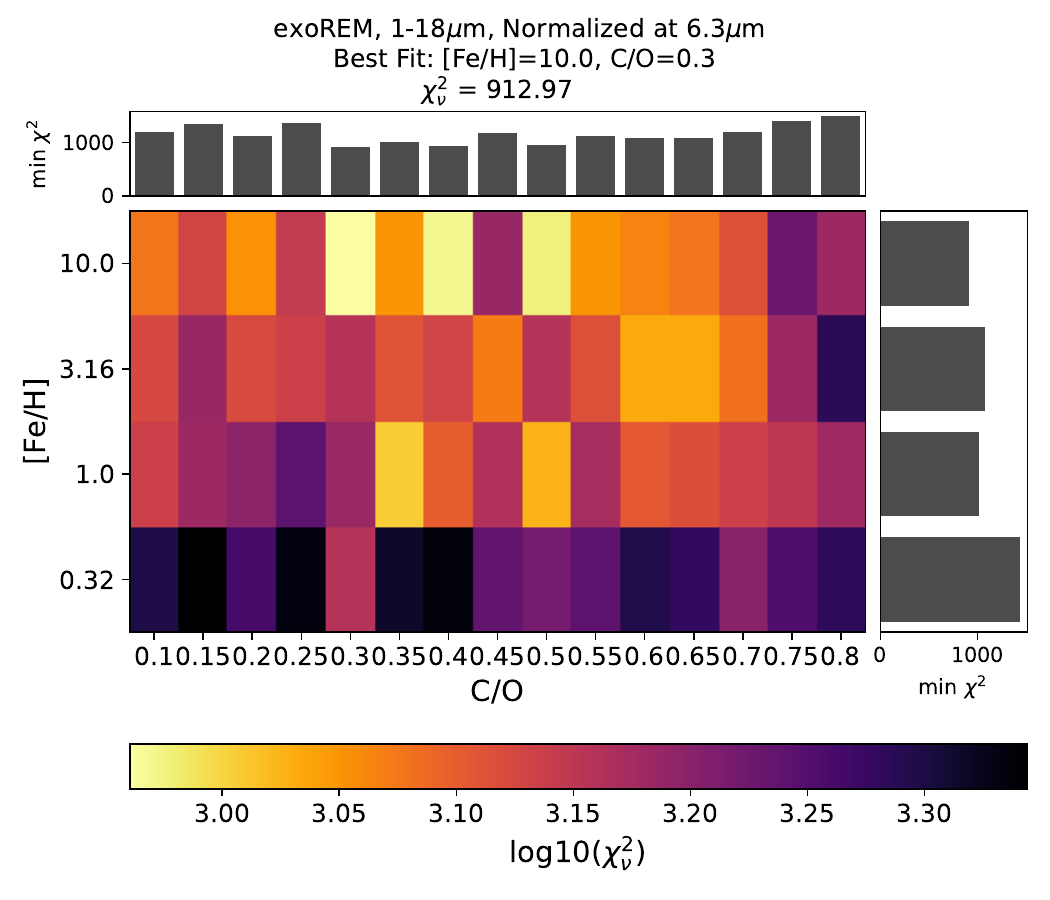}{0.5\textwidth}{(J2148)}
          }
\caption{{$\chi^2$ map comparing [M/H] and C/O from the exoREM grid for fits to J0624 (left) and J2148 (right). Bar charts above and to the right of the $\chi^2$ maps show the minimum $\chi^2$ found in that that row or column. Note the lack of a clearly constrained region of parameter space.}\label{fig:chi_comp}}
\end{figure*}

Maps of how $\chi^2$ varies between $\log{(g)}$ and $T_{\rm eff}$ in the normalized flux fits are shown in Figure~\ref{fig:chi_teff_logg}, indicating the temperature can be constrained, while $\log{(g)}$ less so. Interestingly, the same is not clearly true for [M/H] and C/O. Exo-REM was the only grid to directly vary these two parameters and the $\chi^2$ map shown in Figure~\ref{fig:chi_comp} does not show a clear range that is well constrained by the grid, despite this being the best-fitting grid for each object. This suggests equilibrium [M/H] and C/O compositions are inadequately describing the chemistry in these two objects, potentially implying disequilibrium chemistry, non-solar elemental abundances, or both.

\begin{figure*}[h]
\gridline{\fig{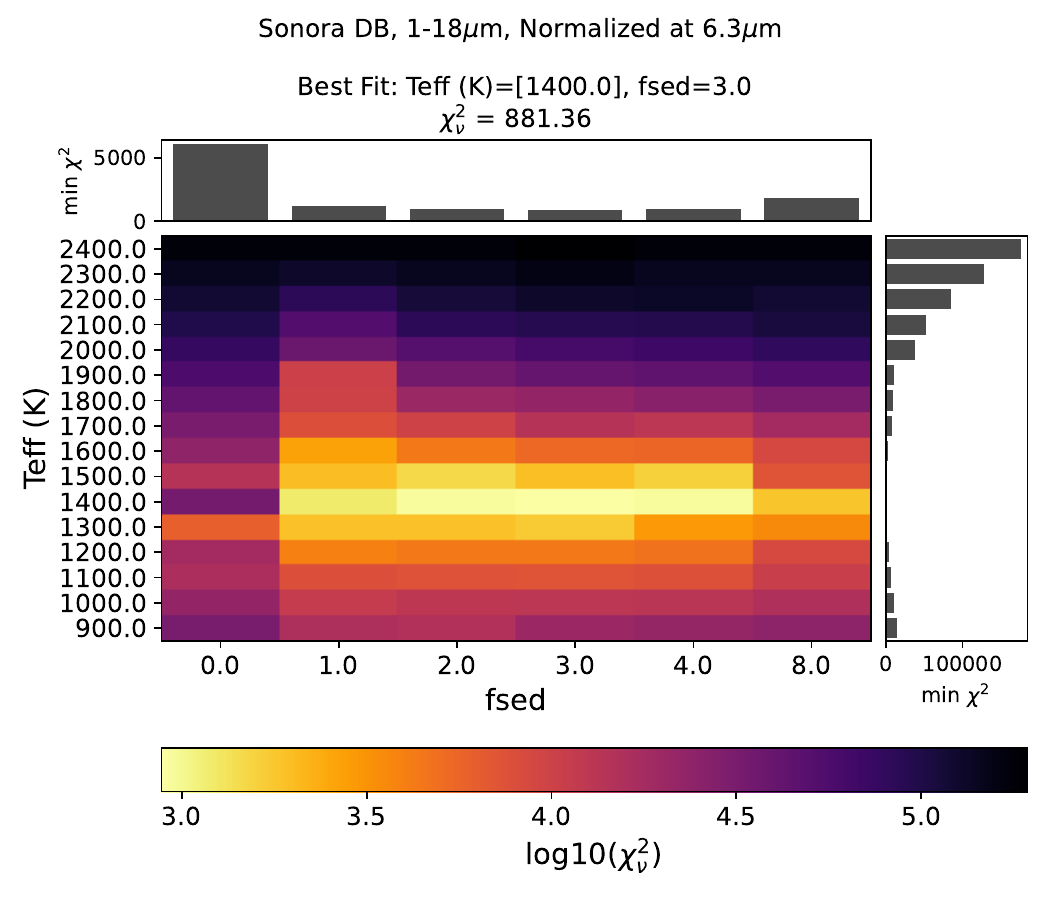}{0.5\textwidth}{(J0624)}
          \fig{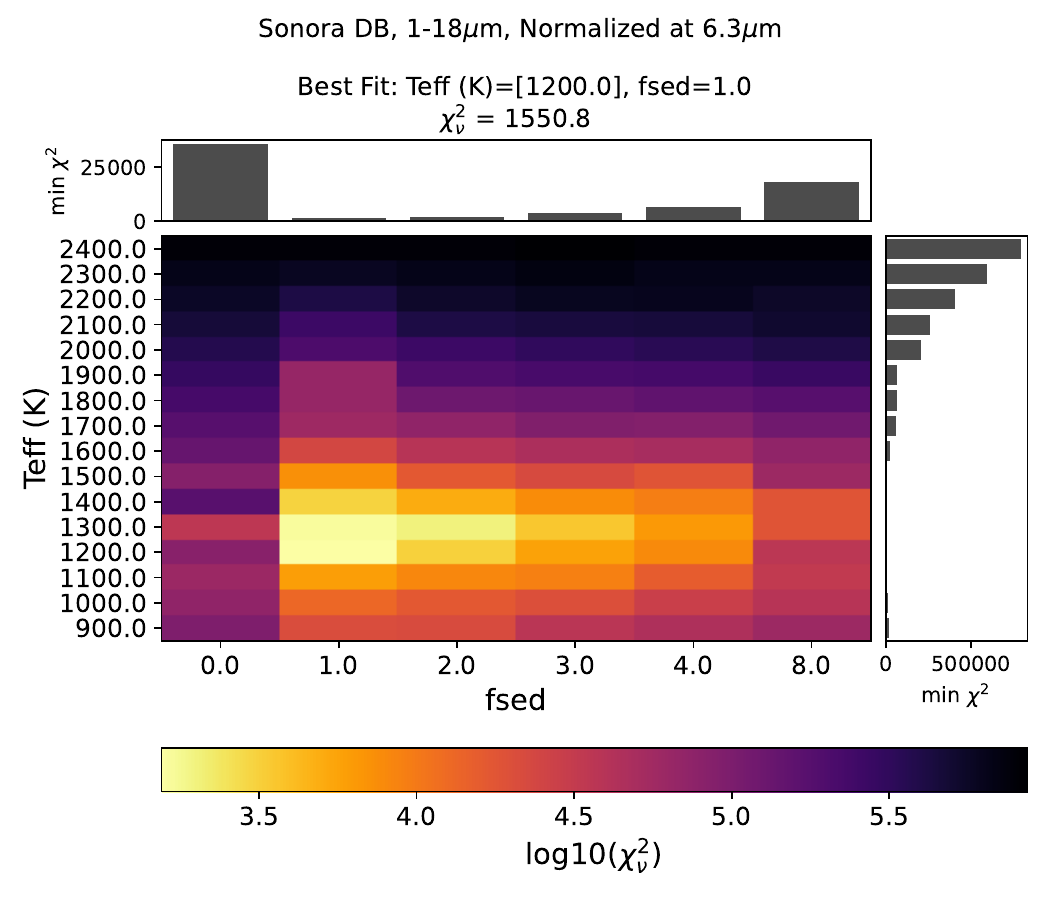}{0.5\textwidth}{(J2148)}
          }    
\caption{T$_{\rm{eff}}$ versus $f_{\rm sed}$ $\chi^2$ map for J0624 (left) and J2148 (right) from Sonora DB. Bar charts above and to the right of the $\chi^2$ maps show the minimum $\chi^2$ found in that that row or column. Fits for J0624 cluster around $f_{\rm sed}=3.0$, while the fits clearly prefer $f_{\rm sed}=1.0$ for J2148. This implies more lofted clouds in J2148 compared to J0624.\label{fig:chi_fsed}}
\end{figure*}

Figure~\ref{fig:chi_fsed} shows the $\chi^2$ map for $T_{\rm eff}$ and $f_{\rm sed}$. For both J0624 and J2148, the Sonora Diamondback grid prefers cloudy models to cloud-free models ($f_{\rm sed}=0$) and the range of preferred $f_{\rm sed}$ for J2148 is markedly lower than for J0624, indicating a more vertically-extended cloud layer. This is also seen in the goodness-of-fit: in both absolute and normalized flux comparisons, a lower $f_{\rm sed}$ is fit for J2148 than for J0624. This interpretation is consistent with a qualitative understanding of the spectra, i.e., J2148 appears redder with a larger MIR spectral feature likely caused by a more extended cloud of smaller grains, as well as with the retrieval results as is described below in Section~\ref{sec:retrievals}.

\section{Low-Resolution Retrieval Methods}\label{sec:retrievals}

\begin{deluxetable}{ll}\label{tab:priors}
\tablecaption{Uniform prior constraints adopted in the pRT retrieval framework.}
\tablehead{
\colhead{Parameter} &
\colhead{Prior}
}
\startdata
$\log g~(\log_{10}[\mathrm{cm~s^{-2}}])$          & $\mathcal{U}(3.0,\ 6.0)$ \\
$R_{\rm pl}~(R_{\rm J})$                          & $\mathcal{U}(0.733,\ 1.222)$ \\
$T_{\rm int}~(\mathrm{K})$                        & $\mathcal{U}(750,\ 3000)$ \\
$T_3$                                             & $\mathcal{U}(0,\ 1)$ \\
$T_2$                                             & $\mathcal{U}(0,\ 1)$ \\
$T_1$                                             & $\mathcal{U}(0,\ 1)$ \\
$\alpha$                                          & $\mathcal{U}(1,\ 2)$ \\
$\log\delta$                                      & $\mathcal{U}(0,\ 1)$ \\
$\log P_{\rm quench}~(\log_{10}[\mathrm{bar}])$   & $\mathcal{U}(-6,\ 3)$ \\
$\mathrm{[M/H]}~(\mathrm{dex})$                  & $\mathcal{U}(-1,\ 2)$ \\
C/O                                               & $\mathcal{U}(0.1,\ 1.5)$ \\
$\sigma_{\rm lnorm}$\tablenotemark{a}                             & $\mathcal{U}(1.05,\ 3.0)$ \\
$b_{\rm hans}$\tablenotemark{b}                                    & $\mathcal{U}(0.001,\ 0.5)$ \\
$\log K_{zz}~(\log_{10}[\mathrm{cm^2~s^{-1}}])$   & $\mathcal{U}(5,\ 13)$ \\
$f_{\rm sed}$                                     & $\mathcal{U}(0.01,\ 10.01)$ \\
$\log X_{\rm cb}$\tablenotemark{c}                        & $\mathcal{U}(-6,\ 0)$ \\
$\log P_{\rm base}~(\log_{10}[\mathrm{bar}])$\tablenotemark{c}   & $\mathcal{U}(-5,\ 3)$ \\
\enddata
\tablenotetext{a}{Used if a log-normal particle size distribution is assumed.}
\tablenotetext{b}{Used if a Hansen particle size distribution is assumed.}
\tablenotetext{c}{$\log X_{\rm cb}$ and $\log P_{\rm base}$ priors were the same for each individual cloud species.}
\end{deluxetable}


To find a better fit to our observations than those provided by the self-consistent modeling, we analyzed the data in a retrieval framework, which includes flexible parameterizations for the temperature-pressure structure, the chemical abundances, and the clouds. We performed a suite of retrievals with petitRADTRANS \citep{molliere:2019,nasedkin:2024,nasedkin:2025} on the dataset from 0.98 to 18$\mu$m at R$\sim$100. We chose the low-resolution version of the data to perform the full-wavelength retrievals for computational efficiency in order to test many retrieval set ups and evaluate the inclusion of different cloud species. Our main goal here is to constrain the cloud species responsible for the MIR absorption seen in each object, as well as the cloud crystallinity, particle sizes, and vertical distribution. Our setup is described below with all priors listed in Table~\ref{tab:priors}.

Each retrieval uses the three-zone temperature profile parameterization designed for self-luminous objects in \cite{molliere:2020}. The deep atmosphere is parameterized by an adiabat with an internal temperature, T$_{\rm int}$. The middle atmosphere is then parametrized by an Eddington approximation where the optical depth, $\tau$, is parameterized as:

\begin{equation}
    \tau = \delta P^\alpha,
\end{equation}

\noindent where $\delta$ and $\alpha$ are free parameters and $P$ is the pressure. The middle atmosphere extends from $\tau=0.1$ down to the radiative convective boundary (determined from an interpolated grid in petitRADTRANS's chemistry table). Finally, the upper atmosphere ($\tau<0.1$) is parameterized by a three-point spline from three temperatures, $T1$, $T2$, $T3$, spread equidistant to the top of the model. 

For the chemistry, we assume chemical equilibrium abundances from an interpolated grid of [M/H] and C/O assuming solar abundances according to \cite{asplund:2009}, with an additional parameter, $P_{\rm quench}$, which quenches the abundances of H$_2$O, CO, CO$_2$, and CH$_4$ below this pressure to their values at $P_{\rm quench}$. This is different than a timescale-based quenching using $K_{\rm{zz}}$ that gives each molecule a different quench pressure based on their chemical conversion timescale \citep[e.g.,][]{prinn:1977}. 

From these abundances, opacities were calculated for H$_2$O \citep{polyansky:2018}, CO \citep{li:2015}, CH$_4$ \citep{yurchenko:2014}, CO$_2$ \citep{yurchenko:2020}, NH$_3$ \citep{coles:2019}, HCN \citep{barber:2014}, H$_2$S \citep{azzam:2016}, FeH \citep{wende:2010}, PH$_3$ \citep{Sousa-Silva:2015}, Na \citep{piskunov:1995}, K \citep{piskunov:1995}, TiO \citep{mckemmish:2019}, VO \citep{mckemmish:2016}, and SiO \citep{barton:2013}. Additionally, H$_2$-H$_2$ \citep{borysow:2002} and H$_2$-He \citep{borysow:1986} collision-induced opacity was added, as well as H$_2$ and He Rayleigh scattering.

As mentioned above, various cloud species and set-ups were tested, summarized in Table~\ref{tab:cloud_species}. We included different combinations of opacity from  Fe(a) \citep{henning:1996}, SiO(a) \citep{palik1985handbook,kitzmann:2018}, SiO$_2$(a,c) \citep{palik1985handbook,henning:1997,zeidler:2013,kitzmann:2018}, MgSiO$_3$(a) \citep{jaeger:2003}, Mg$_2$SiO$_4$(a) \citep{jaeger:2003}, and Al$_2$O$_3$(c) \citep{koike:1995}, with a or c denoting amorphous or crystalline structures, respectively. We also tested SiO$_2$ polymorphs cristobalite and tridymite, which are chemically identical to quartz and amorphous SiO$_2$ but exhibit a different crystalline structure \citep{moran:2024}. In addition to these clouds, we also test MgSiO$_3$(c) \citep{scott:1996,jaeger:1998}, Mg$_2$SiO$_4$(c) \citep{servoin:1973}, Mg$_{0.5}Fe_{0.5}$SiO$_3$(a) \citep{jaeger:1994}, and MgFeSiO$_4$(a) \citep{dorschner:1995,kitzmann:2018b} in our best-fit optimization framework, described below in Section~\ref{sec:CMA_methods}. These cloud species were selected based on their expected dominance of the cloud mass at the temperatures found in L-dwarf atmospheres \citep{wakeford:2017a}. 

All cloud opacity for the retrievals were calculated using the Distributed Hollow Sphere (DHS) approximation \citep{min:2005} as opposed to simple solid sphere Mie scattering.\footnote{We tested our best-fitting retrieval scenarios for each object with the Mie approximation and did not find it was preferred.} We use the 1D condensate \citep{ackerman:2001} model in pRT, where vertical mixing parameterized by an eddy diffusion coefficient, $K_{\rm zz}$, and gravitational settling parameterized by a sedimentation efficiency, $f_{\rm sed}$ compete to determine the vertical extent and particle sizes of the clouds. Our fiducial setup used a single $f_{\rm sed}$ to describe different cloud species, but we also tested species-specific $f_{\rm sed}$ values. The fiducial retrieval setup also assumes crystalline/quartz SiO$_2$ and amorphous Fe, MgSiO$_3$, Mg$_2$SiO$_4$.

The fiducial setup also assumed a log-normal particle size distribution with a free parameter for its width given by $\sigma_{\rm lnorm}$. Additionally, we tested a Hansen particle distribution \citep{hansen:1971}, which has been shown to potentially better describe the particle size distribution in substellar atmospheres \citep{burningham:2021}, as it more heavily weights small, sub-micron particles with a longer tail of less frequent larger particles. For the Hansen distribution we parameterize the width of particle sizes with the $b_{\rm hans}$ parameter, with the particle sizes and vertical distribution then calculated with $K_{\rm{zz}}$ and $f_{\rm sed}$ according to Appendix D of \cite{nasedkin:2024b}. Lastly, the radius and $\log{(g)}$ of the objects are freely fit parameters. The distance is set to the Gaia distances listed in Table~\ref{table:properties}.

The retrieval is run in a nested sampling framework \citep{fer08,fer09} with 4000 live points for J0624 and 5000 live points for J2148.\footnote{The reason for the difference in live points between the two objects is simple human error. Results were found to be consistent at and above 2000 live points.} An error inflation term was also included in the fits that scaled the observed uncertainties to achieve $\chi_\nu^2 \approx 1.0$. This effectively increased the width of the retrieved posteriors to better match the disagreement between the retrieved model and the observations.

\begin{deluxetable*}{lccc}
\tablecaption{Cloud species explored in the retrieval and CMA-ES optimization frameworks.\label{tab:cloud_species}}
\tablehead{
\colhead{Cloud Species} &
\colhead{Retrieval} &
\colhead{CMA-ES} &
\colhead{Optical Constants}
}
\startdata \small
Fe(a)                                      & \checkmark & \checkmark & \citet{pollack:1994} \\
SiO(a)                                     & \checkmark & \checkmark & \citet{palik1985handbook,kitzmann:2018} \\
SiO$_2$(a)                                 & \checkmark & \checkmark & \citet{palik1985handbook,henning:1997,zeidler:2013,kitzmann:2018} \\
SiO$_2$(c) (quartz)                        & \checkmark & \checkmark & \citet{palik1985handbook,henning:1997,zeidler:2013,kitzmann:2018} \\
SiO$_2$(c) (cristobalite)                  & \checkmark     & \checkmark         & \citet{palik1985handbook,koike:2013,moran:2024} \\
SiO$_2$(c) (tridymite)                     & \checkmark     & \checkmark         & \citet{palik1985handbook,sitarz:2000,moran:2024} \\
MgSiO$_3$(a)                               & \checkmark & \checkmark & \citet{jaeger:2003} \\
MgSiO$_3$(c)                               & ---         & \checkmark & \citet{scott:1996,jaeger:1998} \\
Mg$_{0.5}$Fe$_{0.5}$SiO$_3$(a)             & ---         & \checkmark & \citet{jaeger:1994} \\
Mg$_2$SiO$_4$(a)                           & \checkmark & \checkmark & \citet{jaeger:2003} \\
Mg$_2$SiO$_4$(c)                           & ---         & \checkmark & \citet{servoin:1973} \\
MgFeSiO$_4$(a)                             & ---         & \checkmark & \citet{dorschner:1995,kitzmann:2018b} \\
Al$_2$O$_3$(c)                             & \checkmark & \checkmark & \citet{koike:1995} \\
\enddata
\tablecomments{The fiducial retrieval framework uses Fe(c), SiO(a), quartz SiO$_2$(c), MgSiO$_3$(a), and Mg$_2$SiO$_4$(a). Additional retrieval tests include amorphous SiO$_2$ and the SiO$_2$ polymorphs cristobalite and tridymite. The CMA-ES optimization additionally explores crystalline MgSiO$_3$ and Mg$_2$SiO$_4$, Fe-bearing silicates (Mg$_{0.5}$Fe$_{0.5}$SiO$_3$ and MgFeSiO$_4$) in lieu of Mg-only silicates, and all allowed combinations of silicate cloud species atop a deep Fe cloud.}
\end{deluxetable*}

\subsection{CMA-ES Cloud Scenario Optimization}\label{sec:CMA_methods}

With our fiducial full-wavelength nested sampling retrieval framework, individual runs can take up to about a week on a cluster of a few dozen cores. While we run several different cloud scenarios with the full setup, this computational inefficiency makes it difficult to evaluate all possible cloud species permutations. We therefore built a custom parameter optimization framework to more quickly quantify the best fitting cloud scenarios, allowing us to search a wider array of cloud species, crystallinities, particle size distributions, and opacity computation method. 

We use the Covariance Matrix Adaptation Evolution Strategy (CMA-ES) using the \texttt{pycma} package \citep{hansen2019pycma}. CMA-ES is an optimization algorithm designed for non-linear, noisy, multimodal, and high-dimensional parameter spaces with broad application to optimization problems \citep{hansen:2001}. We choose CMA-ES as a more robust algorithm for our potentially complex parameter space compared to simpler local optimization algorithms. For each cloud scenario, we repeated the fit five times, each with different starting conditions, taking the best-fitting result. We verified CMA-ES's behavior in the low-dimension limit by fitting [M/H] and C/O with \texttt{petitRADTRANS} to the observed spectrum of J0624 with both CMA-ES and Nelder-Mead algorithms, obtaining the same solution to better than 1\%.

To further balance computational efficiency, we fit only the 5-18$\mu$m portion of the observed spectra, keeping all non-cloud-related parameters fixed to the best-fit solution found from the fiducial full-wavelength nested sampling retrieval. However, we compared the spectra at a higher resolution (R$\sim$1000) compared to the lower-resolution, full-wavelength retrieval (R$\sim$100), leveraging any additional information on the cloud species from the higher resolution enabled by MIRI/MRS.

Compared to the retrieval setup, we also test 1) both amorphous and crystalline MgSiO$_3$ and Mg$_2$SiO$_4$ (rather than just amorphous), 2) both DHS and Mie scattering opacity computation, and 3) every possible combination SiO$_2$, one of MgSi$O_3$ and Mg$_{0.5}$Fe$_{0.5}$SiO$_3$, and one of Mg$_2$SiO$_4$ and MgFeSiO$_4$ on top of a Fe deep cloud base. Cloud species data sources are listed in Table~\ref{tab:cloud_species}.

\section{Low-Resolution Retrieval Results}\label{sec:retrieval_results}

Figures~\ref{fig:J0624_spec} and \ref{fig:J2148_spec} show the best-fit retrieved model to the observations of J0624 and J2148 using the fiducial setup (see above), alongside the best fits from the self-consistent grids. The fits obtained in the retrieval analysis are much better than the self-consistent grids, thanks to the flexibility in the temperature structure, gas chemistry, and cloud properties. All major spectral features are fit, including the shape of the J, H, and K-bands, which the self-consistent grids struggled to fit. Importantly, the full-wavelength retrieval is able to fit the MIR silicate feature between 8-10~$\mu$m, albeit not perfectly. At the R$\sim$100 resolution used for the full-wavelength retrievals, the best-fit J0624 retrieval achieves a $\chi^2/N_{\rm data}$ of 1648.7 without error inflation, compared to 12,913.2 from the best-fitting grid model from Exo-REM. Meanwhile, the higher-resolution MIR-specific CMA-ES fits were able to fit the cloud absorption even better, with a minimum $\chi^2/N_{\rm data}$ of 50.248. A similar story holds for J2148 where the retrieval obtains $\chi^2/N_{\rm data}=1506.4$ compared to 48,560.0 from Exo-REM. The corresponding best-fitting CMA-ES scenario resulted in $\chi^2/N_{\rm data}$ = 96.836.

In summary, the best-fitting models for J0624 contained small, amorphous SiO$_2$ grains. The CMA-ES fits also clearly preferred the inclusion of iron and magnesium-rich silicate grains as well. Additional Al$_2$O$_3$ grains were suggested by the full-wavelength retrieval, while Fe clouds matched the data well in the CMA-ES fits. Likewise for J2148, small SiO$_2$ grains best explain the MIR spectrum, but quartz rather than amorphous SiO$_2$ grains are favored by both the full-wavelength retrieval and the CMA-ES fits. We describe these results in more detail below.

\begin{deluxetable*}{c|c|c|c|c|c|c}
\tablecaption{Retrieved properties of J0624 from retrievals with alternative cloud properties. \label{table:J0624:rets}}
\tablehead{Parameter & Fiducial & Multi-$f_{\rm sed}$ & No SiO$_2$ & No Mg$_x$SiO$_y$ & No Fe & With Al$_2$O$_3$}
\startdata
Live Points & 4000  & 4000 & 4000 & 4000 & 4000 & 4000 \\
ln Z\tablenotemark{a} & 10642.5 & 10632.9$\pm$0.9 & 10643.3 & 10699.5 & 10673.1 & 10663.3$\pm$0.0 \\
$\chi^2_{\nu}$ & 1648.7 & 1691.1 & 1713.0 & 1526.0 & 1541.5 & 1687.4 \\
\hline
$T_{\mathrm{int}}$ & 2202$\pm$13  & 2182$\pm$15 & 2206$\pm$19 & 2252$\pm$77 & 2170$\pm$16 & 2186$\pm$35 \\
$\log{g}$ & 4.85$\pm$0.05  & 4.85$\pm$0.07 & 4.90$\pm$0.05 & 4.48$\pm$0.04 & 4.78$\pm$0.06 & 4.31$\pm$0.05 \\
$R_{\mathrm{Jup}}$ & 0.897$\pm$0.006 & 0.897$\pm$0.003 & 0.896$\pm$0.005 & 0.924$\pm$0.005 & 0.924$\pm$0.007 & 0.937$\pm$0.006 \\
$M_{\rm Jup}$\tablenotemark{b} & $23 \pm 3$ & $23 \pm 3$ & $26 \pm 3$ & $10 \pm 3$ & $21 \pm 3$ & $7 \pm 3$ \\
\hline
$\log{p_{\mathrm{quench}}}$ & $-3.29\pm$0.92  & $-3.77\pm$0.79 & $-2.57\pm$0.70 & $-0.137\pm$0.035 & $-0.172\pm$0.071 & $-0.358\pm$0.030 \\
$\mathrm{M/H}$ & 0.58$\pm$0.03& 0.63$\pm$0.03 & 0.65$\pm$0.02 & 0.70$\pm$0.04 & 0.70$\pm$0.04 & 0.55$\pm$0.03 \\
$\mathrm{C/O}$ & 0.655$\pm$0.002 & 0.657$\pm$0.002 & 0.658$\pm$0.002 & 0.75$\pm$0.02 & 0.685$\pm$0.007 & 0.71$\pm$0.01 \\
\hline
$\sigma_{\mathrm{lnorm}}$ & 2.38$\pm$0.13  & 2.02$\pm$0.16 & 2.13$\pm$0.19 & 2.74$\pm$0.19 & 2.68$\pm$0.21 & 2.76$\pm$0.16 \\
$\log{k_{\mathrm{zz}}}$ & 7.77$\pm$0.17  & 7.08$\pm$0.13 & 9.72$\pm$0.57 & 7.50$\pm$0.18 & 7.09$\pm$0.13 & 6.86$\pm$0.13 \\
$f_{\mathrm{sed}}$ & 1.34$\pm$0.29  & - & 1.17$\pm$0.27 & 1.19$\pm$0.17 & 2.04$\pm$0.23 & 4.98$\pm$0.60 \\
$f_{\mathrm{sed,SiO_2}}$ & -  & 1.15$\pm$0.24 & - & - & - & - \\
$f_{\mathrm{sed,MgSiO_3}}$ & -  & 4.13$\pm$0.96 & - & - & - & - \\
$f_{\mathrm{sed,Mg_2SiO_4}}$ & - & 1.25$\pm$0.64 & - & - & - & - \\
$f_{\mathrm{sed,Fe}}$ & -  & 4.42$\pm$0.59 & - & - & - & - \\
\hline
$\log{X_{\mathrm{cb,SiO_2}}}$ & -3.07$\pm$0.11  & -3.60$\pm$0.14 & - & -3.84$\pm$0.11 & -3.64$\pm$0.11 & -3.00$\pm$0.14 \\
$\log{P_{\mathrm{base,SiO_2}}}$ & -1.370$\pm$0.085  & -1.06$\pm$0.16 & - & -0.594$\pm$0.033 & -0.823$\pm$0.031 & -1.009$\pm$0.015 \\
$\log{X_{\mathrm{cb,MgSiO_3}}}$ & -3.99$\pm$0.87  & -3.79$\pm$0.69 & -0.35$\pm$0.46 & - & -3.01$\pm$0.12 & -4.89$\pm$0.67 \\
$\log{P_{\mathrm{base,MgSiO_3}}}$ & -3.45$\pm$0.78  & -3.69$\pm$0.84 & -3.78$\pm$0.46 & - & -1.025$\pm$0.023 & -3.2$\pm$1.2 \\
$\log{X_{\mathrm{cb,Mg_2SiO_4}}}$ & -4.90$\pm$0.61  & -4.44$\pm$0.49 & -4.11$\pm$0.92 & - & -4.54$\pm$0.85 & -2.6$\pm$1.5 \\
$\log{P_{\mathrm{base,Mg_2SiO_4}}}$ & -1.85$\pm$0.71 & -3.21$\pm$0.89 & 1.13$\pm$0.98 & - & -3.41$\pm$0.98 & 2.06$\pm$0.52 \\
$\log{X_{\mathrm{cb,Fe}}}$ & -3.34$\pm$0.13  & -3.23$\pm$0.20 & -4.52$\pm$0.46 & -4.28$\pm$0.91 & - & -3.67$\pm$0.82 \\
$\log{P_{\mathrm{base,Fe}}}$ & -2.059$\pm$0.096  & -2.10$\pm$0.15 & 0.9$\pm$2.0 & 1.63$\pm$0.71 & - & 0.76$\pm$0.17 \\
$\log{X_{\mathrm{cb,Al_2O_3}}}$  & - & - & - & - & - & -4.85$\pm$0.71 \\
$\log{P_{\mathrm{base,Al_2O_3}}}$  & - & - & - & - & - & -3.3$\pm$1.1 \\
\enddata
\tablenotetext{a}{The log-Bayesian Evidence of the given retrieval.}
\tablenotetext{b}{Calculated from the retrieved surface gravity and radius.}
\end{deluxetable*}

\begin{deluxetable*}{c|c|c|c|c|c|c}
\tablecaption{Retrieved properties of J0624 from retrievals with different SiO$_2$ polymorphs.
\label{table:J0624:rets:polymorphs}}
\tablehead{ & Fiducial (Quartz) & Amorphous* & Tridymite & Cristobalite & SiO(a) & Am. SiO$_2$ (Hansen)}
\startdata
Live Points & 4000 & 4000 & 4000 & 4000 & 4000 & 4000 \\
ln Z & 10642.5 & 10697.3 & $10678.2 \pm 0.0$ & $10690.0 \pm 0.0$ & 10594.1 & $10643.1 \pm 0.0$ \\
$\chi^2_{\nu}$ & 1648.7 & 1427.4 & 1522.6 & 1684.7 & 2575 & 1717.2 \\
\hline
$T_{\mathrm{int}}$ & $2202\pm13$ & $2179\pm18$ & $2175\pm34$ & $2177\pm15$ & $1588\pm18$ & $2154 \pm 16$ \\
$\log{g}$ & $4.854\pm0.048$ & $4.676\pm0.030$ & $4.384\pm0.047$ & $4.591\pm0.042$ & $4.410\pm0.073$ & $4.705 \pm 0.053$ \\
$R_{\mathrm{pl}}$ & $0.897\pm0.006$ & $0.915\pm0.004$ & $0.924\pm0.005$ & $0.919\pm0.005$ & $0.880\pm0.005$ & $0.880 \pm 0.003$ \\
$M_{\rm Jup}$\tablenotemark{b} & 23 $\pm$ 3 & 16$\pm$ 3&  8 $\pm$ 3& 13$\pm$ 3 &  8$\pm$ 3 &
       16$\pm$ 3\\
\hline
$\log{p_{\mathrm{quench}}}$ & $-3.29\pm0.92$ & $-0.020\pm0.032$ & $-0.250\pm0.031$ & $-0.134\pm0.031$ & $0.96\pm0.21$ & $-4.34 \pm 0.87$ \\
$\mathrm{[M/H]}$ & $0.575\pm0.025$ & $0.770\pm0.026$ & $0.599\pm0.030$ & $0.703\pm0.030$ & $0.724\pm0.041$ & $0.603 \pm 0.028$ \\
$\mathrm{C/O}$ & $0.6551\pm0.0015$ & $0.7505\pm0.0086$ & $0.728\pm0.015$ & $0.728\pm0.013$ & $0.8275\pm0.0043$ & $0.6563 \pm 0.0017$ \\
\hline
$\sigma_{\mathrm{lnorm}}$ & $2.38\pm0.13$ & $2.85\pm0.11$ & $2.859\pm0.094$ & $2.84\pm0.10$ & $1.40\pm0.15$ & - \\
$b_{\mathrm{hans}}$ & - & - & - & - & - & $0.442 \pm 0.037$ \\
$\log{k_{\mathrm{zz}}}$ & $7.77\pm0.17$ & $6.86\pm0.12$ & $7.08\pm0.14$ & $6.709\pm0.080$ & $10.406\pm0.091$ & $7.32 \pm 0.10$ \\
$f_{\mathrm{sed}}$ & $1.34\pm0.29$ & $4.55\pm0.63$ & $4.03\pm0.46$ & $4.50\pm0.47$ & $7.7\pm1.6$ & $2.22 \pm 0.42$ \\
$f_{\mathrm{sed,SiO_2}}$ & - & - & - & - & - & - \\
$f_{\mathrm{sed,MgSiO_3}}$ & - & - & - & - & - & - \\
$f_{\mathrm{sed,Mg_2SiO_4}}$ & - & - & - & - & - & - \\
$f_{\mathrm{sed,Fe}}$ & - & - & - & - & - & - \\
\hline
$\log{X_{\mathrm{cb,SiO}}}$ & - & - & - & - & $-3.1\pm1.3$ & - \\
$\log{P_{\mathrm{base,SiO}}}$ & - & - & - & - & $-0.9\pm2.7$ & - \\
$\log{X_{\mathrm{cb,SiO_2}}}$ & $-3.07\pm0.11$ & $-2.88\pm0.18$ & $-3.31\pm0.16$ & $-2.83\pm0.13$ & - & $-3.23 \pm 0.12$ \\
$\log{P_{\mathrm{base,SiO_2}}}$ & $-1.370\pm0.085$ & $-0.912\pm0.025$ & $-0.685\pm0.024$ & $-1.015\pm0.017$ & - & $-0.98 \pm 0.10$ \\
$\log{X_{\mathrm{cb,MgSiO_3}}}$ & $-3.99\pm0.87$ & $-3.8\pm1.7$ & $-4.2\pm1.1$ & $-4.66\pm0.82$ & $-3.0\pm1.3$ & $-3.82 \pm 0.88$ \\
$\log{P_{\mathrm{base,MgSiO_3}}}$ & $-3.45\pm0.78$ & $1.86\pm0.66$ & $-3.29\pm0.93$ & $-3.4\pm1.0$ & $-0.9\pm2.7$ & $-3.4 \pm 1.0$ \\
$\log{X_{\mathrm{cb,Mg_2SiO_4}}}$ & $-4.90\pm0.61$ & $-3.6\pm1.7$ & $-4.64\pm0.83$ & $-4.68\pm0.81$ & $-3.1\pm1.3$ & $-4.61 \pm 0.70$ \\
$\log{P_{\mathrm{base,Mg_2SiO_4}}}$ & $-1.85\pm0.71$ & $1.87\pm0.64$ & $-3.3\pm1.0$ & $-3.47\pm0.95$ & $-1.0\pm2.7$ & $-3.24 \pm 0.96$ \\
$\log{X_{\mathrm{cb,Fe}}}$ & $-3.34\pm0.13$ & $-4.84\pm0.75$ & $-2.6\pm1.3$ & $-4.93\pm0.69$ & $-3.0 \pm 1.3$ & $-4.86 \pm 0.60$ \\
$\log{P_{\mathrm{base,Fe}}}$ & $-2.059\pm0.096$ & $-3.1\pm1.3$ & $1.21\pm0.34$ & $-3.53\pm0.88$ & $-1.2\pm2.6$ & $-2.68 \pm 0.88$ \\
\enddata
\tablenotetext{a}{The log-Bayesian Evidence of the given retrieval.}
\tablenotetext{b}{Calculated from the retrieved surface gravity and radius.}
\tablenotetext{*}{Highest evidence full-wavelength retrieval scenario for J0624.}
\end{deluxetable*}

\subsection{J0624 Results}

The median-fit parameters and 1$\sigma$ uncertainties from each retrieval of the J0624 observations are listed in Tables~\ref{table:J0624:rets} and \ref{table:J0624:rets:polymorphs}. The former table compares retrievals with different cloud properties including species-specific $f_{\rm sed}$ values, the absence of SiO$_2$, Mg$_x$SiO$_y$ (i.e., magnesium silicates), and Fe, and the presence of Al$_2$O$_3$. Table~\ref{table:J0624:rets:polymorphs} compares additional retrievals computed in order to identify the specific properties of the SiO$_2$ clouds that appear to fit the observations best. Despite the qualitative quality of the fit to the spectra, all retrievals fit for an error inflation term of about 30, implying that the deviation of our residuals of the fits were about 30 times worse than photon noise alone, likely representing systematic inadequacies of our relatively simple 1D modeling approach.

\begin{figure*}[ht]
    \centering
    \includegraphics[width=1.0\linewidth]{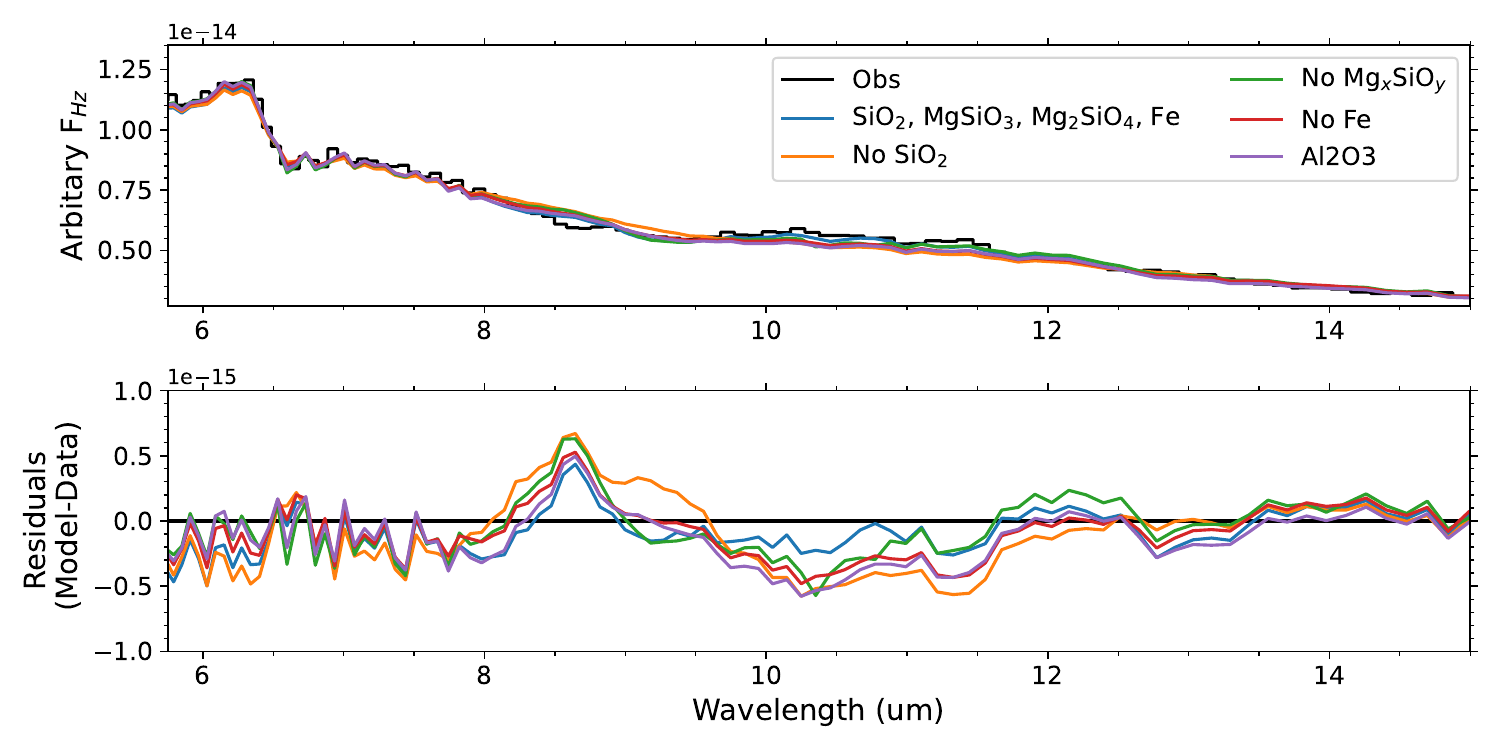}
    \includegraphics[width=1.0\linewidth]{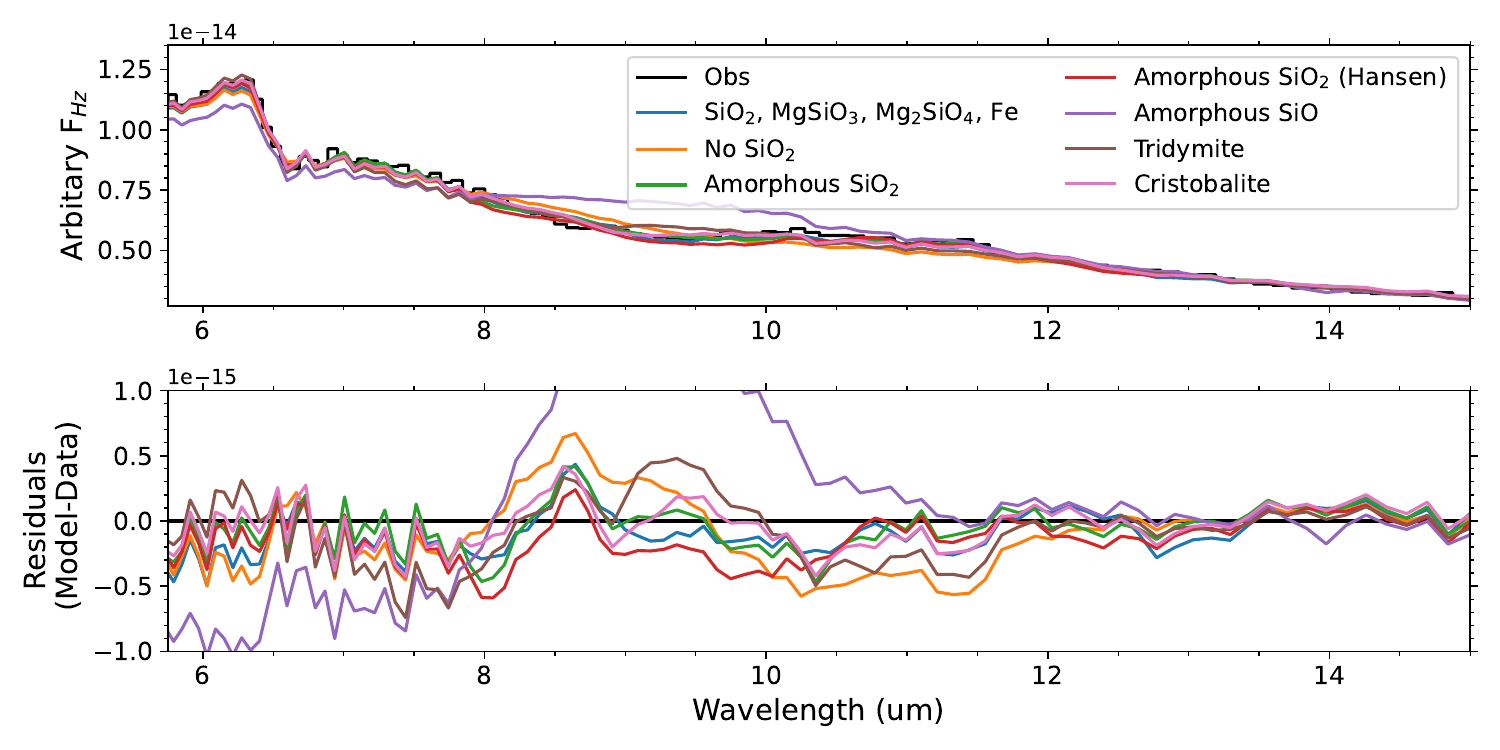}

    \caption{Top: Comparison of the best-fit full-wavelength retrievals of J0624 with different cloud species compared to the R$\sim$100 observations (black). Bottom: Residuals between the retrieved best-fit spectra and the observations.}
    \label{fig:cloud_compare_J0624}
\end{figure*}

\subsubsection{J0624 Bulk Properties}

All retrievals tend to find an elevated [M/H], ranging from 0.548 to 0.770. Similarly, a somewhat super-solar C/O ratio is consistently retrieved, ranging from 0.655 to 0.828, not unlike the T-dwarf retrievals of \cite{zalesky:2022}. Retrieved radii ranged from 0.880 to 0.924~$R_{\rm Jup}$, with $\log(g)_{\rm cgs}$ varying between 4.384 to 4.928. This corresponds to a range of implied masses between 8 and 26~$M_{\rm Jup}$. While the retrieval fits are not outright unphysical, we do not expect a mass lower than the deuterium burning limit and metallicities significantly above solar. The retrieved surface gravity in particular appears to be consistent with the range of best-fit self-consistent models, which ranged from 4.0 to 5.0, with the Sonora models further agreeing with the somewhat elevated metallicity, pushing up against the edge of the grid boundary (see Table~\ref{table:J0624:fits_v2}). We discuss possible biases in J0624's bulk retrieved properties in our comparison to J2148 below (see Section~\ref{sec:J2148_bulk}).

\subsubsection{J0624 Cloud Properties}

For our main goal of identifying the cloud species and properties present in J0624, we compare our various cloud scenarios in terms of both the Bayesian Evidence (ln Z) as well as the least squares of the residuals in the best-fit spectrum ($\chi_\nu^2$). By comparing to our fiducial retrieval, which includes crystalline SiO$_2$, MgSiO$_3$, Mg$_2$SiO$_4$, and Fe clouds with a single $f_{\rm sed}$, we can quantify whether certain cloud species or treatments are statistically justified. For example, the treatment of clouds with species-specific $f_{\rm sed}$ parameters resulted in a \textit{lower} Bayesian evidence (ln B = ln Z$_1$ - ln Z$_2$ = 9.6), implying the additional parameters to control species-specific vertical profiles and size distributions were not statistically justified. Interestingly, this is confirmed with the fact that the best-fit retrieved spectrum was somewhat worse when fitting species-specific $f_{\rm sed}$ parameters, perhaps implying that the higher dimensional retrieval had trouble finding an equivalent best-fit compared to the single-$f_{\rm sed}$ case.

The removal of Fe and magnesium-silicate clouds both resulted in higher Bayesian evidences by $>30$, implying that their inclusion in the retrievals of J0624 was not justified. However, we removed both MgSiO$_3$ and Mg$_2$SiO$_4$ simultaneously; as shown below, the inclusion of magnesium and iron silicates is suggested by fits with the CMA-ES optimization framework (see Section~\ref{sec:J0624:CMA}). The lack of clear evidence for Fe or the combined magnesium-silicate clouds is somewhat surprising as both silicate and iron clouds were necessary to explain the spectrum of J2224 \citep{burningham:2021} and PSO J316 \citep{molliere:2025}. Meanwhile, the inclusion of Al$_2$O$_3$ resulted in a higher evidence by $>20$, implying that its presence was justified by the data. We suspect that the retrieval is using Al$_2$O$_3$ as an additional gray opacity source, preferable over Fe perhaps due to a different implied vertical location, extent, and/or particle size distribution.

The best-fit spectra from these scenarios compared to the observations are shown in Figure~\ref{fig:cloud_compare_J0624}. While the combination of species explored here does not perfectly match the silicate feature, especially the deepest part of the absorption at 8.5~$\mu$m, it is clear that the inclusion of SiO$_2$ greatly improves the fit. In particular, the red edge of the feature is much better matched with SiO$_2$ compared to the spectra without SiO$_2$.

Because the inclusion of SiO$_2$ in the retrievals is highly preferred by the retrieval and somewhat unexpected, we ran additional retrievals to identify additional details of the cloud properties, including different crystal structures and particle size distributions. We also replaced SiO$_2$ with amorphous SiO as recently seen in PSO J316 \citep{molliere:2025}. We run these additional SiO$_2$ tests with the fiducial setup, including Fe, MgSiO$_3$ and Mg$_2$SiO$_4$ in the event that these species become an important component of the fit. These results are listed in Table~\ref{table:J0624:rets:polymorphs}.

Both tridymite and cristobalite are preferred over quartz for J0624's spectrum, with Bayes factors of $\ln B = 35.7$ and 47.5, respectively, aligning with the expected microphysical stability of the crystal structures at J0624's relevant atmospheric pressure and temperatures \citep{moran:2024}. However, we find a strong preference for amorphous SiO$_2$ over all other cloud types or crystalline structures, with a Bayes factor of $\ln B=54.7$ over the fiducial quartz retrievals and a reduced $\chi^2$ almost 100 less than the second best-fitting polymorph, tridymite. Compared to the other crystal structures, the amorphous SiO$_2$ provides a better fit between 9 and 9.5\,$\mu$m and longward of 11$\mu$m, but regions of disagreement still exist, particularly around 8.5 and 10.5\,$\mu$m.

We next test for a preference between a log-normal and Hansen particle size distribution by running our preferred amorphous SiO$_2$ scenario with a Hansen particle size distribution. We find a significantly worse fit to the data and a significantly lower Bayesian evidence ($\ln B=54.2$) for the Hansen size distribution scenario. Figure~\ref{fig:cloud_compare_J0624} shows that the best-fit Hansen distribution model absorbs too heavily at both the short and long wavelength edges of the silicate absorption feature, despite matching the 8.5$\mu$m better than other models.

\begin{figure*}
\gridline{\fig{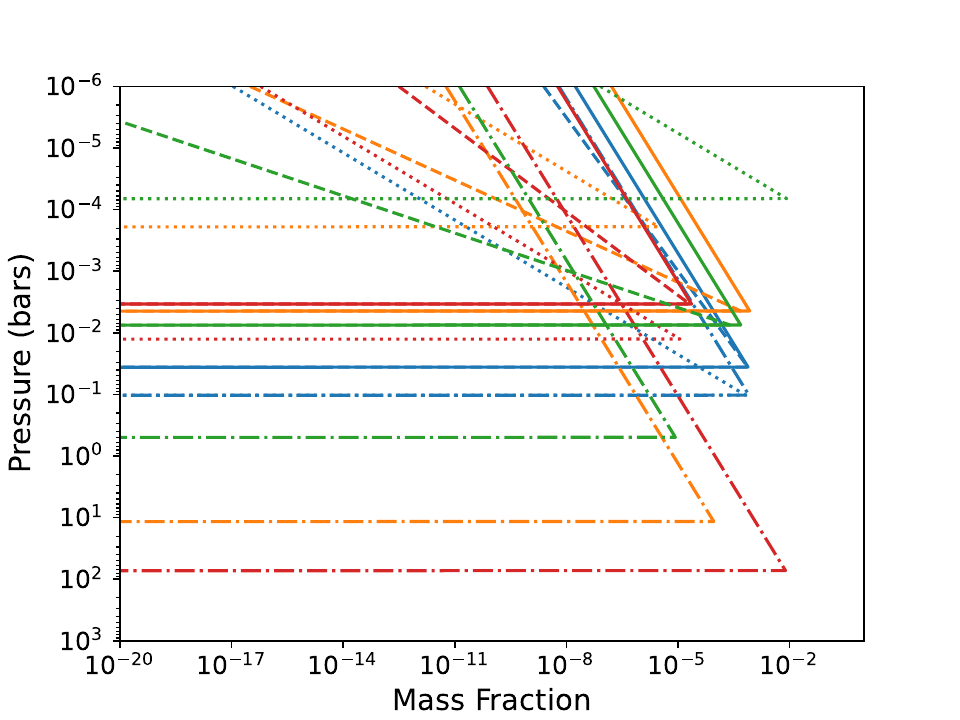}{0.5\textwidth}{(J0624)}
          \fig{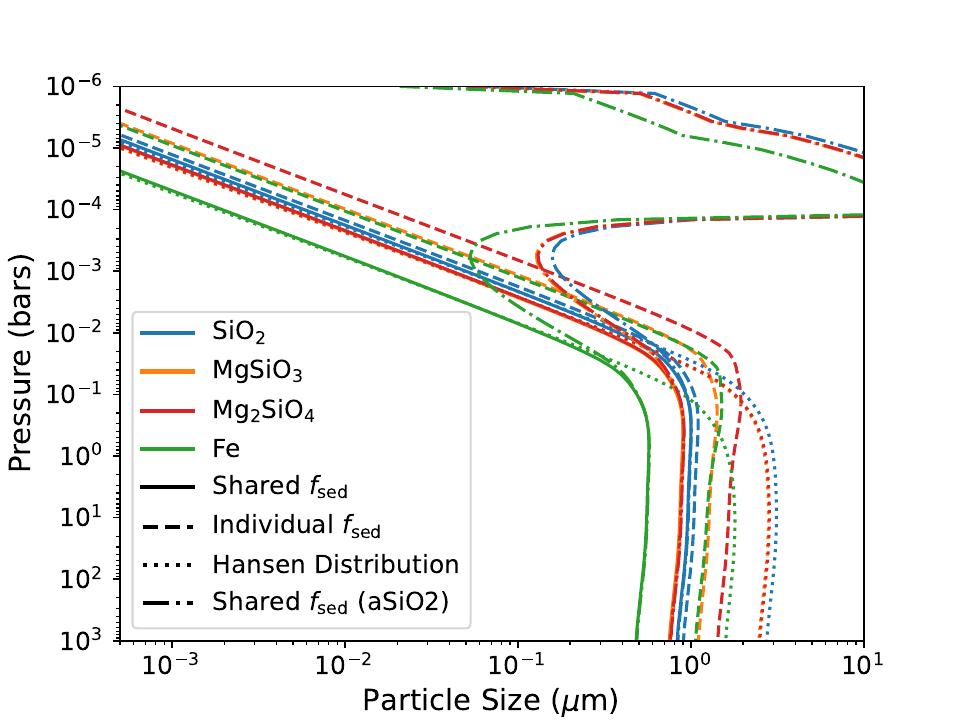}{0.5\textwidth}{(J0624)}
          }    
\gridline{\fig{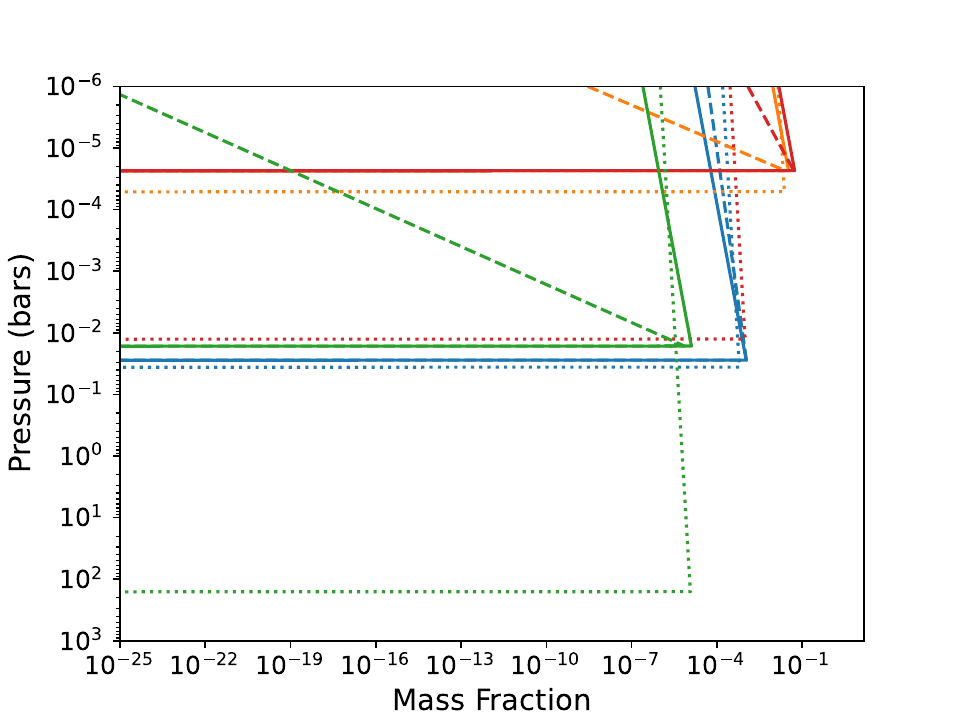}{0.5\textwidth}{(J2148)}
          \fig{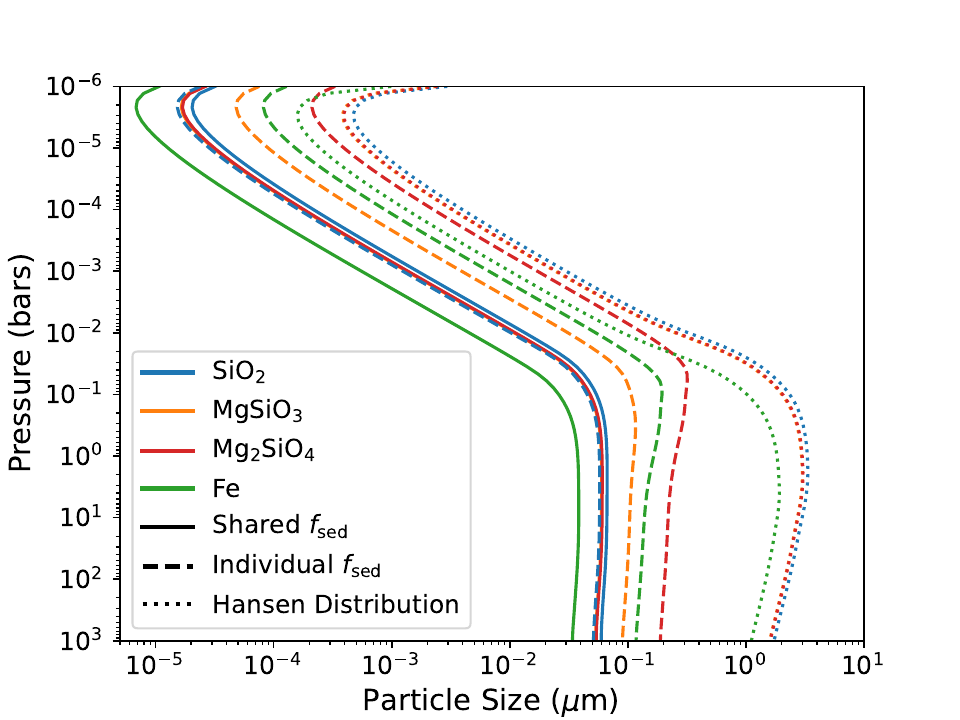}{0.5\textwidth}{(J2148)}
          }  
\caption{Cloud mass fraction (left) and particle size(right) as a function of pressure for J0624 (top) and J2148 (bottom) from the best-fit retrieval model. We show results from both the fiducial retrieval, where one $f_\mathrm{sed}$ is shared between cloud species (solid), as well as from the retrieval where an individual $f_\mathrm{sed}$ was retrieved for each cloud species (see Table~\ref{table:J0624:rets} and \ref{table:J2148:rets}). \label{fig:psizes}}
\end{figure*}

Figure~\ref{fig:psizes} shows the retrieved particle size distributions and vertical profiles for the fiducial retrieval, the preferred amorphous SiO$_2$ retrieval, the retrieval using species-specific $f_{\rm sed}$ values, and the Hansen distribution retrieval. In general, particles between 0.1 and 1.0\,$\mu$m are retrieved. The Hansen distribution allows for larger particles are higher pressures, while preserving the low particle sizes near the IR photosphere.

The SiO$_2$ cloud deck is retrieved between 10 and 100 millibar with the exact location of the other cloud decks dependent on the retrieval setup. Interestingly, the non-SiO$_2$ clouds are placed below the SiO$_2$ clouds in the amorphous SiO$_2$ case, while the opposite is true for the fiducial crystalline SiO$_2$ case.

The large oscillation in the particle size in the amorphous SiO$_2$ retrieval appears to be from a subtle increase in the temperature in the upper atmosphere compared to the fiducial retrieval. The atmosphere at pressures below 10 millibar is poorly constrained by the observations as evidenced by the contribution function (see Figure~\ref{fig:TPs}), so we do not consider the temperature or particle sizes in this regime to be physical.

\paragraph{J0624 CMA-ES Best-Fits}\label{sec:J0624:CMA}

\begin{figure*}
    \centering
    \includegraphics[width=1.0\linewidth]{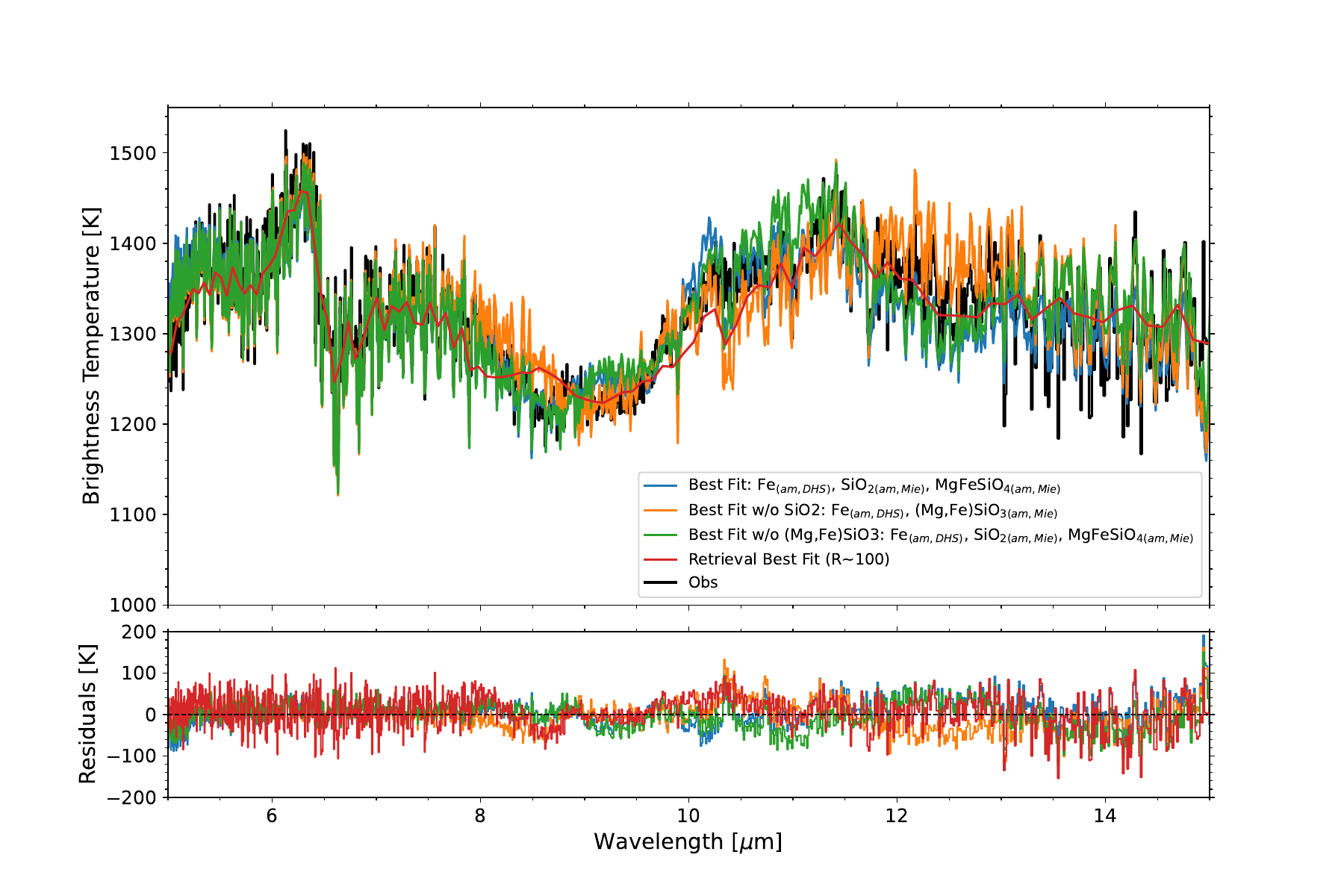}

    \caption{Top: Fits to the R$\sim$1000 mid-IR spectrum of J0624 (black) using CMA-ES. Our best-fit retrieval (computed at R$\sim$100) is shown in red, which included amorphous Fe, SiO$_2$, MgSiO$_3$, and Mg$_2$SiO$_4$ using DHS for the opacity computation. We show the best fitting scenarios without SiO$_2$ and without magnesium-iron silicates for comparison. Bottom: Residuals between the observations and each model fit.}
    \label{fig:cloud_cma_j0624}
\end{figure*}

To search more broadly for the best-fitting cloud scenario, we performed a sweep of cloud composition, crystallinity, particle size distribution, and opacity computation method with the CMA-ES algorithm framework. Figure~\ref{fig:cloud_cma_j0624} shows the best fitting model scenarios compared to the data, as well as compared to the preferred scenario from the full-wavelnegth retrievals, in brightness temperature space. We compare the brightness temperatures as this removes the general blackbody shape, helping to isolate the shape of the absorption features. Appendix Table~\ref{tab:J0624:CMA} ranks the various combinations that were tried based on their respective $\chi^2/N_{\rm data}$ goodness-of-fit. 

The best-fitting CMA-ES scenarios contain both amorphous SiO$_2$ and iron-rich pyroxene or olivine. In agreement with the retrieval, the CMA-ES fits generally indicate the presence of amorphous SiO$_2$ cloud particles over crystalline particles. Additionally, the difference in goodness-of-fit between the Hansen and log-normal particle size distributions was insignificant. The best-fitting model without SiO$_2$ clouds has a $\chi^2/N_{\rm data}$ about 4.5 worse than the best-fitting model with SiO$_2$ clouds (55.25 versus 50.74, respectively), particularly driven by a poor match to the blue edge of the silicate absorption ($\approx7.5 - 8.5 \mu$m). However, in addition to SiO$_2$, Mg$_0.5$Fe$_0.5$SiO$_3$ or MgFeSiO$_4$ is also suggested by the fit; the best-fit is improved from $\chi^2/N_{\rm data} = 59.24$ to 50.74 with the inclusion of these silicates. While iron-rich silicates are slightly preferred, iron-poor silicates provide comparable fits with $\chi^2/N_{\rm data} = 53.60$. The inclusion of these other silicates improves the fit to the red edge of the silicate feature, particularly between 8.5 and 9.5\,$\mu$m. The worst fitting 29 spectra all contained no SiO$_2$ and the three worst-of-the-worst contained only the Fe base and Mg$_2$SiO$_4$.

\subsubsection{J0624 TP Profile}\label{sec:J0624:TP}

\begin{figure*}
\gridline{\fig{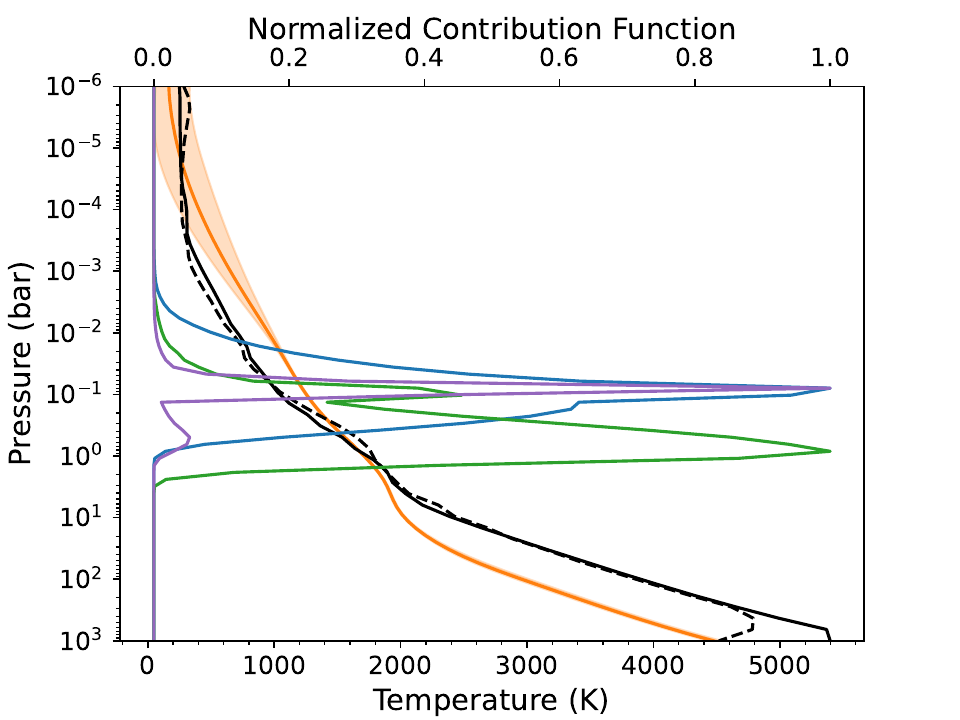}{0.5\textwidth}{(J0624)}
          \fig{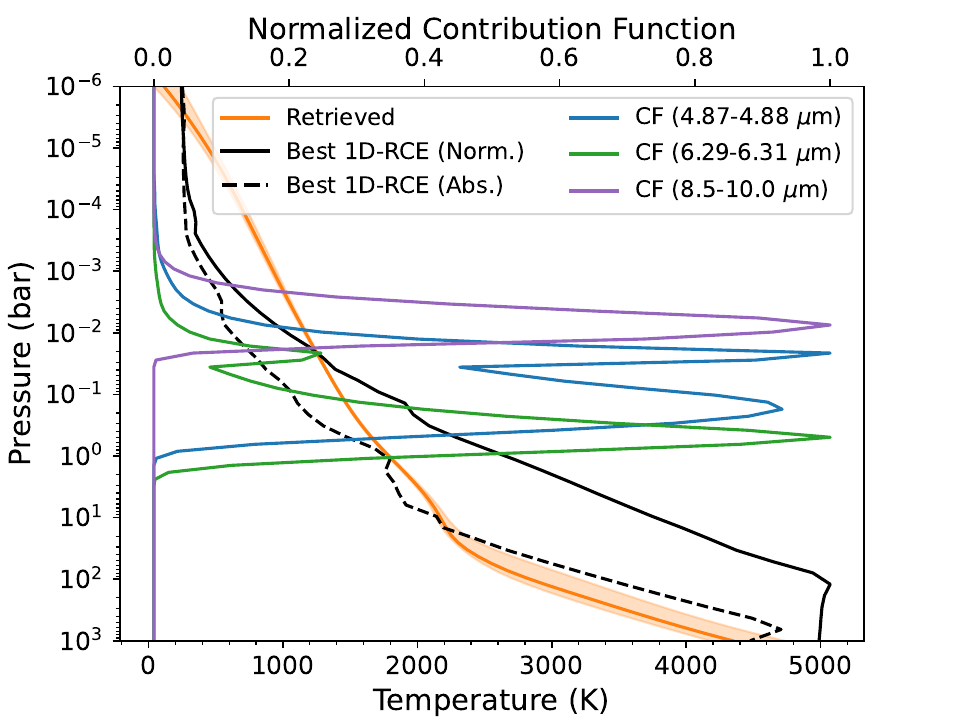}{0.5\textwidth}{(J2148)}
          }
\caption{{Retrieved (yellow) versus best-fit 1D-RCE (black) pressure-temperature profiles of J0624 (left) and J2148 (right). The best-fit 1D-RCE models are from exoREM (see \ref{table:J0624:fits_v2} and \ref{table:J2148:fits_v2}). We show the best-fit TP profiles after both normalizing the data at 6.3 $\mu$m (solid) and using the absolute fluxes, assuming 1 Jupiter radius for the brown dwarf radius. We also show contribution functions (CF) for the retrieved model at probing CO (blue), the continuum (green), and the silicate cloud feature (purple). \label{fig:TPs}}}
\end{figure*}

In Figure~\ref{fig:TPs}, we compare the model temperature profiles. Compared to the best-fit 1D-RCE temperature profile, the retrieved TP profile is significantly steeper, i.e., $dT/dP$ is less for the retrieved TP profile. The TP profile parameterization of \cite{molliere:2020} qualitatively recreates the general shape of the 1D-RCE profiles with a deep adiabat and a radiative photosphere around 1 bar. However, at all observed pressures, the retrieved profile remains steeper. In general, the pressures probed by our observations vary between 10 mbar and 1 bar according to the contribution function. Therefore, the TP profile above and below those pressures is constrained mainly through the parameterization rather than empirical observation. 

\begin{deluxetable*}{c|c|c|c|c|c|c}
\tablecaption{Retrieved properties of J2148 from retrievals with alternative cloud properties. \label{table:J2148:rets}}
\tablehead{Parameter & Fiducial & Multi-$f_{\rm sed}$ & No SiO2 & No Mg$_x$SiO$_y$ & No Fe & With Al$_2$O$_3$}
\startdata
Live Points & 5000 & 5000 & 5000 & 5000 & 5000 & 5000 \\
ln Z\tablenotemark{a} & 10491.1 & 10500.1 & 10454.1 & 10495.1 & 10487.6 & 10491.5 \\
$\chi^2_{\nu}$ & 1506.4 & 1610.8 & 1781.0 & 1686.9 & 1688.1 & 1484.5 \\
\hline
$T_{\mathrm{int}}$ & 2353$\pm$16 & 2615$\pm$57 & 2523$\pm$30 & 2292$\pm$19 & 2375$\pm$24 & 2304$\pm$18 \\
$\log{g}$ & 4.389$\pm$0.053 & 4.011$\pm$0.061 & 4.214$\pm$0.053 & 4.352$\pm$0.060 & 4.241$\pm$0.074 & 4.554$\pm$0.050 \\
$R_{\mathrm{Jup}}$ & 1.0480$\pm$0.0043 & 1.0593$\pm$0.0035 & 1.1119$\pm$0.0065 & 1.0365$\pm$0.0032 & 1.0525$\pm$0.0052 & 1.0851$\pm$0.0046 \\
$M_{\rm Jup}$\tablenotemark{b} & 11 $\pm$ 3 & 5 $\pm$ 3&  8 $\pm$ 3&  10 $\pm$ 3& 8 $\pm$ 3&
       17 $\pm$ 3\\
\hline
$\log{p_{\mathrm{quench}}}$ & -5.32$\pm$0.37 & -5.23$\pm$0.52 & -3.81$\pm$0.85 & -5.15$\pm$0.49 & -4.72$\pm$0.81 & -4.98$\pm$0.68 \\
$\mathrm{M/H}$ & 0.424$\pm$0.024 & 0.354$\pm$0.028 & 0.439$\pm$0.027 & 0.397$\pm$0.030 & 0.404$\pm$0.034 & 0.440$\pm$0.023 \\
$\mathrm{C/O}$ & 0.6444$\pm$0.0051 & 0.6560$\pm$0.0022 & 0.6639$\pm$0.0018 & 0.6477$\pm$0.0044 & 0.6471$\pm$0.0046 & 0.6242$\pm$0.0051 \\
\hline
$\sigma_{\mathrm{lnorm}}$ & 2.937$\pm$0.047 & 2.959$\pm$0.032 & 2.870$\pm$0.086 & 2.945$\pm$0.044 & 2.952$\pm$0.042 & 2.84$\pm$0.10 \\
$\log{k_{\mathrm{zz}}}$ & 7.51$\pm$0.15 & 7.438$\pm$0.088 & 8.473$\pm$0.071 & 7.61$\pm$0.10 & 7.054$\pm$0.061 & 7.567$\pm$0.092 \\
$f_{\mathrm{sed}}$ & 0.61$\pm$0.18 & - & 0.74$\pm$0.11 & 0.75$\pm$0.13 & 0.90$\pm$0.17 & 0.460$\pm$0.091 \\
$f_{\mathrm{sed,SiO_2}}$ & - & 0.418$\pm$0.094 & - & - & - & - \\
$f_{\mathrm{sed,MgSiO_3}}$ & - & 4.1$\pm$1.1 & - & - & - & - \\
$f_{\mathrm{sed,Mg_2SiO_4}}$ & - & 1.9$\pm$1.1 & - & - & - & - \\
$f_{\mathrm{sed,Fe}}$ & - & 4.29$\pm$0.51 & - & - & - & - \\
\hline
$\log{X_{\mathrm{cb,SiO_2}}}$ & -2.924$\pm$0.055 & -3.392$\pm$0.074 & - & -2.696$\pm$0.081 & -3.14$\pm$0.14 & -2.968$\pm$0.062 \\
$\log{P_{\mathrm{base,SiO_2}}}$ & -1.495$\pm$0.030 & -1.386$\pm$0.030 & - & -1.727$\pm$0.033 & -1.296$\pm$0.070 & -1.377$\pm$0.028 \\
$\log{X_{\mathrm{cb,MgSiO_3}}}$ & -3.49$\pm$0.87 & -3.0$\pm$1.3 & -1.674$\pm$0.060 & - & -3.84$\pm$0.81 & -3.83$\pm$0.74 \\
$\log{P_{\mathrm{base,MgSiO_3}}}$ & -3.72$\pm$0.77 & 2.05$\pm$0.56 & -2.756$\pm$0.026 & - & -3.79$\pm$0.85 & -3.75$\pm$0.69 \\
$\log{X_{\mathrm{cb,Mg_2SiO_4}}}$ & -1.72$\pm$0.52 & -1.36$\pm$0.41 & -4.09$\pm$0.33 & - & -2.63$\pm$0.68 & -1.84$\pm$0.41 \\
$\log{P_{\mathrm{base,Mg_2SiO_4}}}$ & -4.17$\pm$0.52 & -4.41$\pm$0.41 & 1.61$\pm$0.41 & - & -3.20$\pm$0.79 & -3.88$\pm$0.43 \\
$\log{X_{\mathrm{cb,Fe}}}$ & -4.76$\pm$0.37 & -1.14$\pm$0.77 & -4.036$\pm$0.096 & -3.99$\pm$0.42 & - & -4.73$\pm$0.17 \\
$\log{P_{\mathrm{base,Fe}}}$ & -1.95$\pm$0.48 & 1.05$\pm$0.15 & -1.25$\pm$0.12 & -2.79$\pm$0.41 & - & -2.03$\pm$0.25 \\
$\log{X_{\mathrm{cb,Al_2O_3}}}$ & - & - & - & - & - & -4.820$\pm$0.093 \\
$\log{P_{\mathrm{base,Al_2O_3}}}$ & - & - & - & - & - & -0.260$\pm$0.099 \\
\enddata
\tablenotetext{a}{The log-Bayesian Evidence of the given retrieval.}
\tablenotetext{b}{Calculated from the retrieved surface gravity and radius.}
\end{deluxetable*}

\begin{deluxetable*}{c|c|c|c|c|c|c}
\tablecaption{Retrieved properties of J2148 from retrievals with different SiO$_2$ polymorphs.\label{table:J2148:rets:polymorphs}}
\tablehead{ & Fiducial (Quartz) & Amorphous & Tridymite & Cristobalite & SiO$_2$ (Hansen)* & SiO(a)}
\startdata
Live Points & 5000 & 5000 & 5000 & 5000 & 5000 & 5000 \\
ln Z & 10491.1 & 10479.4 & $10456.4\pm0.0$ & $10475.8\pm0.0$ & $10505.5\pm0.0$ & 10381.4 \\
$\chi^2_{\nu}$ & 1506.4 & 1679.7 & 1782.2 & 1628.7 & 1469.3 & 2566.2 \\
\hline
$T_{\mathrm{int}}$ & $2353\pm16$ & $2602\pm57$ & $2708\pm98$ & $2456\pm72$ & $2544\pm57$ & $1727\pm21$ \\
$\log{g}$ & $4.389\pm0.053$ & $4.015\pm0.058$ & $4.075\pm0.062$ & $4.062\pm0.066$ & $4.266\pm0.061$ & $4.376\pm0.070$ \\
$R_{\mathrm{Jup}}$ & $1.05\pm0.004$ & $1.10\pm0.005$ & $1.16\pm0.006$ & $1.10\pm0.004$ & $1.11\pm0.004$ & $1.15 \pm 0.01$ \\
$M_{\rm Jup}$\tablenotemark{b} & 11$\pm$ 3&  5$\pm$ 3&  6$\pm$ 3& 6$\pm$ 3&  9$\pm$ 3&
       13$\pm$ 3 \\
\hline
$\log{p_{\mathrm{quench}}}$ & $-5.32\pm0.37$ & $-1.97\pm0.25$ & $-2.82\pm0.85$ & $-1.95\pm0.22$ & $-3.0\pm1.2$ & $-4.1\pm1.2$ \\
$\mathrm{[M/H]}$ & $0.424\pm0.024$ & $0.272\pm0.029$ & $0.402\pm0.031$ & $0.306\pm0.028$ & $0.411\pm0.027$ & $0.533\pm0.038$ \\
$\mathrm{C/O}$ & $0.6444\pm0.0051$ & $0.6552\pm0.0020$ & $0.6583\pm0.0019$ & $0.6604\pm0.0018$ & $0.6496\pm0.0037$ & $0.6143\pm0.0058$ \\
\hline
$\mathrm{b}_{\mathrm{hans}}$ & - & - & - & - & $0.145\pm0.045$ & - \\
$\sigma_{\mathrm{lnorm}}$ & $2.937\pm0.047$ & $2.922\pm0.055$ & $1.415\pm0.092$ & $2.934\pm0.046$ & - & $2.21\pm0.20$ \\
$\log{k_{\mathrm{zz}}}$ & $7.51\pm0.15$ & $7.794\pm0.073$ & $8.281\pm0.093$ & $7.211\pm0.060$ & $7.787\pm0.081$ & $10.58\pm0.13$ \\
$f_{\mathrm{sed}}$ & $0.61\pm0.18$ & $0.271\pm0.046$ & $0.170\pm0.036$ & $1.50\pm0.16$ & $0.183\pm0.042$ & $0.451\pm0.076$ \\
$f_{\mathrm{sed,SiO_2}}$ & - & - & - & - & - & - \\
$f_{\mathrm{sed,MgSiO_3}}$ & - & - & - & - & - & - \\
$f_{\mathrm{sed,Mg_2SiO_4}}$ & - & - & - & - & - & - \\
$f_{\mathrm{sed,Fe}}$ & - & - & - & - & - & - \\
\hline
$\log{X_{\mathrm{cb,SiO}}}$ & - & - & - & - & - & $-3.05\pm0.61$ \\
$\log{P_{\mathrm{base,SiO}}}$ & - & - & - & - & - & $-3.54\pm0.63$ \\
$\log{X_{\mathrm{cb,SiO_2}}}$ & $-2.924\pm0.055$ & $-3.995\pm0.089$ & $-4.101\pm0.089$ & $-3.193\pm0.089$ & $-3.209\pm0.071$ & - \\
$\log{P_{\mathrm{base,SiO_2}}}$ & $-1.495\pm0.030$ & $-1.377\pm0.068$ & $-1.952\pm0.034$ & $-1.961\pm0.028$ & $-1.385\pm0.032$ & - \\
$\log{X_{\mathrm{cb,MgSiO_3}}}$ & $-3.49\pm0.87$ & $-3.24\pm0.53$ & $-2.821\pm0.070$ & $-4.26\pm0.48$ & $-3.54\pm0.97$ & $-0.032 \pm 0.028$ \\
$\log{P_{\mathrm{base,MgSiO_3}}}$ & $-3.72\pm0.77$ & $-4.03\pm0.52$ & $-2.076\pm0.030$ & $-4.08\pm0.52$ & $-3.78\pm0.78$ & $-4.60 \pm 0.11$ \\
$\log{X_{\mathrm{cb,Mg_2SiO_4}}}$ & $-1.72\pm0.52$ & $-3.781\pm0.080$ & $-4.922\pm0.063$ & $-3.217\pm0.074$ & $-2.970\pm0.066$ & $-3.92 \pm 0.70$ \\
$\log{P_{\mathrm{base,Mg_2SiO_4}}}$ & $-4.17\pm0.52$ & $-1.292\pm0.071$ & $2.57\pm0.31$ & $-1.502\pm0.028$ & $-1.846\pm0.032$ & $-2.8 \pm 1.2$ \\
$\log{X_{\mathrm{cb,Fe}}}$ & $-4.76\pm0.37$ & $-3.91\pm0.31$ & $-4.841\pm0.081$ & $-3.35\pm0.25$ & $-4.79\pm0.14$ & $-4.5\pm0.35$ \\
$\log{P_{\mathrm{base,Fe}}}$ & $-1.95\pm0.48$ & $-3.35\pm0.29$ & $0.81\pm0.29$ & $-3.54\pm0.21$ & $2.39\pm0.29$ & $-2.1\pm 1.1$ \\
\enddata
\tablenotetext{a}{The log-Bayesian Evidence of the given retrieval.}
\tablenotetext{b}{Calculated from the retrieved surface gravity and radius.}
\tablenotetext{*}{Highest evidence full-wavelength retrieval scenario for J2148.}
\end{deluxetable*}

\subsection{J2148 Results}

As a direct comparison to J0624, we run many of the same fits for J2148. As mentioned above, J2148 has a very similar optical and IR spectral type as J0624 -- just 1 and 1.5 spectral types later, respectively. However, their full IR spectra are quite different, with J2148 being significantly redder. This suggests that J2148 is likely cloudier than J0624, perhaps due to the equator-on inclination at which we observe J2148 \citep{vos:2017,suarez:2023}. Similar to J0624, an error inflation term of about 30 was required for the $\chi^2$ of the fits to be about 1.0.

\begin{figure*}
    \centering
    \includegraphics[width=1.0\linewidth]{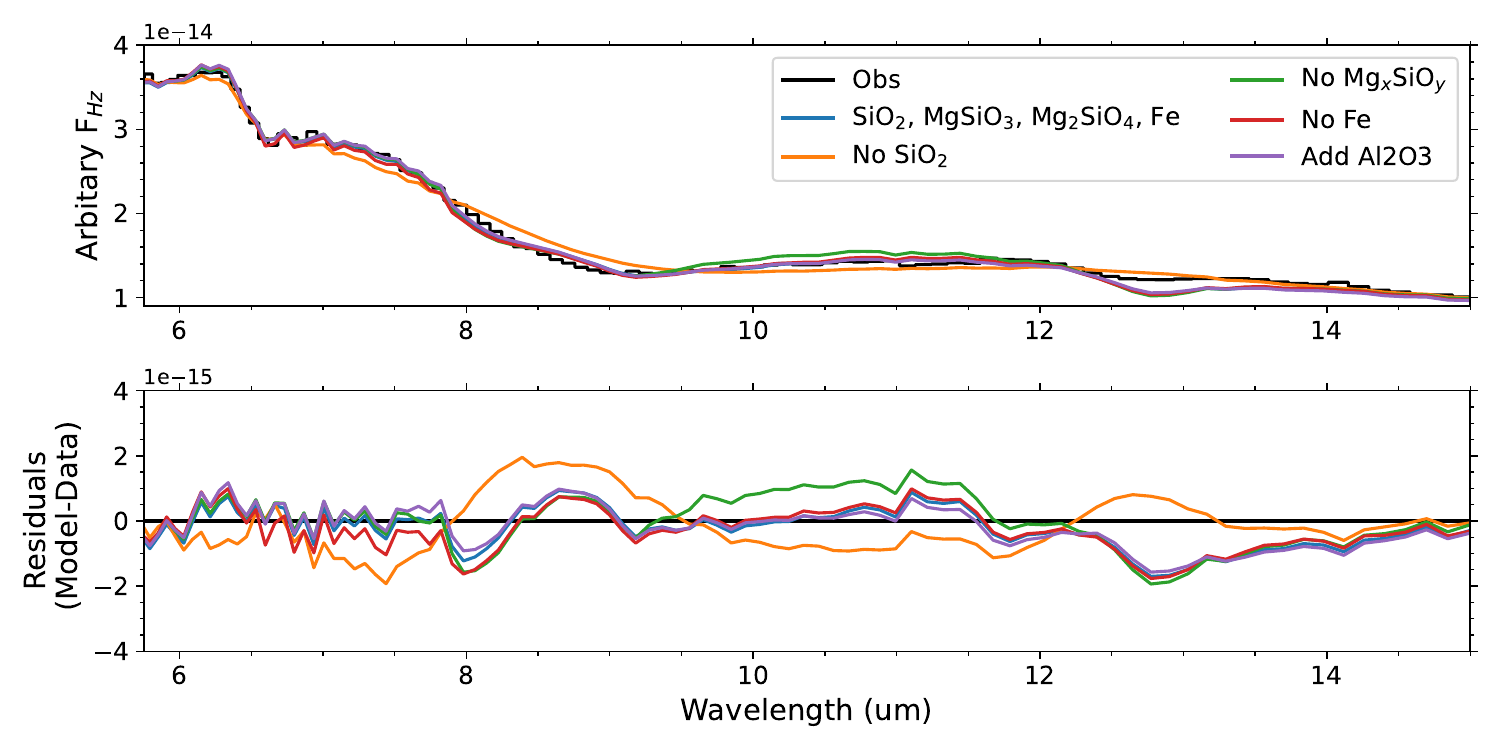}
    \includegraphics[width=1.0\linewidth]{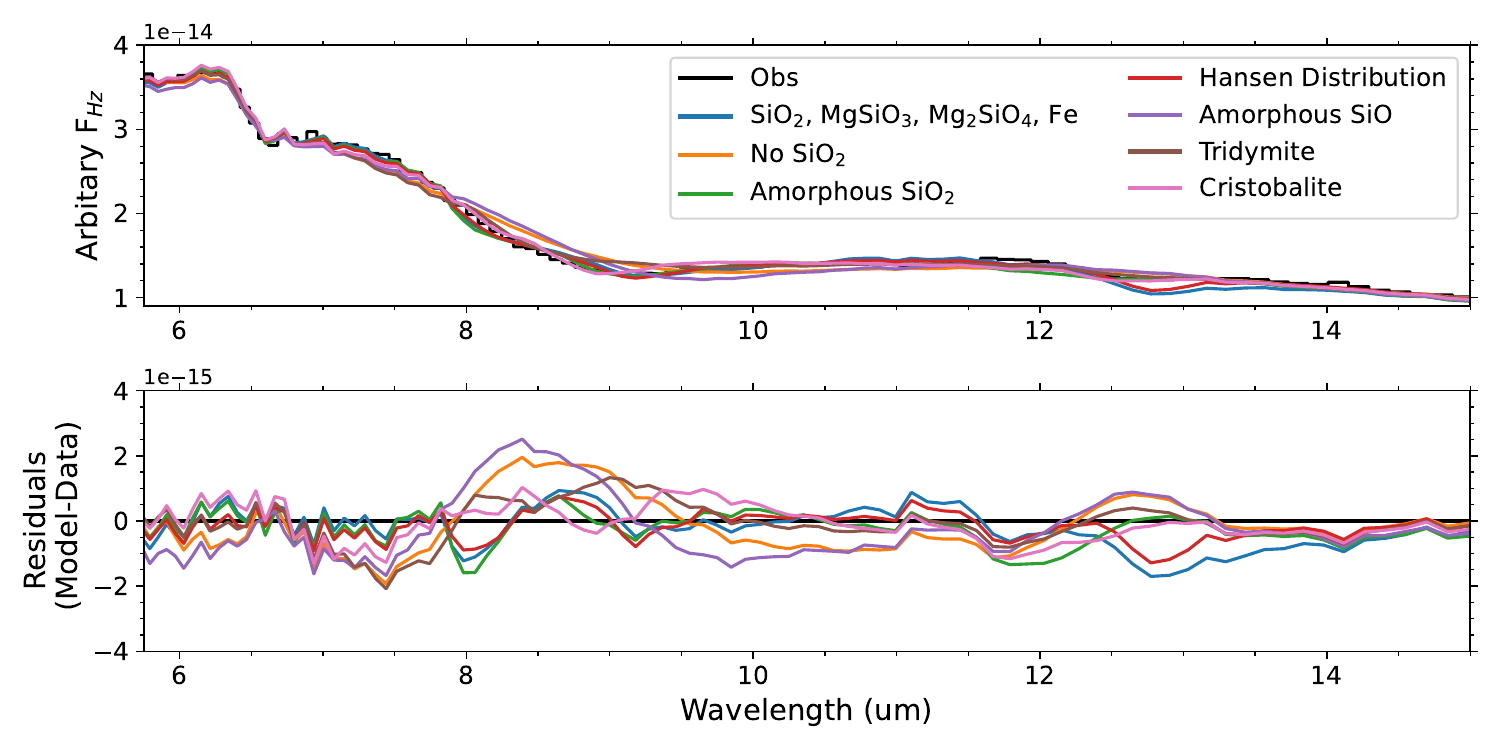}

    \caption{Same as Figure~\ref{fig:cloud_compare_J0624}, but for J2148.}
    \label{fig:cloud_compare_J2148}
\end{figure*}

\subsubsection{J2148 Cloud Properties}

The median-fit parameters and 1$\sigma$ uncertainties from each J2148 retrieval are listed in Tables~\ref{table:J2148:rets} and \ref{table:J2148:rets:polymorphs} with the best-fit spectra from the preferred scenario shown in Figure~\ref{fig:cloud_compare_J2148}. Overall, a similar picture to J0624 is suggested, with SiO$_2$ clouds strongly preferred over magnesium-silicates. The scenario without SiO$_2$ had a significantly worse fit and $\ln{B} = 37.0$. Not including magnesium silicates is moderately preferred at $\ln{B} = 4.0$ (though again see results from the CMA fits below in Section~\ref{sec:J2148:CMA}). Fe clouds are moderately justified with $\ln{B} = 3.5$. Al$_2$O$_3$ is only marginally suggested with $\ln{B} = 0.4$. Fitting for species-specific $f_{\rm sed}$ parameters is justified by the data with $\ln{B} = 9.0$.

Of all SiO$_2$ polymorphs, the fiducial crystalline quartz particles appear to be the most preferred by the retrieval. Amorphous SiO$_2$ has the second-best Bayesian evidence but with $\ln{B} = 11.7$, while the cristobalite scenario has a slightly worse Bayesian evidence and a slightly better fitting $\chi^2$. SiO cloud particles are not favored with $\ln{B} \approx 110$. 

Unlike J0624, a Hansen particle size distribution is preferred over a log-normal size distribution with $\ln{B} = 14.4$. In fact, the Hansen distribution scenario has a higher Bayesian evidence than the species-specific $f_{\rm sed}$ scenario. In particular, the Hansen distribution scenario can better fit the heart of the solid-state cloud feature and the red edge, compared to the fiducial log-normal particle size distribution. Figure~\ref{fig:psizes} shows that while the Hansen distribution scenario has a similar vertical profile for the clouds, the effective particle sizes are significantly larger than the fiducial log-normal scenario. We hypothesize that the asymmetric nature of the Hansen distribution allows more small particles to fit the cloud feature while retaining some larger particles. The species-specific $f_{\rm sed}$ allows for larger magnesium-silicate and Fe particles, while keeping the SiO$_2$ particles small. The retrieval is likely using the small SiO$_2$ particles to fit the MIR silicate feature, while using the other cloud species to adjust the rest of the spectral energy distribution.

\paragraph{J2148 CMA-ES Best-Fits}\label{sec:J2148:CMA}

\begin{figure*}
    \centering
    \includegraphics[width=1.0\linewidth]{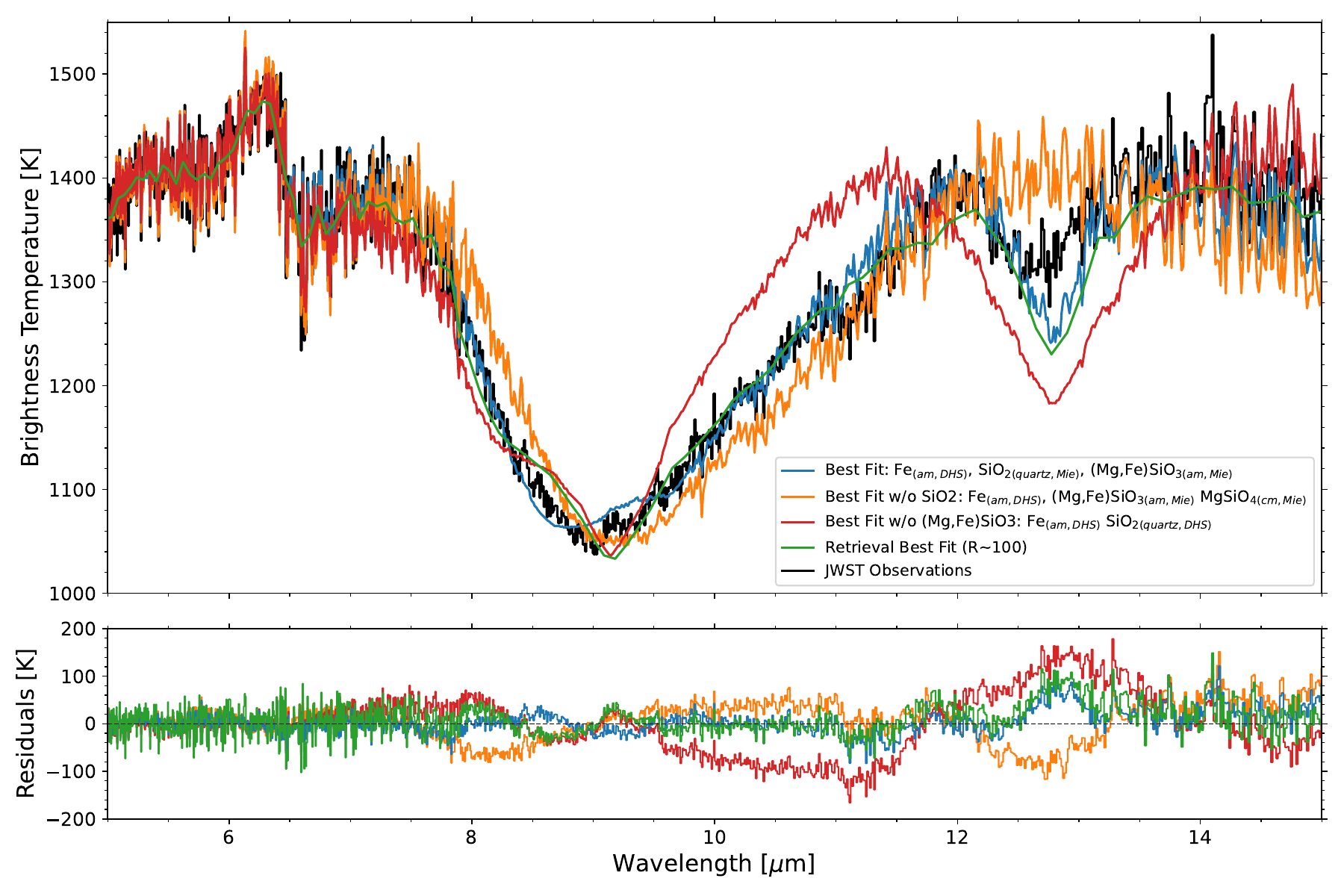}

    \caption{Top: Fits to the R$\sim$1000 mid-IR spectrum of J2148 (black) using CMA-ES. Our best-fit retrieval (computed at R$\sim$100) is shown in red, which included amorphous Fe, SiO$_2$, MgSiO$_3$, and Mg$_2$SiO$_4$ using DHS for the opacity computation. We show the best fitting scenarios with and without SiO$_2$ and MgSiO$_3$. Bottom: Residuals between the observations and each model fit.}
    \label{fig:cloud_cma_j2148}
\end{figure*}

As with J0624, we ran CMA-ES fits to explore a wider range of possible cloud scenarios. Figure~\ref{fig:cloud_cma_j2148} shows the fits to the observations from various best-fit CMA-ES scenarios. Similar to J0624, both SiO$_2$ and a magnesium-iron silicate are required to adequately fit the silicate absorption. In J2148's case, the pyroxene Mg$_{0.5}$Fe$_{0.5}$SiO$_3$ is preferred slightly over olivine MgFeSiO$_4$. The behavior of the fit is similar to J0624, if not even more clear, due to the deeper and sharper silicate feature: without SiO$_2$, the blue edge of the silicate feature cannot be fit, while a magnesium-iron silicate is necessary to fit the red edge of the feature. 

Appendix Table~\ref{tab:J2148:CMA} ranks the various combinations that were trialed based on their respective $\chi^2/N_{\rm data}$ goodness-of-fit. The best-fitting model without SiO$_2$ had a $\chi^2/N_{\rm data}$ value of 115.17, far worse than the best-fitting model with SiO$_2$ at 87.58. Including magnesium silicates was even more important with the best-fitting SiO$_2$ + Fe scenario only reaching $\chi^2/N_{\rm data} = 179.84$. Like with J0624, the worst-fitting scenarios were isolated MgSiO$_4$, plus the Fe base. The bottom 13 worst-fitting models all did not contain SiO$_2$, highlighting its importance to the fit.

As with the retrieval analysis, the best-fit scenario for J2148 included quartz SiO$_2$. Amorphous SiO$_2$ fit significantly worse with $\chi^2/N_{\rm data} = 78.66$ versus the best-fit at $62.22$. Interestingly, cristobalite and tridymite is preferred thermodynamically \citep{moran:2024}, yet each fit worse than quartz at $\chi^2/N_{\rm data} = 85.89$ and 86.507, respectively.

\subsubsection{J2148 Bulk Properties}\label{sec:J2148_bulk}

We also note that J2148 showed a similar metallicity and C/O ratio enhancement as J0624. Specifically, the metallicity ranges from [M/H] = 0.272 to 0.533 and the C/O ratio varies between 0.6143 and 0.6604 across all scenarios tested. While the C/O ratio is close enough to solar to be explained by primordial abundance variations, the enhanced metallicity in both objects is somewhat is surprising. We discuss possible reasons for this elevated metallicity in Section~\ref{sec:disc:substellar}.

In a similar sense, the retrieved surface gravity appear to be somewhat low compared to what we would expect for a field brown dwarf. With $\log{g}$ ranging from 4.01 to 4.55 and radii between 1.04 and 1.15~$R_{\rm Jup}$, our retrievals imply a mass for J2148 between 5 and 17~$M_{\rm Jup}$. Like J0624, this is compared to the inference from \cite{filippazzo:2015} of $\log{g}\approx 5.13$ and $\approx 50 M_{\rm Jup}$. It is not clear what in the observed spectrum is driving the retrieval to prefer larger radii, lower surface gravity, and lower masses, but these biases have commonly appeared for retrievals of self-luminous objects, which we discuss in Section~\ref{sec:disc:substellar}. We further interpret these bulk properties in a comparison to brown dwarf evolution tracks in Section~\ref{sec:disc:evo}.

\subsubsection{J2148 TP Profile}\label{sec:J2148:TP}

Lastly, we find the same steep TP profile for J2148 compared to the best-fit self-consistent 1D-RCE models as we did for J0624 as shown in Figure~\ref{fig:TPs}. This suggests a common reason (or bias) in the retrieval parameterization for a steep profile in cloudy L-dwarfs.  While J2148 more closely matches the 1D-RCE models at pressures above 1 bar, the retrieved profile becomes significantly steeper at pressures below 1 bar. Further investigation of the TP profile with additional retrieval parameterizations is likely necessary to robustly assess the disagreement between retrieval and self-consistent models.

\section{Discussion}\label{sec:disc}

\subsection{Cloud Properties}

\begin{figure*}
    \centering
    \includegraphics[width=1\linewidth]{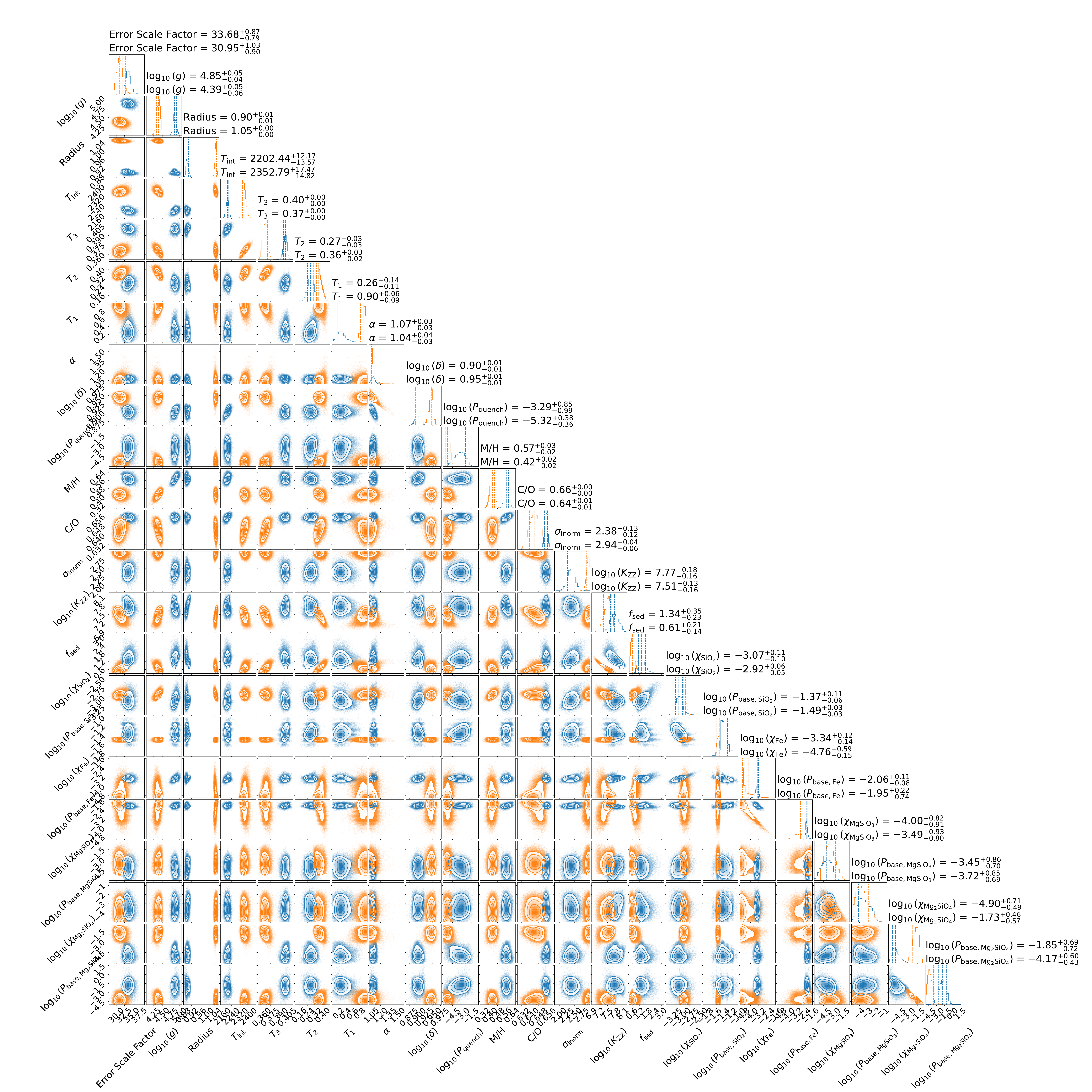}
    \caption{Posterior distributions from our fiducial retrievals of J0624 (top labels, blue) and J2148 (bottom labels, orange).}
    \label{fig:combined_corner}
\end{figure*}

Our observing program (JWST-GO-2288) originally selected J0624 and J2148 due to their close proximity ($\sim10$~pc), their similar spectral type (L5 and L6, respectively), and their seemingly different colors: J2148 appears to be a redder, slightly later-type object compared to J0624. Our hypothesis was that J2148 should be cloudier and that by comparing their entire infrared emission spectra, we could determine if this ``cloudiness" corresponded with changes in particle composition, size, crystallinity, or vertical distribution. 

Figure~\ref{fig:combined_corner} compares the posterior distribution for our fiducial retrievals of each object. Our retrieval results broadly support the picture that J2148 is qualitatively cloudier than J0624 as illustrated in Tables~\ref{table:J0624:rets}-\ref{table:J2148:rets:polymorphs} and Figure~\ref{fig:psizes}. While constraining the total cloud mass is difficult, J2148 appears to have more lofted clouds, made up of smaller particles. The $K_{\rm{zz}}$ measurements for J2148 are generally somewhat higher than J0624, but the most consequential difference is the lower $f_{\rm sed}$ value for J2148. Values for $f_{\rm sed}$ are almost uniformly below 1.0 for J2148, yet are above 1.0 for J0624, reaching $4.55 \pm 0.63$ for the best-fit amorphous SiO$_2$ retrieval. This picture of a lower $f_{\rm sed}$ for J2148 compared to J0624 is further supported by the Sonora Diamondback grid comparisons shown in Figure~\ref{fig:chi_fsed}. This difference in mixing and sedimentation may have tangible effects on the crystallinity of the particles with the amorphous particles in J0624 falling out of the atmosphere before they can anneal into a crystalline structure \citep{fabian:2000,toppani:2006,koike:2013,moran:2024}.

Future observations of other L-dwarfs may shed on the physical processes causing these differences. For example, J2148 is known to be observed equator-on \citep{vos:2017,suarez:2023}, while J0624 is unconstrained and most likely observed at a mid-latitude. Thus, latitudinal differences in atmospheric circulation may be consequential for cloud formation and microphysics. Other JWST programs (e.g., JWST-GO-5474) are observing more cloudy L-dwarfs with differing inclinations and surface gravities, potentially revealing population trends.

As for cloud composition, a picture is already emerging that more simple silicate minerals may be present in substellar atmospheres than previously thought. Early kinetic and chemical equilibrium theoretical studies indicated that iron- and magnesium silicates were most likely to form in the atmospheres of brown dwarfs and exoplanets \citep[e.g.,][]{visscher:2010}, though some studies had noted SiO$_2$ as an important constituent \citep{helling:2006}. Yet, early observations with JWST have revealed a different picture. MIRI/LRS transmission spectroscopy measurements of hot Jupiter WASP-17b were best fit by a SiO$_2$ cloud \citep{grant:2023}. As mentioned above, planetary mass PSO-513 appears to be best-fit by a SiO cloud.

Our observations agree with this picture; both our retrievals and CMA-ES fits of J0624 and J2148 found strong evidence for SiO$_2$ clouds. In both cases, the presence of magnesium silicates was not justified in the retrievals according to the Bayesian evidence, but iron- and magnesium-rich pyroxene were strongly favored in the CMA-ES optimization framework. Meanwhile, the retrieval found evidence for Fe clouds in J2148 and Al$_2$O$_3$ clouds in J0624 (though this is likely just a preference for a deep, gray cloud deck). Together, this implies multiple cloud species and layers are likely present, in agreement with previous analyses \citep{burningham:2021}. Future modeling and observation can help explore whether the preference for simple silicates reflects a elemental abundance variation \citep{calamari:2024} or is microphysical in origin.

The suggested presence of iron-rich silicates in both objects from the CMA-ES optimization is intriguing because such species have previously been unexpected due to the initial expected condensation of Fe into metal and the low oxygen fugacity of substellar atmospheres \citep{ebel:2006,visscher:2010}. While such a detection would be novel and would challenge current models of cloud formation, we note that the precise form of iron and magnesium silicate is not unambiguously constrained despite our high-SNR observations, with small differences between different cloud scenarios. Perhaps non-solar Mg, Si, and Fe abundances \citep{calamari:2024} or microphysical details may result in a wider set of observed cloud species than expected. We suggest future model comparisons of cloudy L-dwarf spectra fit a wide variety of cloud scenarios and species.

Another consideration is whether either brown dwarf we have observed is variable in time. Here, we present a single snapshot of each object, but other cloudy objects have shown significant variability through complex interactions of the atmospheric temperature, chemistry, and clouds \citep{biller:2024,nasedkin:2025,oliverosgomez:2025}. Future time-series observations of these two objects would reveal whether the cloud properties we measure here are stable with time or whether these worlds are more dynamic. A measurement of J0624's inclination via its rotation rate and $v\sin{i}$ would also be helpful for interpreting the difference in clouds between J0624 and J2148.

\subsection{Bulk Properties} \label{sec:disc:substellar}

Our analysis highlights the fact that even with unprecedented wavelength coverage and signal-to-noise, the inference of a substellar object's bulk properties, like radius, mass, structure, and composition, remains a difficult problem. Here, we chose to focus on the identification of the cloud properties and were unable to explore the full range of possible bulk property parameterizations. Most glaringly, our desire to calculate many retrieval scenarios across the full wavelength regime in an efficient manner dictated a reduction in resolving power to $R\sim100$, while our observations were taken at much higher spectral resolution ($R\sim3,000$). In future work, we hope to expand on the analysis of L-dwarf bulk properties now that statistically-justified cloud property scenarios have been identified.

\subsubsection{Metallicity}

Our measurement of super-solar metallicity in both J0624 and J2148 is surprising. While the intrinsic scatter of stellar metallicities does reach up to $[M/H] \approx 0.5$ for young objects \citep{feuillet:2019}, it seems unlikely that two separate field dwarfs would exhibit metallicities around this upper limit by chance. We note that retrieval analyses of cloudy L-dwarf J2224 also indicated a super-solar metallicity of [M/H] = $0.38^{+0.07}_{-0.06}$ and C/O = $0.83^{+0.06}_{-0.07}$ \citep{burningham:2021}. Retrievals of other substellar companions also show elevated metallicities approaching and sometimes exceeding exceeding 10$\times$ solar, including HR~8799 c, d, and e \citep[][but see \citealt{xuan:2026}]{nasedkin:2024b}, HD 72946 B \citep{balmer:2023}, and AF Lep b \citep{balmer:2025}.

Free retrievals of less-cloudy T-dwarfs do not appear to show any significant metallicity enhancement \citep{zalesky:2022}, suggesting either a bias in our retrievals of cloudy brown dwarfs or a bias in retrieval with chemical equilibrium due to disequilibrium processes remaining relevant to L-dwarfs. While we do include a quench pressure parameterization to account for disequilibrium chemistry, only the J0624 retrievals found a $P_{\rm quench}$ that was not above the IR photosphere. This was likely used by the retrieval to help fit the CH$_4$ $q$-branch absorption at 3.3\,$\mu$m seen in J0624. More physically motivated disequilibrium chemistry treatments (i.e., using a time-scale based setup) may be an avenue to more robustly constraining the bulk composition of these objects.

An additional source of bias in the composition constraints may be the choice of temperature profile parameterization: mock L-dwarf retrievals of cloudy L-dwarfs indicated that physically motivated temperature profile parameterizations can bias metallicity measurements by over 0.1 dex compared to more flexible layer-by-layer parameterizations \citep{wang:2022,rowland:2023}. The retrieved TP profiles for both J0624 and J2148 are much steeper than the 1D-RCE models (see Sections~\ref{sec:J0624:TP} and \ref{sec:J2148:TP}), so it is possible that the retrieval increases the molecular abundances to produce features of the right strength to compensate. Future sensitivity analyses may shed light on this issue.

One last possible bias affecting the metallicity measurement may be with the cloud parameterization \citep{balmer:2023,balmer:2025}. The lack of flexibility in setting the particle size and its distribution may be requiring the retrieval to adjust the atmospheric opacity by increasing [M/H] to match the data. This may be especially true in our data, where the retrieval may have have to choose between fitting the MIR silicate feature with small particles and having the correct opacity at the photosphere through large particles.

\subsubsection{Gravity, Radius, and Mass}

Lastly, we consider our retrieved surface gravity, radius, and mass. Retrievals of self-luminous objects have historically run into issues with fitting radius; the retrieval will often use the radius as a dilution factor to adjust the effective temperature up or down. This can easily lead to unphysical retrieval results.

For example, retrievals of cloudy L-dwarf J2224 \citep{burningham:2021} found a radius of 0.75$\pm$0.02, which is significantly lower than the 0.99$\pm$0.08 inferred from the planets distance and photometry \citep{filippazzo:2015}. Our retrievals of J0624 find a radius about 1--2$\sigma$ smaller than in \cite{filippazzo:2015}, while the radii retrieved for J2148 generally agree within 1$\sigma$.

A bigger discrepancy is found in the retrieved surface gravity, which also results in unexpectedly low masses. For J0624, the surface gravity is about 0.3--0.5 dex lower than expected from a normal field dwarf or from \cite{filippazzo:2015}. For J2148, the discrepancy is worse, at about 0.7--0.8 dex in surface gravity. For both objects, this results in masses between 8 and 23 $M_{\rm Jup}$, depending on the retrieval scenario. Below, we evaluate these results compared to evolution models and our empirically-measured flux.

\subsubsection{Evolution Analysis of J0624 and J2148}\label{sec:disc:evo}

With our retrieved radii for J0624 and J2148, we can use the spectrum to calculate a semi-empirical effective temperature. Using these two quantities, we can then compare to evolutionary models to assess the reasonableness of our retrieved surface gravity and mass constraints, as well as to estimate an approximate age. We first integrate the measured flux from our observations between 0.98 and 17.6~$\mu$m. This flux is then scaled by the distance modulus, taking the radius of the preferred retrieval scenario for each object ($R_{\rm Jup} = 0.915 \pm 0.004$ and $1.05 \pm 0.004$ for J0624 and J2148, respectively) and the Gaia distance ($12.191 \pm 0.054$ and $8.086 \pm 0.023$ pc, respectively). From the Stefan-Boltzman law, we obtain a preliminary effective temperature of 1525.5 and 1438.2~K, respectively. We then use this preliminary temperature to calculate a bolometric correction to find the amount of flux estimated to be outside our observed bandpass by integrating a blackbody at our preliminary temperatures between 0.98 and 17.6~$\mu$m. We estimate that we measure 98.1\% and 98.4\% of the flux from each object in our bandpass. We use this factor to estimate a final bolometric flux from which we measure an empirical effective temperature of 1532.8 $\pm$ 7.3~K and 1443.9 $\pm$ 5.7~K for J0624 and J2148, respectively, where the measurement uncertainty is taken to be the difference between the preliminary and bolometrically-corrected temperature.

\begin{figure}
    \centering
    \includegraphics[width=1.0\linewidth]{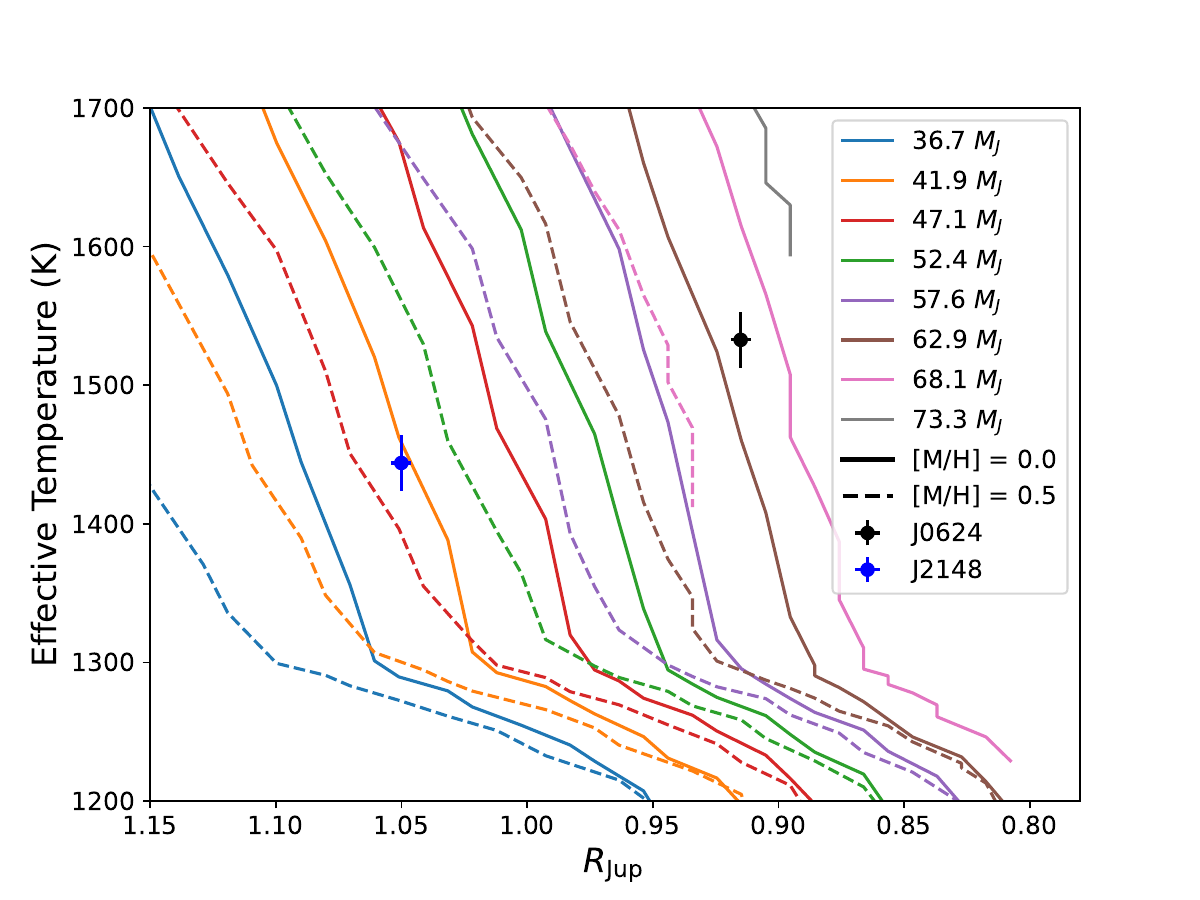}
    \caption{Brown dwarf evolution tracks from \cite{davis:2025} compared to the radii of J0624 and J2148 from the best-fitting retrieval with effective temperature then measured from the flux from each object. Solid lines show tracks for objects with solar metallicity, while dashed lines show tracks for elevated metallicity objects.}
    \label{fig:evo}
\end{figure}

With a measured radius and temperature, we can now compare to brown dwarf evolution tracks to constrain possible ages and masses for the objects. Crucially, this will not depend on the potentially biased surface gravity estimated from the retrieval, though it still depends on the inferred radius. We use the Red Diamondback Sonora substellar atmosphere grid \citep{davis:2025} to find a best matching isochrone in radius versus effective temperature space, shown in Figure~\ref{fig:evo}. We perform this calculation for both the solar metallicity models and, to match our retrieval results, the $[M/H] = 0.5$ elevated metallicity models.

J0624 lies between two solar metallicity isochrones at 0.06$M_{\rm Sun}$ = 62.9$M_{\rm Jup}$ and 0.065$M_{\rm Sun}$ = 68.1$M_{\rm Jup}$, indicating an age between 1.2 and 1.9 Gyr and $\log{g}$ between 5.26 and 5.32 -- near the values found in \cite{filippazzo:2015}. No elevated metallicity track is near J0624, as the 0.07$M_{\rm Sun}$ = 73.3$M_{\rm Jup}$ track ends at a temperature of 1700~K, but J0624's location above the 0.065$M_{\rm Sun}$ would imply and age greater than 2.2 Gyr and $\log{g} > 5.28$.

J2148 lies within 1$\sigma$ of the 0.04$M_{\rm Sun}$ = 41.9$M_{\rm Jup}$ solar metallicity track, corresponding to an age of 0.42 Gyr and $\log{g} = 4.98$. For the elevated metallicity isochrones, J2148 lies between the 0.045$M_{\rm Sun}$ = 47.1$M_{\rm Jup}$ and 0.05$M_{\rm Sun}$ = 52.4$M_{\rm Jup}$ tracks, implying an age between 0.57 and 0.77 Gyr and $\log{g}$ between 5.01 and 5.09. The inferred surface gravity from the isochrones and retrieved radius imply a much higher surface gravity than obtained in the retrievals by more than half a dex. If the surface gravity of J0624 and J2148 were truly as low as the retrievals imply, we would expect them to be very young and much larger if at the same effective temperature. 

\section{Conclusion}\label{sec:conc}

We have presented a modeling analysis of 1-28$\mu$m observations of two cloudy L-dwarfs, J0624 and J2148. We first compared our observations to grids of 1-D radiative-convective equilibrium models from BT-Settl, Sonora Diamondback, and Exo-REM. While these self-consistent models did not have the necessary flexibility to fit our high-signal to noise observations, reasonable effective temperatures and $f_{\rm sed}$ parameters were found. Compositional information in the form of [M/H] and C/O, along with surface gravity, were poorly constrained by the self-consistent models.

We next presented a variety of 1-D atmospheric retrievals and CMA-ES optimization analyses, focusing on testing different cloud property scenarios. Our findings indicate that both J0624 and J2148 have SiO$_2$ clouds. For J2148, these clouds appear to be lofted, sub-$\mu$m crystalline SiO$_2$ particles. Both the retrieval and CMA-ES optimization prefer quartz particles compared to tridymite or cristobalite polymorphs. For J0624, the observations were better fit by slightly larger, more vertically-compact amorphous SiO$_2$ particles. The CMA-ES also showed strong evidence for amorphous iron-rich magnesium silicates, with both Mg$_{0.5}$Fe$_{0.5}$SiO$_3$ and MgFeSiO$_4$ fitting well for both objects. A comparison to more L-dwarfs and other sub-stellar objects may shed light on these compositional and microphysical differences.

Lastly, we discuss other aspects of the retrieval, including our measurement of elevated metallicity and low surface-gravity in both objects. We interpret these measurements in comparison to other brown dwarf measurements, and suggest that remaining retrieval biases may need to be mitigated for more robust bulk property inferences. We also compare our retrieved bulk properties to evolution models, finding reasonable constraints on age and mass if we use the empirically measured flux and our retrieved radii but ignore the retrieved surface gravity, which appear to biased low in both objects.

In future works, we plan to connect J0624 and J2148 to other substellar objects through empirical comparisons. While this work focused specifically on the cloud properties of the objects, we also plan to also understand the bulk properties of J0624 and J2148, like metallicity and the TP structure, compared to the broader population of brown dwarfs through a more comprehensive set of retrievals, specifically testing many of the parameterizations used in this work. We also plan to leverage the high-resolution of the JWST data to measure the isotope ratios \citep[e.g.,][]{gandhi:2023b}.

As mentioned above, further observation of L-dwarfs will help establish trends with temperature, observed latitude, and mass, while also defining the expected intrinsic diversity we may expect from field brown dwarfs. Already, JWST has shown a surprising variety among brown dwarfs in the rich spectra already observed \citep[e.g.,][]{miles:2023,faherty:2024,biller:2024,faherty:2025,molliere:2025,nasedkin:2025,oliverosgomez:2025}. Our observations here highlight the remaining open-questions in our understanding of cloudy L-dwarfs, while simultaneously demonstrating the explanatory power of wide-wavelength JWST observations.

\section{Acknowledgements}

We thank the referee for a constructive and helpful review that improved the manuscript. We thank Evert Nasedkin for sharing quartz and amorphous SiO$_2$ cloud opacities. We thank Jacqueline Faherty for continual encouragement towards publication of these results. This work is based on observations made with the NASA/ESA/CSA James Webb Space Telescope. All of the data presented in this paper were obtained from the Mikulski Archive for Space Telescopes (MAST) at the Space Telescope Science Institute. The specific observations analyzed can be accessed via \dataset[https://doi.org/10.17909/xa61-yf45]{https://doi.org/10.17909/xa61-yf45}. STScI is operated by the Association of Universities for Research in Astronomy, Inc., under NASA contract NAS5–26555. Support to MAST for these data is provided by the NASA Office of Space Science via grant NAG5–7584 and by other grants and contracts. These observations are associated with program JWST-GO-2288. Support for program JWST-GO-2288 was provided by NASA through a grant from the Space Telescope Science Institute, which is operated by the Association of Universities for Research in Astronomy, Inc., under NASA contract NAS 5-03127. This research has made use of the VizieR catalog access tool, CDS, Strasbourg, France.  The original description of the VizieR service was published in \citet{ochsenbein:2000}.  This research has made use of NASA's Astrophysics Data System Bibliographic Services.


\vspace{5mm}
\facilities{JWST(NIRSpec, MIRI)}

\software{astropy \citep{astropy:2018}, matplotlib \citep{hunter:2007}, NumPy \citep{numpy:2020}, iPython \citep{perez:2007}, petitRADTRANS \citep{molliere:2019,nasedkin:2024}
          }

\appendix
\section{Cloud Scenario Best Fits with CMA-ES}

\startlongtable 
\begin{deluxetable*}{lllllll}\label{tab:J0624:CMA}
\tabletypesize{\small}
\tablecaption{J0624 Best-Fit Cloud Scenarios with CMA-ES}
\tablehead{
  \colhead{$\chi^2$/N} &
  \colhead{SiO$_2$} &
  \colhead{MgSiO$_3$} &
  \colhead{Mg$_2$SiO$_4$} &
  \colhead{Mg$_{0.5}$Fe$_{0.5}$SiO$_3$} &
  \colhead{MgFeSiO$_4$} &
  \colhead{Size Dist.}
}
\startdata
  50.737 & Amorph./Mie & \nodata & \nodata & \nodata & Amorph./Mie & Hansen \\
  51.387 & Amorph./Mie & \nodata & \nodata & Amorph./Mie & \nodata & Hansen \\
  52.540 & Amorph./DHS & \nodata & \nodata & Amorph./DHS & \nodata & Log-normal \\
  52.885 & Amorph./Mie & \nodata & \nodata & Amorph./Mie & \nodata & Log-normal \\
  53.604 & Crys./DHS & Crys./DHS & \nodata & \nodata & \nodata & Log-normal \\
  53.756 & Amorph./DHS & \nodata & \nodata & \nodata & Amorph./DHS & Hansen \\
  54.562 & Amorph./DHS & \nodata & \nodata & \nodata & Amorph./DHS & Log-normal \\
  55.229 & Amorph./DHS & Crys./DHS & \nodata & \nodata & \nodata & Hansen \\
  55.250 & \nodata & \nodata & \nodata & Amorph./Mie & Amorph./Mie & Log-normal \\
  55.311 & Crys./DHS & \nodata & \nodata & \nodata & Amorph./DHS & Log-normal \\
  55.424 & Amorph./Mie & \nodata & Amorph./Mie & \nodata & \nodata & Log-normal \\
  55.659 & Crys./DHS & Crys./DHS & \nodata & \nodata & \nodata & Hansen \\
  55.765 & Crys./DHS & \nodata & \nodata & \nodata & Amorph./DHS & Hansen \\
  56.011 & Crys./DHS & \nodata & \nodata & Amorph./DHS & \nodata & Hansen \\
  56.031 & Amorph./DHS & \nodata & Crys./DHS & \nodata & \nodata & Log-normal \\
  56.641 & Crys./DHS & \nodata & \nodata & Amorph./DHS & \nodata & Log-normal \\
  56.685 & Crist./DHS & \nodata & \nodata & \nodata & Amorph./DHS & Hansen \\
  56.728 & Amorph./Mie & \nodata & \nodata & \nodata & Amorph./Mie & Log-normal \\
  56.844 & Crys./Mie & \nodata & \nodata & Amorph./Mie & \nodata & Hansen \\
  57.022 & Amorph./DHS & \nodata & Amorph./DHS & \nodata & \nodata & Log-normal \\
  57.070 & \nodata & \nodata & \nodata & Amorph./Mie & \nodata & Hansen \\
  57.448 & \nodata & \nodata & Crys./Mie & Amorph./Mie & \nodata & Log-normal \\
  57.462 & \nodata & \nodata & \nodata & Amorph./Mie & \nodata & Log-normal \\
  57.612 & Amorph./Mie & \nodata & Crys./Mie & \nodata & \nodata & Hansen \\
  57.752 & Crys./Mie & \nodata & \nodata & \nodata & Amorph./Mie & Hansen \\
  57.945 & \nodata & \nodata & Crys./Mie & Amorph./Mie & \nodata & Hansen \\
  58.180 & Amorph./Mie & \nodata & \nodata & \nodata & \nodata & Log-normal \\
  58.255 & Crys./Mie & Crys./Mie & \nodata & \nodata & \nodata & Hansen \\
  58.730 & Amorph./DHS & \nodata & Crys./DHS & \nodata & \nodata & Hansen \\
  59.210 & Amorph./Mie & \nodata & \nodata & \nodata & \nodata & Hansen \\
  59.232 & Amorph./Mie & \nodata & Amorph./Mie & \nodata & \nodata & Hansen \\
  59.300 & Crys./DHS & \nodata & Amorph./DHS & \nodata & \nodata & Hansen \\
  59.537 & Crys./DHS & \nodata & Crys./DHS & \nodata & \nodata & Log-normal \\
  59.855 & \nodata & \nodata & \nodata & Amorph./Mie & Amorph./Mie & Hansen \\
  60.300 & Trid./DHS & Crys./DHS & \nodata & \nodata & \nodata & Log-normal \\
  60.350 & \nodata & \nodata & Amorph./Mie & Amorph./Mie & \nodata & Hansen \\
  60.546 & Amorph./DHS & Amorph./DHS & \nodata & \nodata & \nodata & Log-normal \\
  60.687 & Amorph./DHS & Crys./DHS & \nodata & \nodata & \nodata & Log-normal \\
  60.738 & Crys./Mie & \nodata & \nodata & \nodata & Amorph./Mie & Log-normal \\
  60.743 & Crist./DHS & \nodata & \nodata & Amorph./DHS & \nodata & Hansen \\
  60.763 & Crys./DHS & \nodata & Crys./DHS & \nodata & \nodata & Hansen \\
  60.806 & Trid./DHS & Crys./DHS & \nodata & \nodata & \nodata & Hansen \\
  60.841 & Amorph./Mie & Amorph./Mie & \nodata & \nodata & \nodata & Log-normal \\
  61.822 & Trid./DHS & \nodata & Amorph./DHS & \nodata & \nodata & Log-normal \\
  61.894 & Crist./DHS & Crys./DHS & \nodata & \nodata & \nodata & Log-normal \\
  61.964 & Amorph./DHS & \nodata & \nodata & \nodata & \nodata & Log-normal \\
  62.016 & Amorph./DHS & \nodata & \nodata & \nodata & \nodata & Hansen \\
  62.057 & Amorph./DHS & \nodata & \nodata & Amorph./DHS & \nodata & Hansen \\
  62.133 & Trid./DHS & \nodata & \nodata & Amorph./DHS & \nodata & Hansen \\
  62.457 & Crys./Mie & \nodata & Amorph./Mie & \nodata & \nodata & Hansen \\
  62.715 & Amorph./Mie & \nodata & Crys./Mie & \nodata & \nodata & Log-normal \\
  62.739 & Trid./DHS & \nodata & \nodata & Amorph./DHS & \nodata & Log-normal \\
  62.810 & Crys./DHS & Amorph./DHS & \nodata & \nodata & \nodata & Hansen \\
  62.847 & Trid./DHS & Amorph./DHS & \nodata & \nodata & \nodata & Hansen \\
  62.936 & Trid./DHS & \nodata & \nodata & \nodata & Amorph./DHS & Log-normal \\
  64.123 & Crist./DHS & \nodata & Amorph./DHS & \nodata & \nodata & Log-normal \\
  64.468 & Trid./DHS & \nodata & \nodata & \nodata & \nodata & Log-normal \\
  64.650 & Amorph./Mie & Crys./Mie & \nodata & \nodata & \nodata & Hansen \\
  64.894 & Crist./DHS & \nodata & \nodata & \nodata & \nodata & Log-normal \\
  65.238 & Crist./DHS & \nodata & \nodata & \nodata & \nodata & Hansen \\
  65.327 & Crys./DHS & \nodata & \nodata & \nodata & \nodata & Hansen \\
  65.663 & Trid./DHS & \nodata & Amorph./DHS & \nodata & \nodata & Hansen \\
  65.702 & Trid./DHS & \nodata & Crys./DHS & \nodata & \nodata & Log-normal \\
  65.950 & Amorph./Mie & Crys./Mie & \nodata & \nodata & \nodata & Log-normal \\
  65.973 & Crist./DHS & \nodata & Amorph./DHS & \nodata & \nodata & Hansen \\
  66.095 & Crys./DHS & \nodata & \nodata & \nodata & \nodata & Log-normal \\
  66.258 & Crist./DHS & \nodata & Crys./DHS & \nodata & \nodata & Log-normal \\
  66.260 & Crist./DHS & \nodata & Crys./DHS & \nodata & \nodata & Hansen \\
  66.269 & Trid./DHS & Amorph./DHS & \nodata & \nodata & \nodata & Log-normal \\
  66.488 & Crist./DHS & Amorph./DHS & \nodata & \nodata & \nodata & Hansen \\
  67.019 & \nodata & Amorph./Mie & \nodata & \nodata & Amorph./Mie & Log-normal \\
  67.370 & Crys./Mie & \nodata & Crys./Mie & \nodata & \nodata & Hansen \\
  67.677 & \nodata & Amorph./Mie & \nodata & \nodata & Amorph./Mie & Hansen \\
  68.040 & Trid./DHS & \nodata & \nodata & \nodata & Amorph./DHS & Hansen \\
  68.377 & \nodata & Amorph./Mie & Crys./Mie & \nodata & \nodata & Log-normal \\
  68.482 & Trid./DHS & \nodata & \nodata & \nodata & \nodata & Hansen \\
  68.584 & Crist./DHS & Crys./DHS & \nodata & \nodata & \nodata & Hansen \\
  68.970 & Crys./Mie & Amorph./Mie & \nodata & \nodata & \nodata & Log-normal \\
  68.977 & \nodata & Amorph./DHS & Crys./DHS & \nodata & \nodata & Log-normal \\
  69.097 & \nodata & Amorph./Mie & Amorph./Mie & \nodata & \nodata & Log-normal \\
  69.173 & \nodata & Amorph./Mie & Crys./Mie & \nodata & \nodata & Hansen \\
  69.275 & Crys./Mie & \nodata & \nodata & \nodata & \nodata & Log-normal \\
  69.622 & Crys./Mie & Amorph./Mie & \nodata & \nodata & \nodata & Hansen \\
  70.148 & \nodata & Amorph./DHS & \nodata & \nodata & \nodata & Hansen \\
  70.174 & Amorph./Mie & Amorph./Mie & \nodata & \nodata & \nodata & Hansen \\
  70.468 & Crist./DHS & Amorph./DHS & \nodata & \nodata & \nodata & Log-normal \\
  70.622 & \nodata & Amorph./DHS & \nodata & \nodata & Amorph./DHS & Log-normal \\
  70.626 & \nodata & Amorph./Mie & \nodata & \nodata & \nodata & Log-normal \\
  70.886 & \nodata & Amorph./DHS & \nodata & \nodata & \nodata & Log-normal \\
  71.199 & \nodata & Amorph./DHS & Amorph./DHS & \nodata & \nodata & Hansen \\
  72.365 & Crist./DHS & \nodata & \nodata & Amorph./DHS & \nodata & Log-normal \\
  72.545 & Crys./Mie & \nodata & \nodata & \nodata & \nodata & Hansen \\
  72.860 & \nodata & Amorph./Mie & Amorph./Mie & \nodata & \nodata & Hansen \\
  72.898 & Amorph./DHS & Amorph./DHS & \nodata & \nodata & \nodata & Hansen \\
  72.936 & \nodata & \nodata & \nodata & Amorph./DHS & Amorph./DHS & Hansen \\
  73.077 & Crys./Mie & \nodata & Amorph./Mie & \nodata & \nodata & Log-normal \\
  73.195 & \nodata & \nodata & Crys./DHS & Amorph./DHS & \nodata & Log-normal \\
  73.500 & \nodata & \nodata & \nodata & Amorph./DHS & \nodata & Hansen \\
  73.642 & \nodata & \nodata & Amorph./DHS & Amorph./DHS & \nodata & Hansen \\
  73.683 & \nodata & Amorph./DHS & \nodata & \nodata & Amorph./DHS & Hansen \\
  73.981 & Crys./Mie & \nodata & \nodata & Amorph./Mie & \nodata & Log-normal \\
  74.421 & Crys./Mie & Crys./Mie & \nodata & \nodata & \nodata & Log-normal \\
  74.721 & \nodata & Amorph./DHS & Amorph./DHS & \nodata & \nodata & Log-normal \\
  75.023 & Crys./Mie & \nodata & Crys./Mie & \nodata & \nodata & Log-normal \\
  75.484 & \nodata & \nodata & \nodata & Amorph./DHS & Amorph./DHS & Log-normal \\
  75.514 & \nodata & \nodata & Crys./DHS & Amorph./DHS & \nodata & Hansen \\
  75.910 & \nodata & Amorph./Mie & \nodata & \nodata & \nodata & Hansen \\
  77.093 & \nodata & \nodata & \nodata & Amorph./DHS & \nodata & Log-normal \\
  77.235 & Crist./DHS & \nodata & \nodata & \nodata & Amorph./DHS & Log-normal \\
  79.859 & Trid./DHS & \nodata & Crys./DHS & \nodata & \nodata & Hansen \\
  84.127 & \nodata & \nodata & Amorph./Mie & Amorph./Mie & \nodata & Log-normal \\
  85.124 & \nodata & \nodata & Amorph./DHS & Amorph./DHS & \nodata & Log-normal \\
  95.541 & Crys./DHS & Amorph./DHS & \nodata & \nodata & \nodata & Log-normal \\
  95.790 & Amorph./DHS & \nodata & Amorph./DHS & \nodata & \nodata & Hansen \\
  97.610 & Crys./DHS & \nodata & Amorph./DHS & \nodata & \nodata & Log-normal \\
  102.093 & \nodata & Crys./Mie & Crys./Mie & \nodata & \nodata & Log-normal \\
  102.274 & \nodata & Crys./Mie & \nodata & \nodata & Amorph./Mie & Hansen \\
  103.003 & \nodata & Crys./Mie & \nodata & \nodata & \nodata & Log-normal \\
  104.257 & \nodata & \nodata & Amorph./Mie & \nodata & \nodata & Log-normal \\
  104.867 & \nodata & \nodata & Amorph./Mie & \nodata & \nodata & Hansen \\
  104.951 & \nodata & Crys./Mie & Amorph./Mie & \nodata & \nodata & Log-normal \\
  105.702 & \nodata & Crys./Mie & Amorph./Mie & \nodata & \nodata & Hansen \\
  106.602 & \nodata & Crys./Mie & \nodata & \nodata & Amorph./Mie & Log-normal \\
  108.333 & \nodata & Crys./DHS & \nodata & \nodata & Amorph./DHS & Log-normal \\
  109.525 & \nodata & Crys./DHS & \nodata & \nodata & \nodata & Log-normal \\
  110.293 & \nodata & Crys./Mie & \nodata & \nodata & \nodata & Hansen \\
  110.617 & \nodata & \nodata & \nodata & \nodata & Amorph./Mie & Hansen \\
  111.228 & \nodata & Crys./DHS & Crys./DHS & \nodata & \nodata & Hansen \\
  111.843 & \nodata & \nodata & \nodata & \nodata & Amorph./DHS & Hansen \\
  112.110 & \nodata & Crys./DHS & \nodata & \nodata & Amorph./DHS & Hansen \\
  113.496 & \nodata & \nodata & \nodata & \nodata & Amorph./DHS & Log-normal \\
  115.284 & \nodata & \nodata & \nodata & \nodata & Amorph./Mie & Log-normal \\
  115.706 & \nodata & Crys./DHS & Crys./DHS & \nodata & \nodata & Log-normal \\
  116.000 & \nodata & Crys./DHS & Amorph./DHS & \nodata & \nodata & Log-normal \\
  118.428 & \nodata & Amorph./DHS & Crys./DHS & \nodata & \nodata & Hansen \\
  122.498 & \nodata & Crys./DHS & \nodata & \nodata & \nodata & Hansen \\
  123.261 & \nodata & Crys./Mie & Crys./Mie & \nodata & \nodata & Hansen \\
  125.609 & \nodata & \nodata & Amorph./DHS & \nodata & \nodata & Log-normal \\
  132.254 & \nodata & \nodata & Crys./DHS & \nodata & \nodata & Log-normal \\
  135.958 & \nodata & \nodata & Amorph./DHS & \nodata & \nodata & Hansen \\
  137.260 & \nodata & Crys./DHS & Amorph./DHS & \nodata & \nodata & Hansen \\
  140.005 & \nodata & \nodata & Crys./DHS & \nodata & \nodata & Hansen \\
  155.167 & \nodata & \nodata & Crys./Mie & \nodata & \nodata & Log-normal \\
  161.814 & \nodata & \nodata & Crys./Mie & \nodata & \nodata & Hansen \\
\enddata
\tablecomments{The size distribution column applies to all cloud species in that model. Each dust species column gives crystallinity/opacity treatment. SiO$_2$ crystallinity: Qtz.=Quartz, Trid.=Tridymite, Crist.=Cristobalite, Amorph.=Amorphous. Other species: Crys.=Crystalline, Amorph.=Amorphous. Opacity: Pure Mie or DHS (Distribution of Hollow Spheres) \citep{min:2005}.}
\end{deluxetable*}

\startlongtable
\begin{deluxetable*}{lllllll}\label{tab:J2148:CMA}
\tabletypesize{\small}
\tablecaption{J2148 Best-Fit Cloud Scenarios with CMA-ES}
\tablehead{
  \colhead{$\chi^2$/N} &
  \colhead{SiO$_2$} &
  \colhead{MgSiO$_3$} &
  \colhead{Mg$_2$SiO$_4$} &
  \colhead{Mg$_{0.5}$Fe$_{0.5}$SiO$_3$} &
  \colhead{MgFeSiO$_4$} &
  \colhead{Size Dist.}
}
\startdata
  62.215 & Crys./Mie & \nodata & \nodata & Amorph./Mie & \nodata & Log-normal \\
  67.559 & Crys./DHS & \nodata & \nodata & \nodata & Amorph./DHS & Log-normal \\
  70.995 & Crys./Mie & \nodata & \nodata & Amorph./Mie & \nodata & Hansen \\
  78.660 & Amorph./DHS & \nodata & Amorph./DHS & \nodata & \nodata & Log-normal \\
  82.605 & Crys./Mie & Crys./Mie & \nodata & \nodata & \nodata & Hansen \\
  83.615 & Amorph./Mie & \nodata & \nodata & \nodata & Amorph./Mie & Hansen \\
  84.379 & Crys./Mie & \nodata & \nodata & \nodata & Amorph./Mie & Log-normal \\
  85.891 & Crist./DHS & \nodata & \nodata & \nodata & Amorph./DHS & Log-normal \\
  86.028 & Amorph./Mie & \nodata & Amorph./Mie & \nodata & \nodata & Log-normal \\
  86.507 & Trid./DHS & \nodata & Amorph./DHS & \nodata & \nodata & Log-normal \\
  87.584 & Crys./DHS & \nodata & Amorph./DHS & \nodata & \nodata & Log-normal \\
  88.029 & Crys./DHS & \nodata & Crys./DHS & \nodata & \nodata & Hansen \\
  90.317 & Crys./Mie & \nodata & Amorph./Mie & \nodata & \nodata & Log-normal \\
  91.451 & Amorph./DHS & \nodata & Amorph./DHS & \nodata & \nodata & Hansen \\
  92.469 & Crys./DHS & \nodata & \nodata & \nodata & Amorph./DHS & Hansen \\
  92.790 & Crist./DHS & \nodata & \nodata & Amorph./DHS & \nodata & Log-normal \\
  93.209 & Crys./DHS & Crys./DHS & \nodata & \nodata & \nodata & Log-normal \\
  94.413 & Crist./DHS & Crys./DHS & \nodata & \nodata & \nodata & Log-normal \\
  94.784 & Amorph./DHS & Crys./DHS & \nodata & \nodata & \nodata & Log-normal \\
  95.184 & Crys./Mie & Crys./Mie & \nodata & \nodata & \nodata & Log-normal \\
  95.349 & Trid./DHS & Amorph./DHS & \nodata & \nodata & \nodata & Log-normal \\
  97.830 & Crist./DHS & \nodata & Crys./DHS & \nodata & \nodata & Hansen \\
  98.803 & Amorph./DHS & \nodata & \nodata & \nodata & Amorph./DHS & Log-normal \\
  99.114 & Trid./DHS & Crys./DHS & \nodata & \nodata & \nodata & Log-normal \\
  99.392 & Crist./DHS & \nodata & Crys./DHS & \nodata & \nodata & Log-normal \\
  100.072 & Crys./Mie & \nodata & Amorph./Mie & \nodata & \nodata & Hansen \\
  102.674 & Amorph./Mie & Crys./Mie & \nodata & \nodata & \nodata & Hansen \\
  104.126 & Amorph./DHS & \nodata & Crys./DHS & \nodata & \nodata & Hansen \\
  104.470 & Crist./DHS & Amorph./DHS & \nodata & \nodata & \nodata & Log-normal \\
  104.619 & Amorph./DHS & Amorph./DHS & \nodata & \nodata & \nodata & Hansen \\
  105.102 & Crys./DHS & \nodata & Crys./DHS & \nodata & \nodata & Log-normal \\
  105.809 & Crys./DHS & \nodata & Amorph./DHS & \nodata & \nodata & Hansen \\
  107.536 & Amorph./DHS & Amorph./DHS & \nodata & \nodata & \nodata & Log-normal \\
  108.739 & Crys./Mie & Amorph./Mie & \nodata & \nodata & \nodata & Log-normal \\
  109.184 & Crys./Mie & Amorph./Mie & \nodata & \nodata & \nodata & Hansen \\
  109.347 & Amorph./DHS & \nodata & Crys./DHS & \nodata & \nodata & Log-normal \\
  111.137 & Crys./DHS & \nodata & \nodata & Amorph./DHS & \nodata & Hansen \\
  111.437 & Crys./Mie & \nodata & Crys./Mie & \nodata & \nodata & Log-normal \\
  113.354 & Amorph./Mie & Crys./Mie & \nodata & \nodata & \nodata & Log-normal \\
  114.121 & Amorph./DHS & \nodata & \nodata & Amorph./DHS & \nodata & Log-normal \\
  115.166 & \nodata & \nodata & Crys./Mie & Amorph./Mie & \nodata & Log-normal \\
  116.485 & Amorph./Mie & \nodata & \nodata & Amorph./Mie & \nodata & Log-normal \\
  117.363 & Crys./Mie & \nodata & \nodata & \nodata & Amorph./Mie & Hansen \\
  118.091 & Amorph./Mie & \nodata & Amorph./Mie & \nodata & \nodata & Hansen \\
  119.325 & \nodata & \nodata & \nodata & Amorph./Mie & \nodata & Log-normal \\
  120.224 & Crist./DHS & \nodata & \nodata & Amorph./DHS & \nodata & Hansen \\
  122.587 & \nodata & \nodata & \nodata & Amorph./Mie & Amorph./Mie & Hansen \\
  123.240 & Trid./DHS & \nodata & \nodata & Amorph./DHS & \nodata & Log-normal \\
  124.302 & Crys./DHS & Crys./DHS & \nodata & \nodata & \nodata & Hansen \\
  124.627 & Trid./DHS & Crys./DHS & \nodata & \nodata & \nodata & Hansen \\
  126.050 & Crist./DHS & \nodata & Amorph./DHS & \nodata & \nodata & Hansen \\
  127.819 & Crys./DHS & Amorph./DHS & \nodata & \nodata & \nodata & Hansen \\
  129.106 & Amorph./DHS & \nodata & \nodata & Amorph./DHS & \nodata & Hansen \\
  129.905 & Crist./DHS & \nodata & \nodata & \nodata & Amorph./DHS & Hansen \\
  131.024 & Crys./DHS & \nodata & \nodata & Amorph./DHS & \nodata & Log-normal \\
  131.130 & Trid./DHS & Amorph./DHS & \nodata & \nodata & \nodata & Hansen \\
  133.840 & Amorph./Mie & Amorph./Mie & \nodata & \nodata & \nodata & Log-normal \\
  134.296 & Amorph./Mie & \nodata & \nodata & \nodata & Amorph./Mie & Log-normal \\
  134.365 & \nodata & \nodata & \nodata & Amorph./Mie & \nodata & Hansen \\
  134.586 & Trid./DHS & \nodata & Crys./DHS & \nodata & \nodata & Log-normal \\
  134.856 & Trid./DHS & \nodata & \nodata & \nodata & Amorph./DHS & Log-normal \\
  135.347 & \nodata & \nodata & Amorph./Mie & Amorph./Mie & \nodata & Log-normal \\
  138.579 & Trid./DHS & \nodata & Amorph./DHS & \nodata & \nodata & Hansen \\
  140.463 & \nodata & \nodata & \nodata & Amorph./Mie & Amorph./Mie & Log-normal \\
  140.774 & Amorph./Mie & \nodata & \nodata & Amorph./Mie & \nodata & Hansen \\
  142.144 & Amorph./DHS & \nodata & \nodata & \nodata & Amorph./DHS & Hansen \\
  147.685 & Trid./DHS & \nodata & Crys./DHS & \nodata & \nodata & Hansen \\
  147.702 & \nodata & Amorph./Mie & \nodata & \nodata & Amorph./Mie & Hansen \\
  150.534 & \nodata & Amorph./DHS & \nodata & \nodata & \nodata & Log-normal \\
  152.248 & \nodata & Amorph./DHS & Amorph./DHS & \nodata & \nodata & Hansen \\
  152.373 & Trid./DHS & \nodata & \nodata & \nodata & \nodata & Log-normal \\
  152.957 & \nodata & Amorph./DHS & Crys./DHS & \nodata & \nodata & Log-normal \\
  153.325 & \nodata & Amorph./Mie & \nodata & \nodata & Amorph./Mie & Log-normal \\
  153.338 & Crist./DHS & Crys./DHS & \nodata & \nodata & \nodata & Hansen \\
  154.696 & \nodata & Amorph./Mie & \nodata & \nodata & \nodata & Log-normal \\
  154.872 & \nodata & Amorph./Mie & Amorph./Mie & \nodata & \nodata & Log-normal \\
  156.664 & \nodata & Amorph./DHS & \nodata & \nodata & Amorph./DHS & Log-normal \\
  156.862 & \nodata & Amorph./DHS & Amorph./DHS & \nodata & \nodata & Log-normal \\
  157.982 & \nodata & \nodata & Crys./Mie & Amorph./Mie & \nodata & Hansen \\
  160.093 & Amorph./DHS & Crys./DHS & \nodata & \nodata & \nodata & Hansen \\
  160.776 & \nodata & Amorph./DHS & \nodata & \nodata & \nodata & Hansen \\
  161.500 & Crys./DHS & Amorph./DHS & \nodata & \nodata & \nodata & Log-normal \\
  162.181 & \nodata & Amorph./Mie & Amorph./Mie & \nodata & \nodata & Hansen \\
  169.229 & \nodata & \nodata & Amorph./Mie & Amorph./Mie & \nodata & Hansen \\
  169.459 & Amorph./Mie & Amorph./Mie & \nodata & \nodata & \nodata & Hansen \\
  171.833 & \nodata & Amorph./Mie & \nodata & \nodata & \nodata & Hansen \\
  172.523 & Trid./DHS & \nodata & \nodata & Amorph./DHS & \nodata & Hansen \\
  176.491 & Trid./DHS & \nodata & \nodata & \nodata & Amorph./DHS & Hansen \\
  176.665 & Crys./Mie & \nodata & \nodata & \nodata & \nodata & Log-normal \\
  179.174 & \nodata & Amorph./DHS & \nodata & \nodata & Amorph./DHS & Hansen \\
  179.551 & Crys./Mie & \nodata & Crys./Mie & \nodata & \nodata & Hansen \\
  179.837 & Crys./DHS & \nodata & \nodata & \nodata & \nodata & Log-normal \\
  183.048 & Amorph./Mie & \nodata & Crys./Mie & \nodata & \nodata & Log-normal \\
  183.222 & \nodata & Amorph./Mie & Crys./Mie & \nodata & \nodata & Hansen \\
  184.128 & Crys./Mie & \nodata & \nodata & \nodata & \nodata & Hansen \\
  184.355 & \nodata & Amorph./DHS & Crys./DHS & \nodata & \nodata & Hansen \\
  188.645 & \nodata & Amorph./Mie & Crys./Mie & \nodata & \nodata & Log-normal \\
  192.797 & Crys./DHS & \nodata & \nodata & \nodata & \nodata & Hansen \\
  195.641 & Crist./DHS & \nodata & \nodata & \nodata & \nodata & Hansen \\
  206.376 & Crist./DHS & \nodata & Amorph./DHS & \nodata & \nodata & Log-normal \\
  212.921 & Amorph./Mie & \nodata & Crys./Mie & \nodata & \nodata & Hansen \\
  214.778 & Amorph./Mie & \nodata & \nodata & \nodata & \nodata & Log-normal \\
  215.660 & Amorph./DHS & \nodata & \nodata & \nodata & \nodata & Log-normal \\
  216.787 & Crist./DHS & Amorph./DHS & \nodata & \nodata & \nodata & Hansen \\
  219.783 & \nodata & \nodata & \nodata & Amorph./DHS & Amorph./DHS & Log-normal \\
  222.494 & \nodata & \nodata & \nodata & Amorph./DHS & \nodata & Hansen \\
  225.831 & \nodata & \nodata & Amorph./DHS & Amorph./DHS & \nodata & Log-normal \\
  225.937 & Amorph./Mie & \nodata & \nodata & \nodata & \nodata & Hansen \\
  229.182 & Amorph./DHS & \nodata & \nodata & \nodata & \nodata & Hansen \\
  232.635 & Trid./DHS & \nodata & \nodata & \nodata & \nodata & Hansen \\
  236.479 & \nodata & \nodata & \nodata & Amorph./DHS & Amorph./DHS & Hansen \\
  238.503 & Crist./DHS & \nodata & \nodata & \nodata & \nodata & Log-normal \\
  244.593 & \nodata & \nodata & Crys./DHS & Amorph./DHS & \nodata & Log-normal \\
  250.764 & \nodata & \nodata & \nodata & Amorph./DHS & \nodata & Log-normal \\
  263.212 & \nodata & \nodata & Amorph./DHS & Amorph./DHS & \nodata & Hansen \\
  271.914 & \nodata & \nodata & Crys./DHS & Amorph./DHS & \nodata & Hansen \\
  283.048 & \nodata & Crys./Mie & Crys./Mie & \nodata & \nodata & Log-normal \\
  284.089 & \nodata & Crys./Mie & \nodata & \nodata & \nodata & Log-normal \\
  284.522 & \nodata & Crys./Mie & \nodata & \nodata & Amorph./Mie & Log-normal \\
  284.613 & \nodata & Crys./Mie & Crys./Mie & \nodata & \nodata & Hansen \\
  286.274 & \nodata & Crys./Mie & Amorph./Mie & \nodata & \nodata & Log-normal \\
  289.366 & \nodata & Crys./Mie & Amorph./Mie & \nodata & \nodata & Hansen \\
  293.927 & \nodata & Crys./Mie & \nodata & \nodata & \nodata & Hansen \\
  299.543 & \nodata & Crys./Mie & \nodata & \nodata & Amorph./Mie & Hansen \\
  320.940 & \nodata & Crys./DHS & Amorph./DHS & \nodata & \nodata & Log-normal \\
  323.772 & \nodata & Crys./DHS & \nodata & \nodata & Amorph./DHS & Log-normal \\
  324.471 & \nodata & Crys./DHS & Crys./DHS & \nodata & \nodata & Log-normal \\
  325.552 & \nodata & Crys./DHS & \nodata & \nodata & \nodata & Log-normal \\
  332.576 & \nodata & Crys./DHS & \nodata & \nodata & \nodata & Hansen \\
  335.634 & \nodata & Crys./DHS & Crys./DHS & \nodata & \nodata & Hansen \\
  342.593 & \nodata & Crys./DHS & Amorph./DHS & \nodata & \nodata & Hansen \\
  348.238 & \nodata & \nodata & Amorph./Mie & \nodata & \nodata & Hansen \\
  350.193 & \nodata & \nodata & Amorph./Mie & \nodata & \nodata & Log-normal \\
  371.081 & \nodata & Crys./DHS & \nodata & \nodata & Amorph./DHS & Hansen \\
  445.358 & \nodata & \nodata & Amorph./DHS & \nodata & \nodata & Log-normal \\
  452.032 & \nodata & \nodata & Amorph./DHS & \nodata & \nodata & Hansen \\
  457.398 & \nodata & \nodata & \nodata & \nodata & Amorph./Mie & Log-normal \\
  457.770 & \nodata & \nodata & \nodata & \nodata & Amorph./Mie & Hansen \\
  506.010 & \nodata & \nodata & \nodata & \nodata & Amorph./DHS & Log-normal \\
  506.337 & \nodata & \nodata & \nodata & \nodata & Amorph./DHS & Hansen \\
  513.695 & \nodata & \nodata & Crys./Mie & \nodata & \nodata & Log-normal \\
  533.954 & \nodata & \nodata & Crys./Mie & \nodata & \nodata & Hansen \\
  557.006 & \nodata & \nodata & Crys./DHS & \nodata & \nodata & Log-normal \\
  559.020 & \nodata & \nodata & Crys./DHS & \nodata & \nodata & Hansen \\
\enddata
\tablecomments{The size distribution column applies to all cloud species in that model. Each dust species column gives crystallinity/opacity treatment. SiO$_2$ crystallinity: Qtz.=Quartz, Trid.=Tridymite, Crist.=Cristobalite, Amorph.=Amorphous. Other species: Crys.=Crystalline, Amorph.=Amorphous. Opacity: Pure Mie or DHS (Distribution of Hollow Spheres) \citep{min:2005}.}
\end{deluxetable*}

\bibliography{ms_new}{}

@Article{ackerman:2001,
  Title                    = {{Precipitating Condensation Clouds in Substellar Atmospheres}},
  Author                   = {{Ackerman}, A.~S. and {Marley}, M.~S.},
  Journal                  = {\apj},
  Year                     = {2001},

  Month                    = aug,
  Pages                    = {872-884},
  Volume                   = {556},
  Adsurl                   = {http://adsabs.harvard.edu/abs/2001ApJ...556..872A},
  Doi                      = {10.1086/321540},
  Eprint                   = {arXiv:astro-ph/0103423}
}

@Article{apai:2013,
  author        = {{Apai}, D. and {Radigan}, J. and {Buenzli}, E. and {Burrows}, A. and {Reid}, I.~N. and {Jayawardhana}, R.},
  journal       = {\apj},
  title         = {{HST Spectral Mapping of L/T Transition Brown Dwarfs Reveals Cloud Thickness Variations}},
  year          = {2013},
  month         = may,
  pages         = {121},
  volume        = {768},
  adsurl        = {http://adsabs.harvard.edu/abs/2013ApJ...768..121A},
  archiveprefix = {arXiv},
  doi           = {10.1088/0004-637X/768/2/121},
  eid           = {121},
  eprint        = {1303.4151},
  primaryclass  = {astro-ph.EP},
}

@Article{asplund:2009,
  author        = {{Asplund}, M. and {Grevesse}, N. and {Sauval}, A.~J. and {Scott}, P.},
  journal       = {\araa},
  title         = {{The Chemical Composition of the Sun}},
  year          = {2009},
  month         = sep,
  pages         = {481-522},
  volume        = {47},
  adsurl        = {http://adsabs.harvard.edu/abs/2009ARA\%26A..47..481A},
  archiveprefix = {arXiv},
  doi           = {10.1146/annurev.astro.46.060407.145222},
  eprint        = {0909.0948},
  primaryclass  = {astro-ph.SR},
}

@ARTICLE{cushing:2006,
       author = {{Cushing}, Michael C. and {Roellig}, Thomas L. and {Marley}, Mark S. and
         {Saumon}, D. and {Leggett}, S.~K. and {Kirkpatrick}, J. Davy and
         {Wilson}, John C. and {Sloan}, G.~C. and {Mainzer}, Amy K. and
         {Van Cleve}, Jeff E. and {Houck}, James R.},
        title = "{A Spitzer Infrared Spectrograph Spectral Sequence of M, L, and T Dwarfs}",
      journal = {\apj},
         year = 2006,
        month = sep,
       volume = {648},
       number = {1},
        pages = {614-628},
          doi = {10.1086/505637},
archivePrefix = {arXiv},
       eprint = {astro-ph/0605639},
 primaryClass = {astro-ph},
       adsurl = {https://ui.adsabs.harvard.edu/abs/2006ApJ...648..614C}
}

@Article{cushing:2005,
  Title                    = {{An Infrared Spectroscopic Sequence of M, L, and T Dwarfs}},
  Author                   = {{Cushing}, M.~C. and {Rayner}, J.~T. and {Vacca}, W.~D.},
  Journal                  = {\apj},
  Year                     = {2005},

  Month                    = apr,
  Pages                    = {1115-1140},
  Volume                   = {623},
  Adsurl                   = {http://adsabs.harvard.edu/abs/2005ApJ...623.1115C},
  Doi                      = {10.1086/428040},
  Eprint                   = {arXiv:astro-ph/0412313}
}

@Article{kirkpatrick:2010,
  Title                    = {{Discoveries from a Near-infrared Proper Motion Survey Using Multi-epoch Two Micron All-Sky Survey Data}},
  Author                   = {{Kirkpatrick}, J.~D. and {Looper}, D.~L. and {Burgasser}, A.~J. and {Schurr}, S.~D. and {Cutri}, R.~M. and {Cushing}, M.~C. and {Cruz}, K.~L. and {Sweet}, A.~C. and {Knapp}, G.~R. and {Barman}, T.~S. and {Bochanski}, J.~J. and {Roellig}, T.~L. and {McLean}, I.~S. and {McGovern}, M.~R. and {Rice}, E.~L.},
  Journal                  = {\apjs},
  Year                     = {2010},

  Month                    = sep,
  Pages                    = {100-146},
  Volume                   = {190},
  Adsurl                   = {http://adsabs.harvard.edu/abs/2010ApJS..190..100K},
  Archiveprefix            = {arXiv},
  Doi                      = {10.1088/0067-0049/190/1/100},
  Eprint                   = {1008.3591},
  Primaryclass             = {astro-ph.SR}
}

@Article{kitzmann:2018,
  author        = {{Kitzmann}, D. and {Heng}, K. and {Rimmer}, P.~B. and {Hoeijmakers}, H.~J. and {Tsai}, S.-M. and {Malik}, M. and {Lendl}, M. and {Deitrick}, R. and {Demory}, B.-O.},
  journal       = {\apj},
  title         = {{The Peculiar Atmospheric Chemistry of KELT-9b}},
  year          = {2018},
  month         = aug,
  pages         = {183},
  volume        = {863},
  adsurl        = {http://adsabs.harvard.edu/abs/2018ApJ...863..183K},
  archiveprefix = {arXiv},
  doi           = {10.3847/1538-4357/aace5a},
  eid           = {183},
  eprint        = {1804.07137},
  primaryclass  = {astro-ph.EP},
}

@Article{looper:2008,
  Title                    = {{Discovery of a T Dwarf Binary with the Largest Known J-Band Flux Reversal}},
  Author                   = {{Looper}, D.~L. and {Gelino}, C.~R. and {Burgasser}, A.~J. and {Kirkpatrick}, J.~D.},
  Journal                  = {\apj},
  Year                     = {2008},

  Month                    = oct,
  Pages                    = {1183-1192},
  Volume                   = {685},
  Adsurl                   = {http://adsabs.harvard.edu/abs/2008ApJ...685.1183L},
  Archiveprefix            = {arXiv},
  Doi                      = {10.1086/590382},
  Eprint                   = {0803.0544}
}

@Article{prinn:1977,
  Title                    = {{Carbon monoxide on Jupiter and implications for atmospheric convection}},
  Author                   = {{Prinn}, R.~G. and {Barshay}, S.~S.},
  Journal                  = {Science},
  Year                     = {1977},

  Month                    = dec,
  Pages                    = {1031-1034},
  Volume                   = {198},
  Adsurl                   = {http://adsabs.harvard.edu/abs/1977Sci...198.1031P},
  Doi                      = {10.1126/science.198.4321.1031}
}

@Article{reid:2008,
  Title                    = {{Meeting the Cool Neighbors. X. Ultracool Dwarfs from the 2MASS All-Sky Data Release}},
  Author                   = {{Reid}, I.~N. and {Cruz}, K.~L. and {Kirkpatrick}, J.~D. and {Allen}, P.~R. and {Mungall}, F. and {Liebert}, J. and {Lowrance}, P. and {Sweet}, A.},
  Journal                  = {\aj},
  Year                     = {2008},

  Month                    = sep,
  Pages                    = {1290-1311},
  Volume                   = {136},
  Adsurl                   = {http://adsabs.harvard.edu/abs/2008AJ....136.1290R},
  Doi                      = {10.1088/0004-6256/136/3/1290}
}

@Article{saumon:2008,
  Title                    = {{The Evolution of L and T Dwarfs in Color-Magnitude Diagrams}},
  Author                   = {{Saumon}, D. and {Marley}, M.~S.},
  Journal                  = {\apj},
  Year                     = {2008},

  Month                    = dec,
  Pages                    = {1327-1344},
  Volume                   = {689},
  Adsurl                   = {http://adsabs.harvard.edu/abs/2008ApJ...689.1327S},
  Archiveprefix            = {arXiv},
  Doi                      = {10.1086/592734},
  Eprint                   = {0808.2611}
}

@Article{stephens:2009,
  Title                    = {{The 0.8-14.5 {$\mu$}m Spectra of Mid-L to Mid-T Dwarfs: Diagnostics of Effective Temperature, Grain Sedimentation, Gas Transport, and Surface Gravity}},
  Author                   = {{Stephens}, D.~C. and {Leggett}, S.~K. and {Cushing}, M.~C. and {Marley}, M.~S. and {Saumon}, D. and {Geballe}, T.~R. and {Golimowski}, D.~A. and {Fan}, X. and {Noll}, K.~S.},
  Journal                  = {\apj},
  Year                     = {2009},

  Month                    = sep,
  Pages                    = {154-170},
  Volume                   = {702},
  Adsurl                   = {http://adsabs.harvard.edu/abs/2009ApJ...702..154S},
  Archiveprefix            = {arXiv},
  Doi                      = {10.1088/0004-637X/702/1/154},
  Eprint                   = {0906.2991},
  Primaryclass             = {astro-ph.SR}
}

@Article{tsuji:2003,
  Title                    = {{Transition from L to T Dwarfs on the Color-Magnitude Diagram}},
  Author                   = {{Tsuji}, T. and {Nakajima}, T.},
  Journal                  = {\apjl},
  Year                     = {2003},

  Month                    = mar,
  Pages                    = {L151-L154},
  Volume                   = {585},
  Adsurl                   = {http://adsabs.harvard.edu/abs/2003ApJ...585L.151T},
  Doi                      = {10.1086/374388},
  Eprint                   = {arXiv:astro-ph/0302169}
}

@Article{wakeford:2017a,
  Title                    = {{High-temperature condensate clouds in super-hot Jupiter atmospheres}},
  Author                   = {{Wakeford}, H.~R. and {Visscher}, C. and {Lewis}, N.~K. and {Kataria}, T. and {Marley}, M.~S. and {Fortney}, J.~J. and {Mandell}, A.~M.},
  Journal                  = {\mnras},
  Year                     = {2017},

  Month                    = feb,
  Pages                    = {4247-4254},
  Volume                   = {464},
  Adsurl                   = {http://adsabs.harvard.edu/abs/2017MNRAS.464.4247W},
  Archiveprefix            = {arXiv},
  Doi                      = {10.1093/mnras/stw2639},
  Eprint                   = {1610.03325},
  Primaryclass             = {astro-ph.EP}
}

@Article{allard:2001,
  author   = {{Allard}, F. and {Hauschildt}, P.~H. and {Alexander}, D.~R. and {Tamanai}, A. and {Schweitzer}, A.},
  title    = {{The Limiting Effects of Dust in Brown Dwarf Model Atmospheres}},
  journal  = {\apj},
  year     = {2001},
  volume   = {556},
  pages    = {357-372},
  month    = jul,
  adsurl   = {http://adsabs.harvard.edu/abs/2001ApJ...556..357A},
  doi      = {10.1086/321547},
  eprint   = {astro-ph/0104256},
}

@Article{borysow:2002,
  author   = {{Borysow}, A.},
  title    = {{Collision-induced absorption coefficients of H$_{2}$ pairs at temperatures from 60 K to 1000 K}},
  journal  = {\aap},
  year     = {2002},
  volume   = {390},
  pages    = {779-782},
  month    = aug,
  adsurl   = {http://adsabs.harvard.edu/abs/2002A%26A...390..779B},
  doi      = {10.1051/0004-6361:20020555},
}

@Article{borysow:1986,
  author   = {{Borysow}, A. and {Frommhold}, L.},
  title    = {{Theoretical collision-induced rototranslational absorption spectra for the outer planets - H2-CH4 pairs}},
  journal  = {\apj},
  year     = {1986},
  volume   = {304},
  pages    = {849-865},
  month    = may,
  adsurl   = {http://adsabs.harvard.edu/abs/1986ApJ...304..849B},
  doi      = {10.1086/164221},
}

@Article{manjavacas:2018,
  author        = {{Manjavacas}, E. and {Apai}, D. and {Zhou}, Y. and {Lew}, B.~W.~P. and {Schneider}, G. and {Metchev}, S. and {Miles-P{\'a}ez}, P.~A. and {Radigan}, J. and {Marley}, M.~S. and {Cowan}, N. and {Karalidi}, T. and {Burgasser}, A.~J. and {Bedin}, L.~R. and {Lowrance}, P.~J. and {Kauffmann}, P.},
  title         = {{Cloud Atlas: Hubble Space Telescope Near-infrared Spectral Library of Brown Dwarfs, Planetary-mass Companions, and Hot Jupiters}},
  journal       = {\aj},
  year          = {2019},
  volume        = {157},
  pages         = {101},
  month         = mar,
  adsurl        = {http://adsabs.harvard.edu/abs/2019AJ....157..101M},
  archiveprefix = {arXiv},
  doi           = {10.3847/1538-3881/aaf88f},
  eid           = {101},
  eprint        = {1812.03963},
  primaryclass  = {astro-ph.SR},
}

@Article{mckemmish:2019,
  author        = {{McKemmish}, Laura K. and {Masseron}, Thomas and {Hoeijmakers}, H. Jens and {P{\'e}rez-Mesa}, V{\'\i}ctor and {Grimm}, Simon L. and {Yurchenko}, Sergei N. and {Tennyson}, Jonathan},
  title         = {{ExoMol molecular line lists - XXXIII. The spectrum of Titanium Oxide}},
  journal       = {\mnras},
  year          = {2019},
  volume        = {488},
  number        = {2},
  pages         = {2836-2854},
  month         = {Sep},
  adsurl        = {https://ui.adsabs.harvard.edu/abs/2019MNRAS.488.2836M},
  archiveprefix = {arXiv},
  doi           = {10.1093/mnras/stz1818},
  eprint        = {1905.04587},
  primaryclass  = {astro-ph.SR},
}

@Article{molliere:2019,
  author        = {{Molli{\`e}re}, P. and {Wardenier}, J.~P. and {van Boekel}, R. and {Henning}, Th. and {Molaverdikhani}, K. and {Snellen}, I.~A.~G.},
  title         = {{petitRADTRANS. A Python radiative transfer package for exoplanet characterization and retrieval}},
  journal       = {\aap},
  year          = {2019},
  volume        = {627},
  pages         = {A67},
  month         = {Jul},
  adsurl        = {https://ui.adsabs.harvard.edu/abs/2019A&A...627A..67M},
  archiveprefix = {arXiv},
  doi           = {10.1051/0004-6361/201935470},
  eid           = {A67},
  eprint        = {1904.11504},
  primaryclass  = {astro-ph.EP},
}

@Article{marley:2010,
  author        = {{Marley}, Mark S. and {Saumon}, Didier and {Goldblatt}, Colin},
  title         = {{A Patchy Cloud Model for the L to T Dwarf Transition}},
  journal       = {\apjl},
  year          = {2010},
  volume        = {723},
  number        = {1},
  pages         = {L117-L121},
  month         = nov,
  adsurl        = {https://ui.adsabs.harvard.edu/abs/2010ApJ...723L.117M},
  archiveprefix = {arXiv},
  doi           = {10.1088/2041-8205/723/1/L117},
  eprint        = {1009.6217},
  primaryclass  = {astro-ph.SR},
}

@Article{hunter:2007,
  author    = {Hunter, J. D.},
  title     = {Matplotlib: A 2D graphics environment},
  journal   = {Computing in Science \& Engineering},
  year      = {2007},
  volume    = {9},
  number    = {3},
  pages     = {90--95},
  doi       = {10.1109/MCSE.2007.55},
  publisher = {IEEE COMPUTER SOC},
}

@Article{perez:2007,
  author  = {Perez,Fernando and Granger,Brian E.},
  journal = {Computing in Science \& Engineering},
  title   = {IPython: A System for Interactive Scientific Computing},
  year    = {2007},
  number  = {3},
  pages   = {21-29},
  volume  = {9},
  doi     = {10.1109/MCSE.2007.53},
  eprint  = {https://aip.scitation.org/doi/pdf/10.1109/MCSE.2007.53},
  url     = {https://aip.scitation.org/doi/abs/10.1109/MCSE.2007.53},
}

@ARTICLE{numpy:2020,
  author  = {Harris, Charles R. and Millman, K. Jarrod and van der Walt, StÃ©fan J and Gommers, Ralf and Virtanen, Pauli and Cournapeau, David and Wieser, Eric and Taylor, Julian and Berg, Sebastian and Smith, Nathaniel J. and Kern, Robert and Picus, Matti and Hoyer, Stephan and van Kerkwijk, Marten H. and Brett, Matthew and Haldane, Allan and FernÃ¡ndez del RÃ­o, Jaime and Wiebe, Mark and Peterson, Pearu and GÃ©rard-Marchant, Pierre and Sheppard, Kevin and Reddy, Tyler and Weckesser, Warren and Abbasi, Hameer and Gohlke, Christoph and Oliphant, Travis E.},
  title   = {Array programming with {NumPy}},
  journal = {Nature},
  year    = {2020},
  volume  = {585},
  pages   = {357â€“362},
  doi     = {10.1038/s41586-020-2649-2}
}

@ARTICLE{ochsenbein:2000,
       author = {{Ochsenbein}, F. and {Bauer}, P. and {Marcout}, J.},
        title = "{The VizieR database of astronomical catalogues}",
      journal = {\aaps},
         year = 2000,
        month = apr,
       volume = {143},
        pages = {23-32},
          doi = {10.1051/aas:2000169},
archivePrefix = {arXiv},
       eprint = {astro-ph/0002122},
 primaryClass = {astro-ph},
       adsurl = {https://ui.adsabs.harvard.edu/abs/2000A&AS..143...23O}
}

@ARTICLE{pollack:1994,
       author = {{Pollack}, James B. and {Hollenbach}, David and {Beckwith}, Steven and
         {Simonelli}, Damon P. and {Roush}, Ted and {Fong}, Wesley},
        title = "{Composition and Radiative Properties of Grains in Molecular Clouds and Accretion Disks}",
      journal = {\apj},
         year = 1994,
        month = feb,
       volume = {421},
        pages = {615},
          doi = {10.1086/173677},
       adsurl = {https://ui.adsabs.harvard.edu/abs/1994ApJ...421..615P}
}

@ARTICLE{min:2005,
       author = {{Min}, M. and {Hovenier}, J.~W. and {de Koter}, A. and
         {Waters}, L.~B.~F.~M. and {Dominik}, C.},
        title = "{The composition and size distribution of the dust in the coma of Comet Hale Bopp}",
      journal = {\icarus},
         year = 2005,
        month = dec,
       volume = {179},
       number = {1},
        pages = {158-173},
          doi = {10.1016/j.icarus.2005.05.015},
archivePrefix = {arXiv},
       eprint = {astro-ph/0505603},
 primaryClass = {astro-ph},
       adsurl = {https://ui.adsabs.harvard.edu/abs/2005Icar..179..158M}
}

@ARTICLE{molliere:2020,
       author = {{Molli{\`e}re}, P. and {Stolker}, T. and {Lacour}, S. and
         {Otten}, G.~P.~P.~L. and {Shangguan}, J. and {Charnay}, B. and
         {Molyarova}, T. and {Nowak}, M. and {Henning}, Th. and
         {Marleau}, G. -D. and {Semenov}, D.~A. and {van Dishoeck}, E. and
         {Eisenhauer}, F. and {Garcia}, P. and {Garcia Lopez}, R. and
         {Girard}, J.~H. and {Greenbaum}, A.~Z. and {Hinkley}, S. and
         {Kervella}, P. and {Kreidberg}, L. and {Maire}, A. -L. and
         {Nasedkin}, E. and {Pueyo}, L. and {Snellen}, I.~A.~G. and {Vigan}, A. and
         {Wang}, J. and {de Zeeuw}, P.~T. and {Zurlo}, A.},
        title = "{Retrieving scattering clouds and disequilibrium chemistry in the atmosphere of HR 8799e}",
      journal = {\aap},
         year = 2020,
        month = aug,
       volume = {640},
          eid = {A131},
        pages = {A131},
          doi = {10.1051/0004-6361/202038325},
archivePrefix = {arXiv},
       eprint = {2006.09394},
 primaryClass = {astro-ph.EP},
       adsurl = {https://ui.adsabs.harvard.edu/abs/2020A&A...640A.131M}
}

@ARTICLE{mckemmish:2016,
       author = {{McKemmish}, Laura K. and {Yurchenko}, Sergei N. and {Tennyson}, Jonathan},
        title = "{ExoMol line lists - XVIII. The high-temperature spectrum of VO}",
      journal = {\mnras},
         year = 2016,
        month = nov,
       volume = {463},
       number = {1},
        pages = {771-793},
          doi = {10.1093/mnras/stw1969},
archivePrefix = {arXiv},
       eprint = {1609.06120},
 primaryClass = {astro-ph.SR},
       adsurl = {https://ui.adsabs.harvard.edu/abs/2016MNRAS.463..771M}
}

@ARTICLE{burningham:2021,
       author = {{Burningham}, Ben and {Faherty}, Jacqueline K. and {Gonzales}, Eileen C. and {Marley}, Mark S. and {Visscher}, Channon and {Lupu}, Roxana and {Gaarn}, Josefine and {Bieger}, Michelle Fabienne and {Freedman}, Richard and {Saumon}, Didier},
        title = "{Cloud busting: Enstatite and quartz clouds in the atmosphere of 2M2224-0158}",
      journal = {\mnras},
         year = 2021,
        month = may,
          doi = {10.1093/mnras/stab1361},
archivePrefix = {arXiv},
       eprint = {2105.04268},
 primaryClass = {astro-ph.SR},
       adsurl = {https://ui.adsabs.harvard.edu/abs/2021MNRAS.tmp.1380B}
}

@ARTICLE{visscher:2010,
       author = {{Visscher}, Channon and {Lodders}, Katharina and {Fegley}, Bruce, Jr.},
        title = "{Atmospheric Chemistry in Giant Planets, Brown Dwarfs, and Low-mass Dwarf Stars. III. Iron, Magnesium, and Silicon}",
      journal = {\apj},
         year = 2010,
        month = jun,
       volume = {716},
       number = {2},
        pages = {1060-1075},
          doi = {10.1088/0004-637X/716/2/1060},
archivePrefix = {arXiv},
       eprint = {1001.3639},
 primaryClass = {astro-ph.EP},
       adsurl = {https://ui.adsabs.harvard.edu/abs/2010ApJ...716.1060V}
}

@ARTICLE{helling:2006,
       author = {{Helling}, Ch. and {Woitke}, P.},
        title = "{Dust in brown dwarfs. V. Growth and evaporation of dirty dust grains}",
      journal = {\aap},
         year = 2006,
        month = aug,
       volume = {455},
       number = {1},
        pages = {325-338},
          doi = {10.1051/0004-6361:20054598},
       adsurl = {https://ui.adsabs.harvard.edu/abs/2006A&A...455..325H}
}

@ARTICLE{Gaia:2023,
       author = {{Gaia Collaboration} and {Vallenari}, A. and {Brown}, A.~G.~A. and {Prusti}, T. and {de Bruijne}, J.~H.~J. and {Arenou}, F. and {Babusiaux}, C. and {Biermann}, M. and {Creevey}, O.~L. and {Ducourant}, C. and {Evans}, D.~W. and {Eyer}, L. and {Guerra}, R. and {Hutton}, A. and {Jordi}, C. and {Klioner}, S.~A. and {Lammers}, U.~L. and {Lindegren}, L. and {Luri}, X. and {Mignard}, F. and {Panem}, C. and {Pourbaix}, D. and {Randich}, S. and {Sartoretti}, P. and {Soubiran}, C. and {Tanga}, P. and {Walton}, N.~A. and {Bailer-Jones}, C.~A.~L. and {Bastian}, U. and {Drimmel}, R. and {Jansen}, F. and {Katz}, D. and {Lattanzi}, M.~G. and {van Leeuwen}, F. and {Bakker}, J. and {Cacciari}, C. and {Casta{\~n}eda}, J. and {De Angeli}, F. and {Fabricius}, C. and {Fouesneau}, M. and {Fr{\'e}mat}, Y. and {Galluccio}, L. and {Guerrier}, A. and {Heiter}, U. and {Masana}, E. and {Messineo}, R. and {Mowlavi}, N. and {Nicolas}, C. and {Nienartowicz}, K. and {Pailler}, F. and {Panuzzo}, P. and {Riclet}, F. and {Roux}, W. and {Seabroke}, G.~M. and {Sordo}, R. and {Th{\'e}venin}, F. and {Gracia-Abril}, G. and {Portell}, J. and {Teyssier}, D. and {Altmann}, M. and {Andrae}, R. and {Audard}, M. and {Bellas-Velidis}, I. and {Benson}, K. and {Berthier}, J. and {Blomme}, R. and {Burgess}, P.~W. and {Busonero}, D. and {Busso}, G. and {C{\'a}novas}, H. and {Carry}, B. and {Cellino}, A. and {Cheek}, N. and {Clementini}, G. and {Damerdji}, Y. and {Davidson}, M. and {de Teodoro}, P. and {Nu{\~n}ez Campos}, M. and {Delchambre}, L. and {Dell'Oro}, A. and {Esquej}, P. and {Fern{\'a}ndez-Hern{\'a}ndez}, J. and {Fraile}, E. and {Garabato}, D. and {Garc{\'\i}a-Lario}, P. and {Gosset}, E. and {Haigron}, R. and {Halbwachs}, J. -L. and {Hambly}, N.~C. and {Harrison}, D.~L. and {Hern{\'a}ndez}, J. and {Hestroffer}, D. and {Hodgkin}, S.~T. and {Holl}, B. and {Jan{\ss}en}, K. and {Jevardat de Fombelle}, G. and {Jordan}, S. and {Krone-Martins}, A. and {Lanzafame}, A.~C. and {L{\"o}ffler}, W. and {Marchal}, O. and {Marrese}, P.~M. and {Moitinho}, A. and {Muinonen}, K. and {Osborne}, P. and {Pancino}, E. and {Pauwels}, T. and {Recio-Blanco}, A. and {Reyl{\'e}}, C. and {Riello}, M. and {Rimoldini}, L. and {Roegiers}, T. and {Rybizki}, J. and {Sarro}, L.~M. and {Siopis}, C. and {Smith}, M. and {Sozzetti}, A. and {Utrilla}, E. and {van Leeuwen}, M. and {Abbas}, U. and {{\'A}brah{\'a}m}, P. and {Abreu Aramburu}, A. and {Aerts}, C. and {Aguado}, J.~J. and {Ajaj}, M. and {Aldea-Montero}, F. and {Altavilla}, G. and {{\'A}lvarez}, M.~A. and {Alves}, J. and {Anders}, F. and {Anderson}, R.~I. and {Anglada Varela}, E. and {Antoja}, T. and {Baines}, D. and {Baker}, S.~G. and {Balaguer-N{\'u}{\~n}ez}, L. and {Balbinot}, E. and {Balog}, Z. and {Barache}, C. and {Barbato}, D. and {Barros}, M. and {Barstow}, M.~A. and {Bartolom{\'e}}, S. and {Bassilana}, J. -L. and {Bauchet}, N. and {Becciani}, U. and {Bellazzini}, M. and {Berihuete}, A. and {Bernet}, M. and {Bertone}, S. and {Bianchi}, L. and {Binnenfeld}, A. and {Blanco-Cuaresma}, S. and {Blazere}, A. and {Boch}, T. and {Bombrun}, A. and {Bossini}, D. and {Bouquillon}, S. and {Bragaglia}, A. and {Bramante}, L. and {Breedt}, E. and {Bressan}, A. and {Brouillet}, N. and {Brugaletta}, E. and {Bucciarelli}, B. and {Burlacu}, A. and {Butkevich}, A.~G. and {Buzzi}, R. and {Caffau}, E. and {Cancelliere}, R. and {Cantat-Gaudin}, T. and {Carballo}, R. and {Carlucci}, T. and {Carnerero}, M.~I. and {Carrasco}, J.~M. and {Casamiquela}, L. and {Castellani}, M. and {Castro-Ginard}, A. and {Chaoul}, L. and {Charlot}, P. and {Chemin}, L. and {Chiaramida}, V. and {Chiavassa}, A. and {Chornay}, N. and {Comoretto}, G. and {Contursi}, G. and {Cooper}, W.~J. and {Cornez}, T. and {Cowell}, S. and {Crifo}, F. and {Cropper}, M. and {Crosta}, M. and {Crowley}, C. and {Dafonte}, C. and {Dapergolas}, A. and {David}, M. and {David}, P. and {de Laverny}, P. and {De Luise}, F. and {De March}, R. and {De Ridder}, J. and {de Souza}, R. and {de Torres}, A. and {del Peloso}, E.~F. and {del Pozo}, E. and {Delbo}, M. and {Delgado}, A. and {Delisle}, J. -B. and {Demouchy}, C. and {Dharmawardena}, T.~E. and {Di Matteo}, P. and {Diakite}, S. and {Diener}, C. and {Distefano}, E. and {Dolding}, C. and {Edvardsson}, B. and {Enke}, H. and {Fabre}, C. and {Fabrizio}, M. and {Faigler}, S. and {Fedorets}, G. and {Fernique}, P. and {Fienga}, A. and {Figueras}, F. and {Fournier}, Y. and {Fouron}, C. and {Fragkoudi}, F. and {Gai}, M. and {Garcia-Gutierrez}, A. and {Garcia-Reinaldos}, M. and {Garc{\'\i}a-Torres}, M. and {Garofalo}, A. and {Gavel}, A. and {Gavras}, P. and {Gerlach}, E. and {Geyer}, R. and {Giacobbe}, P. and {Gilmore}, G. and {Girona}, S. and {Giuffrida}, G. and {Gomel}, R. and {Gomez}, A. and {Gonz{\'a}lez-N{\'u}{\~n}ez}, J. and {Gonz{\'a}lez-Santamar{\'\i}a}, I. and {Gonz{\'a}lez-Vidal}, J.~J. and {Granvik}, M. and {Guillout}, P. and {Guiraud}, J. and {Guti{\'e}rrez-S{\'a}nchez}, R. and {Guy}, L.~P. and {Hatzidimitriou}, D. and {Hauser}, M. and {Haywood}, M. and {Helmer}, A. and {Helmi}, A. and {Sarmiento}, M.~H. and {Hidalgo}, S.~L. and {Hilger}, T. and {H{\l}adczuk}, N. and {Hobbs}, D. and {Holland}, G. and {Huckle}, H.~E. and {Jardine}, K. and {Jasniewicz}, G. and {Jean-Antoine Piccolo}, A. and {Jim{\'e}nez-Arranz}, {\'O}. and {Jorissen}, A. and {Juaristi Campillo}, J. and {Julbe}, F. and {Karbevska}, L. and {Kervella}, P. and {Khanna}, S. and {Kontizas}, M. and {Kordopatis}, G. and {Korn}, A.~J. and {K{\'o}sp{\'a}l}, {\'A}. and {Kostrzewa-Rutkowska}, Z. and {Kruszy{\'n}ska}, K. and {Kun}, M. and {Laizeau}, P. and {Lambert}, S. and {Lanza}, A.~F. and {Lasne}, Y. and {Le Campion}, J. -F. and {Lebreton}, Y. and {Lebzelter}, T. and {Leccia}, S. and {Leclerc}, N. and {Lecoeur-Taibi}, I. and {Liao}, S. and {Licata}, E.~L. and {Lindstr{\o}m}, H.~E.~P. and {Lister}, T.~A. and {Livanou}, E. and {Lobel}, A. and {Lorca}, A. and {Loup}, C. and {Madrero Pardo}, P. and {Magdaleno Romeo}, A. and {Managau}, S. and {Mann}, R.~G. and {Manteiga}, M. and {Marchant}, J.~M. and {Marconi}, M. and {Marcos}, J. and {Marcos Santos}, M.~M.~S. and {Mar{\'\i}n Pina}, D. and {Marinoni}, S. and {Marocco}, F. and {Marshall}, D.~J. and {Martin Polo}, L. and {Mart{\'\i}n-Fleitas}, J.~M. and {Marton}, G. and {Mary}, N. and {Masip}, A. and {Massari}, D. and {Mastrobuono-Battisti}, A. and {Mazeh}, T. and {McMillan}, P.~J. and {Messina}, S. and {Michalik}, D. and {Millar}, N.~R. and {Mints}, A. and {Molina}, D. and {Molinaro}, R. and {Moln{\'a}r}, L. and {Monari}, G. and {Mongui{\'o}}, M. and {Montegriffo}, P. and {Montero}, A. and {Mor}, R. and {Mora}, A. and {Morbidelli}, R. and {Morel}, T. and {Morris}, D. and {Muraveva}, T. and {Murphy}, C.~P. and {Musella}, I. and {Nagy}, Z. and {Noval}, L. and {Oca{\~n}a}, F. and {Ogden}, A. and {Ordenovic}, C. and {Osinde}, J.~O. and {Pagani}, C. and {Pagano}, I. and {Palaversa}, L. and {Palicio}, P.~A. and {Pallas-Quintela}, L. and {Panahi}, A. and {Payne-Wardenaar}, S. and {Pe{\~n}alosa Esteller}, X. and {Penttil{\"a}}, A. and {Pichon}, B. and {Piersimoni}, A.~M. and {Pineau}, F. -X. and {Plachy}, E. and {Plum}, G. and {Poggio}, E. and {Pr{\v{s}}a}, A. and {Pulone}, L. and {Racero}, E. and {Ragaini}, S. and {Rainer}, M. and {Raiteri}, C.~M. and {Rambaux}, N. and {Ramos}, P. and {Ramos-Lerate}, M. and {Re Fiorentin}, P. and {Regibo}, S. and {Richards}, P.~J. and {Rios Diaz}, C. and {Ripepi}, V. and {Riva}, A. and {Rix}, H. -W. and {Rixon}, G. and {Robichon}, N. and {Robin}, A.~C. and {Robin}, C. and {Roelens}, M. and {Rogues}, H.~R.~O. and {Rohrbasser}, L. and {Romero-G{\'o}mez}, M. and {Rowell}, N. and {Royer}, F. and {Ruz Mieres}, D. and {Rybicki}, K.~A. and {Sadowski}, G. and {S{\'a}ez N{\'u}{\~n}ez}, A. and {Sagrist{\`a} Sell{\'e}s}, A. and {Sahlmann}, J. and {Salguero}, E. and {Samaras}, N. and {Sanchez Gimenez}, V. and {Sanna}, N. and {Santove{\~n}a}, R. and {Sarasso}, M. and {Schultheis}, M. and {Sciacca}, E. and {Segol}, M. and {Segovia}, J.~C. and {S{\'e}gransan}, D. and {Semeux}, D. and {Shahaf}, S. and {Siddiqui}, H.~I. and {Siebert}, A. and {Siltala}, L. and {Silvelo}, A. and {Slezak}, E. and {Slezak}, I. and {Smart}, R.~L. and {Snaith}, O.~N. and {Solano}, E. and {Solitro}, F. and {Souami}, D. and {Souchay}, J. and {Spagna}, A. and {Spina}, L. and {Spoto}, F. and {Steele}, I.~A. and {Steidelm{\"u}ller}, H. and {Stephenson}, C.~A. and {S{\"u}veges}, M. and {Surdej}, J. and {Szabados}, L. and {Szegedi-Elek}, E. and {Taris}, F. and {Taylor}, M.~B. and {Teixeira}, R. and {Tolomei}, L. and {Tonello}, N. and {Torra}, F. and {Torra}, J. and {Torralba Elipe}, G. and {Trabucchi}, M. and {Tsounis}, A.~T. and {Turon}, C. and {Ulla}, A. and {Unger}, N. and {Vaillant}, M.~V. and {van Dillen}, E. and {van Reeven}, W. and {Vanel}, O. and {Vecchiato}, A. and {Viala}, Y. and {Vicente}, D. and {Voutsinas}, S. and {Weiler}, M. and {Wevers}, T. and {Wyrzykowski}, {\L}. and {Yoldas}, A. and {Yvard}, P. and {Zhao}, H. and {Zorec}, J. and {Zucker}, S. and {Zwitter}, T.},
        title = "{Gaia Data Release 3. Summary of the content and survey properties}",
      journal = {\aap},
         year = 2023,
        month = jun,
       volume = {674},
          eid = {A1},
        pages = {A1},
          doi = {10.1051/0004-6361/202243940},
archivePrefix = {arXiv},
       eprint = {2208.00211},
 primaryClass = {astro-ph.GA},
       adsurl = {https://ui.adsabs.harvard.edu/abs/2023A&A...674A...1G}
}

@BOOK{cutri:2003,
       author = {{Cutri}, R.~M. and {Skrutskie}, M.~F. and {van Dyk}, S. and {Beichman}, C.~A. and {Carpenter}, J.~M. and {Chester}, T. and {Cambresy}, L. and {Evans}, T. and {Fowler}, J. and {Gizis}, J. and {Howard}, E. and {Huchra}, J. and {Jarrett}, T. and {Kopan}, E.~L. and {Kirkpatrick}, J.~D. and {Light}, R.~M. and {Marsh}, K.~A. and {McCallon}, H. and {Schneider}, S. and {Stiening}, R. and {Sykes}, M. and {Weinberg}, M. and {Wheaton}, W.~A. and {Wheelock}, S. and {Zacarias}, N.},
        title = "{2MASS All Sky Catalog of point sources.}",
         year = 2003,
       adsurl = {https://ui.adsabs.harvard.edu/abs/2003tmc..book.....C}
}

@ARTICLE{filippazzo:2015,
       author = {{Filippazzo}, Joseph C. and {Rice}, Emily L. and {Faherty}, Jacqueline and {Cruz}, Kelle L. and {Van Gordon}, Mollie M. and {Looper}, Dagny L.},
        title = "{Fundamental Parameters and Spectral Energy Distributions of Young and Field Age Objects with Masses Spanning the Stellar to Planetary Regime}",
      journal = {\apj},
         year = 2015,
        month = sep,
       volume = {810},
       number = {2},
          eid = {158},
        pages = {158},
          doi = {10.1088/0004-637X/810/2/158},
archivePrefix = {arXiv},
       eprint = {1508.01767},
 primaryClass = {astro-ph.SR},
       adsurl = {https://ui.adsabs.harvard.edu/abs/2015ApJ...810..158F}
}

@ARTICLE{vos:2017,
       author = {{Vos}, Johanna M. and {Allers}, Katelyn N. and {Biller}, Beth A.},
        title = "{The Viewing Geometry of Brown Dwarfs Influences Their Observed Colors and Variability Amplitudes}",
      journal = {\apj},
         year = 2017,
        month = jun,
       volume = {842},
       number = {2},
          eid = {78},
        pages = {78},
          doi = {10.3847/1538-4357/aa73cf},
archivePrefix = {arXiv},
       eprint = {1705.06045},
 primaryClass = {astro-ph.SR},
       adsurl = {https://ui.adsabs.harvard.edu/abs/2017ApJ...842...78V}
}

@ARTICLE{astropy:2018,
       author = {{Astropy Collaboration} and {Price-Whelan}, A.~M. and {Sip{\H{o}}cz}, B.~M. and {G{\"u}nther}, H.~M. and {Lim}, P.~L. and {Crawford}, S.~M. and {Conseil}, S. and {Shupe}, D.~L. and {Craig}, M.~W. and {Dencheva}, N. and {Ginsburg}, A. and {VanderPlas}, J.~T. and {Bradley}, L.~D. and {P{\'e}rez-Su{\'a}rez}, D. and {de Val-Borro}, M. and {Aldcroft}, T.~L. and {Cruz}, K.~L. and {Robitaille}, T.~P. and {Tollerud}, E.~J. and {Ardelean}, C. and {Babej}, T. and {Bach}, Y.~P. and {Bachetti}, M. and {Bakanov}, A.~V. and {Bamford}, S.~P. and {Barentsen}, G. and {Barmby}, P. and {Baumbach}, A. and {Berry}, K.~L. and {Biscani}, F. and {Boquien}, M. and {Bostroem}, K.~A. and {Bouma}, L.~G. and {Brammer}, G.~B. and {Bray}, E.~M. and {Breytenbach}, H. and {Buddelmeijer}, H. and {Burke}, D.~J. and {Calderone}, G. and {Cano Rodr{\'\i}guez}, J.~L. and {Cara}, M. and {Cardoso}, J.~V.~M. and {Cheedella}, S. and {Copin}, Y. and {Corrales}, L. and {Crichton}, D. and {D'Avella}, D. and {Deil}, C. and {Depagne}, {\'E}. and {Dietrich}, J.~P. and {Donath}, A. and {Droettboom}, M. and {Earl}, N. and {Erben}, T. and {Fabbro}, S. and {Ferreira}, L.~A. and {Finethy}, T. and {Fox}, R.~T. and {Garrison}, L.~H. and {Gibbons}, S.~L.~J. and {Goldstein}, D.~A. and {Gommers}, R. and {Greco}, J.~P. and {Greenfield}, P. and {Groener}, A.~M. and {Grollier}, F. and {Hagen}, A. and {Hirst}, P. and {Homeier}, D. and {Horton}, A.~J. and {Hosseinzadeh}, G. and {Hu}, L. and {Hunkeler}, J.~S. and {Ivezi{\'c}}, {\v{Z}}. and {Jain}, A. and {Jenness}, T. and {Kanarek}, G. and {Kendrew}, S. and {Kern}, N.~S. and {Kerzendorf}, W.~E. and {Khvalko}, A. and {King}, J. and {Kirkby}, D. and {Kulkarni}, A.~M. and {Kumar}, A. and {Lee}, A. and {Lenz}, D. and {Littlefair}, S.~P. and {Ma}, Z. and {Macleod}, D.~M. and {Mastropietro}, M. and {McCully}, C. and {Montagnac}, S. and {Morris}, B.~M. and {Mueller}, M. and {Mumford}, S.~J. and {Muna}, D. and {Murphy}, N.~A. and {Nelson}, S. and {Nguyen}, G.~H. and {Ninan}, J.~P. and {N{\"o}the}, M. and {Ogaz}, S. and {Oh}, S. and {Parejko}, J.~K. and {Parley}, N. and {Pascual}, S. and {Patil}, R. and {Patil}, A.~A. and {Plunkett}, A.~L. and {Prochaska}, J.~X. and {Rastogi}, T. and {Reddy Janga}, V. and {Sabater}, J. and {Sakurikar}, P. and {Seifert}, M. and {Sherbert}, L.~E. and {Sherwood-Taylor}, H. and {Shih}, A.~Y. and {Sick}, J. and {Silbiger}, M.~T. and {Singanamalla}, S. and {Singer}, L.~P. and {Sladen}, P.~H. and {Sooley}, K.~A. and {Sornarajah}, S. and {Streicher}, O. and {Teuben}, P. and {Thomas}, S.~W. and {Tremblay}, G.~R. and {Turner}, J.~E.~H. and {Terr{\'o}n}, V. and {van Kerkwijk}, M.~H. and {de la Vega}, A. and {Watkins}, L.~L. and {Weaver}, B.~A. and {Whitmore}, J.~B. and {Woillez}, J. and {Zabalza}, V. and {Astropy Contributors}},
        title = "{The Astropy Project: Building an Open-science Project and Status of the v2.0 Core Package}",
      journal = {\aj},
         year = 2018,
        month = sep,
       volume = {156},
       number = {3},
          eid = {123},
        pages = {123},
          doi = {10.3847/1538-3881/aabc4f},
archivePrefix = {arXiv},
       eprint = {1801.02634},
 primaryClass = {astro-ph.IM},
       adsurl = {https://ui.adsabs.harvard.edu/abs/2018AJ....156..123A}
}

@ARTICLE{grant:2023,
       author = {{Grant}, David and {Lewis}, Nikole K. and {Wakeford}, Hannah R. and {Batalha}, Natasha E. and {Glidden}, Ana and {Goyal}, Jayesh and {Mullens}, Elijah and {MacDonald}, Ryan J. and {May}, Erin M. and {Seager}, Sara and {Stevenson}, Kevin B. and {Valenti}, Jeff A. and {Visscher}, Channon and {Alderson}, Lili and {Allen}, Natalie H. and {Ca{\~n}as}, Caleb I. and {Col{\'o}n}, Knicole and {Clampin}, Mark and {Espinoza}, N{\'e}stor and {Gressier}, Am{\'e}lie and {Huang}, Jingcheng and {Lin}, Zifan and {Long}, Douglas and {Louie}, Dana R. and {Pe{\~n}a-Guerrero}, Maria and {Ranjan}, Sukrit and {Sotzen}, Kristin S. and {Valentine}, Daniel and {Anderson}, Jay and {Balmer}, William O. and {Bellini}, Andrea and {Hoch}, Kielan K.~W. and {Kammerer}, Jens and {Libralato}, Mattia and {Mountain}, C. Matt and {Perrin}, Marshall D. and {Pueyo}, Laurent and {Rickman}, Emily and {Rebollido}, Isabel and {Sohn}, Sangmo Tony and {van der Marel}, Roeland P. and {Watkins}, Laura L.},
        title = "{JWST-TST DREAMS: Quartz Clouds in the Atmosphere of WASP-17b}",
      journal = {\apjl},
         year = 2023,
        month = oct,
       volume = {956},
       number = {2},
          eid = {L29},
        pages = {L29},
          doi = {10.3847/2041-8213/acfc3b},
archivePrefix = {arXiv},
       eprint = {2310.08637},
 primaryClass = {astro-ph.EP},
       adsurl = {https://ui.adsabs.harvard.edu/abs/2023ApJ...956L..29G}
}

@ARTICLE{miles:2023,
       author = {{Miles}, Brittany E. and {Biller}, Beth A. and {Patapis}, Polychronis and {Worthen}, Kadin and {Rickman}, Emily and {Hoch}, Kielan K.~W. and {Skemer}, Andrew and {Perrin}, Marshall D. and {Whiteford}, Niall and {Chen}, Christine H. and {Sargent}, B. and {Mukherjee}, Sagnick and {Morley}, Caroline V. and {Moran}, Sarah E. and {Bonnefoy}, Mickael and {Petrus}, Simon and {Carter}, Aarynn L. and {Choquet}, Elodie and {Hinkley}, Sasha and {Ward-Duong}, Kimberly and {Leisenring}, Jarron M. and {Millar-Blanchaer}, Maxwell A. and {Pueyo}, Laurent and {Ray}, Shrishmoy and {Sallum}, Steph and {Stapelfeldt}, Karl R. and {Stone}, Jordan M. and {Wang}, Jason J. and {Absil}, Olivier and {Balmer}, William O. and {Boccaletti}, Anthony and {Bonavita}, Mariangela and {Booth}, Mark and {Bowler}, Brendan P. and {Chauvin}, Gael and {Christiaens}, Valentin and {Currie}, Thayne and {Danielski}, Camilla and {Fortney}, Jonathan J. and {Girard}, Julien H. and {Grady}, Carol A. and {Greenbaum}, Alexandra Z. and {Henning}, Thomas and {Hines}, Dean C. and {Janson}, Markus and {Kalas}, Paul and {Kammerer}, Jens and {Kennedy}, Grant M. and {Kenworthy}, Matthew A. and {Kervella}, Pierre and {Lagage}, Pierre-Olivier and {Lew}, Ben W.~P. and {Liu}, Michael C. and {Macintosh}, Bruce and {Marino}, Sebastian and {Marley}, Mark S. and {Marois}, Christian and {Matthews}, Elisabeth C. and {Matthews}, Brenda C. and {Mawet}, Dimitri and {McElwain}, Michael W. and {Metchev}, Stanimir and {Meyer}, Michael R. and {Molliere}, Paul and {Pantin}, Eric and {Quirrenbach}, Andreas and {Rebollido}, Isabel and {Ren}, Bin B. and {Schneider}, Glenn and {Vasist}, Malavika and {Wyatt}, Mark C. and {Zhou}, Yifan and {Briesemeister}, Zackery W. and {Bryan}, Marta L. and {Calissendorff}, Per and {Cantalloube}, Faustine and {Cugno}, Gabriele and {De Furio}, Matthew and {Dupuy}, Trent J. and {Factor}, Samuel M. and {Faherty}, Jacqueline K. and {Fitzgerald}, Michael P. and {Franson}, Kyle and {Gonzales}, Eileen C. and {Hood}, Callie E. and {Howe}, Alex R. and {Kraus}, Adam L. and {Kuzuhara}, Masayuki and {Lagrange}, Anne-Marie and {Lawson}, Kellen and {Lazzoni}, Cecilia and {Liu}, Pengyu and {Llop-Sayson}, Jorge and {Lloyd}, James P. and {Martinez}, Raquel A. and {Mazoyer}, Johan and {Quanz}, Sascha P. and {Redai}, Jea Adams and {Samland}, Matthias and {Schlieder}, Joshua E. and {Tamura}, Motohide and {Tan}, Xianyu and {Uyama}, Taichi and {Vigan}, Arthur and {Vos}, Johanna M. and {Wagner}, Kevin and {Wolff}, Schuyler G. and {Ygouf}, Marie and {Zhang}, Xi and {Zhang}, Keming and {Zhang}, Zhoujian},
        title = "{The JWST Early-release Science Program for Direct Observations of Exoplanetary Systems II: A 1 to 20 {\ensuremath{\mu}}m Spectrum of the Planetary-mass Companion VHS 1256-1257 b}",
      journal = {\apjl},
         year = 2023,
        month = mar,
       volume = {946},
       number = {1},
          eid = {L6},
        pages = {L6},
          doi = {10.3847/2041-8213/acb04a},
archivePrefix = {arXiv},
       eprint = {2209.00620},
 primaryClass = {astro-ph.EP},
       adsurl = {https://ui.adsabs.harvard.edu/abs/2023ApJ...946L...6M}
}

@ARTICLE{suarez:2022,
       author = {{Su{\'a}rez}, Genaro and {Metchev}, Stanimir},
        title = "{Ultracool dwarfs observed with the Spitzer infrared spectrograph - II. Emergence and sedimentation of silicate clouds in L dwarfs, and analysis of the full M5-T9 field dwarf spectroscopic sample}",
      journal = {\mnras},
         year = 2022,
        month = jul,
       volume = {513},
       number = {4},
        pages = {5701-5726},
          doi = {10.1093/mnras/stac1205},
archivePrefix = {arXiv},
       eprint = {2205.00168},
 primaryClass = {astro-ph.SR},
       adsurl = {https://ui.adsabs.harvard.edu/abs/2022MNRAS.513.5701S}
}

@ARTICLE{nasedkin:2024,
       author = {{Nasedkin}, Evert and {Molli{\`e}re}, Paul and {Blain}, Doriann},
        title = "{Atmospheric Retrievals with petitRADTRANS}",
      journal = {The Journal of Open Source Software},
         year = 2024,
        month = apr,
       volume = {9},
       number = {96},
          eid = {5875},
        pages = {5875},
          doi = {10.21105/joss.05875},
archivePrefix = {arXiv},
       eprint = {2309.06755},
 primaryClass = {astro-ph.EP},
       adsurl = {https://ui.adsabs.harvard.edu/abs/2024JOSS....9.5875N}
}

@ARTICLE{wende:2010,
       author = {{Wende}, S. and {Reiners}, A. and {Seifahrt}, A. and {Bernath}, P.~F.},
        title = "{CRIRES spectroscopy and empirical line-by-line identification of FeH molecular absorption in an M dwarf}",
      journal = {\aap},
         year = 2010,
        month = nov,
       volume = {523},
          eid = {A58},
        pages = {A58},
          doi = {10.1051/0004-6361/201015220},
archivePrefix = {arXiv},
       eprint = {1007.4116},
 primaryClass = {astro-ph.SR},
       adsurl = {https://ui.adsabs.harvard.edu/abs/2010A&A...523A..58W}
}

@ARTICLE{li:2015,
       author = {{Li}, Gang and {Gordon}, Iouli E. and {Rothman}, Laurence S. and {Tan}, Yan and {Hu}, Shui-Ming and {Kassi}, Samir and {Campargue}, Alain and {Medvedev}, Emile S.},
        title = "{Rovibrational Line Lists for Nine Isotopologues of the CO Molecule in the X $^{1}${\ensuremath{\Sigma}}$^{+}$ Ground Electronic State}",
      journal = {\apjs},
         year = 2015,
        month = jan,
       volume = {216},
       number = {1},
          eid = {15},
        pages = {15},
          doi = {10.1088/0067-0049/216/1/15},
       adsurl = {https://ui.adsabs.harvard.edu/abs/2015ApJS..216...15L}
}

@ARTICLE{polyansky:2018,
       author = {{Polyansky}, Oleg L. and {Kyuberis}, Aleksandra A. and {Zobov}, Nikolai F. and {Tennyson}, Jonathan and {Yurchenko}, Sergei N. and {Lodi}, Lorenzo},
        title = "{ExoMol molecular line lists XXX: a complete high-accuracy line list for water}",
      journal = {\mnras},
         year = 2018,
        month = oct,
       volume = {480},
       number = {2},
        pages = {2597-2608},
          doi = {10.1093/mnras/sty1877},
archivePrefix = {arXiv},
       eprint = {1807.04529},
 primaryClass = {astro-ph.EP},
       adsurl = {https://ui.adsabs.harvard.edu/abs/2018MNRAS.480.2597P}
}

@ARTICLE{fer08,
       author = {{Feroz}, F. and {Hobson}, M.~P.},
        title = "{Multimodal nested sampling: an efficient and robust alternative to Markov Chain Monte Carlo methods for astronomical data analyses}",
      journal = {\mnras},
         year = 2008,
        month = feb,
       volume = {384},
       number = {2},
        pages = {449-463},
          doi = {10.1111/j.1365-2966.2007.12353.x},
archivePrefix = {arXiv},
       eprint = {0704.3704},
 primaryClass = {astro-ph},
       adsurl = {https://ui.adsabs.harvard.edu/abs/2008MNRAS.384..449F}
}

@ARTICLE{fer09,
       author = {{Feroz}, F. and {Hobson}, M.~P. and {Bridges}, M.},
        title = "{MULTINEST: an efficient and robust Bayesian inference tool for cosmology and particle physics}",
      journal = {\mnras},
         year = 2009,
        month = oct,
       volume = {398},
       number = {4},
        pages = {1601-1614},
          doi = {10.1111/j.1365-2966.2009.14548.x},
archivePrefix = {arXiv},
       eprint = {0809.3437},
 primaryClass = {astro-ph},
       adsurl = {https://ui.adsabs.harvard.edu/abs/2009MNRAS.398.1601F}
}

@ARTICLE{yurchenko:2014,
       author = {{Yurchenko}, Sergei N. and {Tennyson}, Jonathan},
        title = "{ExoMol line lists - IV. The rotation-vibration spectrum of methane up to 1500 K}",
      journal = {\mnras},
         year = 2014,
        month = may,
       volume = {440},
       number = {2},
        pages = {1649-1661},
          doi = {10.1093/mnras/stu326},
archivePrefix = {arXiv},
       eprint = {1401.4852},
 primaryClass = {astro-ph.EP},
       adsurl = {https://ui.adsabs.harvard.edu/abs/2014MNRAS.440.1649Y}
}

@ARTICLE{Sousa-Silva:2015,
       author = {{Sousa-Silva}, Clara and {Al-Refaie}, Ahmed F. and {Tennyson}, Jonathan and {Yurchenko}, Sergei N.},
        title = "{ExoMol line lists - VII. The rotation-vibration spectrum of phosphine up to 1500 K}",
      journal = {\mnras},
         year = 2015,
        month = jan,
       volume = {446},
       number = {3},
        pages = {2337-2347},
          doi = {10.1093/mnras/stu2246},
archivePrefix = {arXiv},
       eprint = {1410.2917},
 primaryClass = {astro-ph.EP},
       adsurl = {https://ui.adsabs.harvard.edu/abs/2015MNRAS.446.2337S}
}

@ARTICLE{barber:2014,
       author = {{Barber}, R.~J. and {Strange}, J.~K. and {Hill}, C. and {Polyansky}, O.~L. and {Mellau}, G. Ch. and {Yurchenko}, S.~N. and {Tennyson}, Jonathan},
        title = "{ExoMol line lists - III. An improved hot rotation-vibration line list for HCN and HNC}",
      journal = {\mnras},
         year = 2014,
        month = jan,
       volume = {437},
       number = {2},
        pages = {1828-1835},
          doi = {10.1093/mnras/stt2011},
archivePrefix = {arXiv},
       eprint = {1311.1328},
 primaryClass = {astro-ph.SR},
       adsurl = {https://ui.adsabs.harvard.edu/abs/2014MNRAS.437.1828B}
}

@ARTICLE{morley:2024,
       author = {{Morley}, Caroline V. and {Mukherjee}, Sagnick and {Marley}, Mark S. and {Fortney}, Jonathan J. and {Visscher}, Channon and {Lupu}, Roxana and {Gharib-Nezhad}, Ehsan and {Thorngren}, Daniel and {Freedman}, Richard and {Batalha}, Natasha},
        title = "{The Sonora Substellar Atmosphere Models. III. Diamondback: Atmospheric Properties, Spectra, and Evolution for Warm Cloudy Substellar Objects}",
      journal = {\apj},
         year = 2024,
        month = nov,
       volume = {975},
       number = {1},
          eid = {59},
        pages = {59},
          doi = {10.3847/1538-4357/ad71d5},
archivePrefix = {arXiv},
       eprint = {2402.00758},
 primaryClass = {astro-ph.SR},
       adsurl = {https://ui.adsabs.harvard.edu/abs/2024ApJ...975...59M}
}

@software{baudino:2024,
       author = {{Baudino}, Jean-Loup and {B{\'e}zard}, Bruno and {Charnay}, Benjamin and {Doriann}, Blain},
        title = "{Exo-REM: 1D self-consistent radiative-equilibrium model for exoplanetary atmospheres}",
 howpublished = {Astrophysics Source Code Library, record ascl:2410.012},
         year = 2024,
        month = oct,
          eid = {ascl:2410.012},
       adsurl = {https://ui.adsabs.harvard.edu/abs/2024ascl.soft10012B}
}

@INPROCEEDINGS{allard:2012,
       author = {{Allard}, F. and {Homeier}, D. and {Freytag}, B. and {Sharp}, C.~M.},
        title = "{Atmospheres From Very Low-Mass Stars to Extrasolar Planets}",
    booktitle = {EAS Publications Series},
         year = 2012,
       editor = {{Reyl{\'e}}, C{\'e}line and {Charbonnel}, Corinne and {Schultheis}, Mathias},
       series = {EAS Publications Series},
       volume = {57},
        month = nov,
    publisher = {EDP},
        pages = {3-43},
          doi = {10.1051/eas/1257001},
archivePrefix = {arXiv},
       eprint = {1206.1021},
 primaryClass = {astro-ph.SR},
       adsurl = {https://ui.adsabs.harvard.edu/abs/2012EAS....57....3A}
}

@ARTICLE{petrus:2024,
       author = {{Petrus}, Simon and {Whiteford}, Niall and {Patapis}, Polychronis and {Biller}, Beth A. and {Skemer}, Andrew and {Hinkley}, Sasha and {Su{\'a}rez}, Genaro and {Palma-Bifani}, Paulina and {Morley}, Caroline V. and {Tremblin}, Pascal and {Charnay}, Benjamin and {Vos}, Johanna M. and {Wang}, Jason J. and {Stone}, Jordan M. and {Bonnefoy}, Micka{\"e}l and {Chauvin}, Ga{\"e}l and {Miles}, Brittany E. and {Carter}, Aarynn L. and {Lueber}, Anna and {Helling}, Christiane and {Sutlieff}, Ben J. and {Janson}, Markus and {Gonzales}, Eileen C. and {Hoch}, Kielan K.~W. and {Absil}, Olivier and {Balmer}, William O. and {Boccaletti}, Anthony and {Bonavita}, Mariangela and {Booth}, Mark and {Bowler}, Brendan P. and {Briesemeister}, Zackery W. and {Bryan}, Marta L. and {Calissendorff}, Per and {Cantalloube}, Faustine and {Chen}, Christine H. and {Choquet}, Elodie and {Christiaens}, Valentin and {Cugno}, Gabriele and {Currie}, Thayne and {Danielski}, Camilla and {De Furio}, Matthew and {Dupuy}, Trent J. and {Factor}, Samuel M. and {Faherty}, Jacqueline K. and {Fitzgerald}, Michael P. and {Fortney}, Jonathan J. and {Franson}, Kyle and {Girard}, Julien H. and {Grady}, Carol A. and {Henning}, Thomas and {Hines}, Dean C. and {Hood}, Callie E. and {Howe}, Alex R. and {Kalas}, Paul and {Kammerer}, Jens and {Kennedy}, Grant M. and {Kenworthy}, Matthew A. and {Kervella}, Pierre and {Kim}, Minjae and {Kitzmann}, Daniel and {Kraus}, Adam L. and {Kuzuhara}, Masayuki and {Lagage}, Pierre-Olivier and {Lagrange}, Anne-Marie and {Lawson}, Kellen and {Lazzoni}, Cecilia and {Leisenring}, Jarron M. and {Lew}, Ben W.~P. and {Liu}, Michael C. and {Liu}, Pengyu and {Llop-Sayson}, Jorge and {Lloyd}, James P. and {Macintosh}, Bruce and {M{\^a}lin}, Mathilde and {Manjavacas}, Elena and {Marino}, Sebasti{\'a}n and {Marley}, Mark S. and {Marois}, Christian and {Martinez}, Raquel A. and {Matthews}, Elisabeth C. and {Matthews}, Brenda C. and {Mawet}, Dimitri and {Mazoyer}, Johan and {McElwain}, Michael W. and {Metchev}, Stanimir and {Meyer}, Michael R. and {Millar-Blanchaer}, Maxwell A. and {Molli{\`e}re}, Paul and {Moran}, Sarah E. and {Mukherjee}, Sagnick and {Pantin}, Eric and {Perrin}, Marshall D. and {Pueyo}, Laurent and {Quanz}, Sascha P. and {Quirrenbach}, Andreas and {Ray}, Shrishmoy and {Rebollido}, Isabel and {Adams Redai}, Jea and {Ren}, Bin B. and {Rickman}, Emily and {Sallum}, Steph and {Samland}, Matthias and {Sargent}, Benjamin and {Schlieder}, Joshua E. and {Stapelfeldt}, Karl R. and {Tamura}, Motohide and {Tan}, Xianyu and {Theissen}, Christopher A. and {Uyama}, Taichi and {Vasist}, Malavika and {Vigan}, Arthur and {Wagner}, Kevin and {Ward-Duong}, Kimberly and {Wolff}, Schuyler G. and {Worthen}, Kadin and {Wyatt}, Mark C. and {Ygouf}, Marie and {Zurlo}, Alice and {Zhang}, Xi and {Zhang}, Keming and {Zhang}, Zhoujian and {Zhou}, Yifan},
        title = "{The JWST Early Release Science Program for Direct Observations of Exoplanetary Systems. V. Do Self-consistent Atmospheric Models Represent JWST Spectra? A Showcase with VHS 1256{\textendash}1257 b}",
      journal = {\apjl},
         year = 2024,
        month = may,
       volume = {966},
       number = {1},
          eid = {L11},
        pages = {L11},
          doi = {10.3847/2041-8213/ad3e7c},
archivePrefix = {arXiv},
       eprint = {2312.03852},
 primaryClass = {astro-ph.EP},
       adsurl = {https://ui.adsabs.harvard.edu/abs/2024ApJ...966L..11P}
}

@ARTICLE{asplund:2021,
       author = {{Asplund}, M. and {Amarsi}, A.~M. and {Grevesse}, N.},
        title = "{The chemical make-up of the Sun: A 2020 vision}",
      journal = {\aap},
         year = 2021,
        month = sep,
       volume = {653},
          eid = {A141},
        pages = {A141},
          doi = {10.1051/0004-6361/202140445},
archivePrefix = {arXiv},
       eprint = {2105.01661},
 primaryClass = {astro-ph.SR},
       adsurl = {https://ui.adsabs.harvard.edu/abs/2021A&A...653A.141A}
}

@INPROCEEDINGS{lodders:2010,
       author = {{Lodders}, Katharina},
        title = "{Solar System Abundances of the Elements}",
    booktitle = {Principles and Perspectives in Cosmochemistry},
         year = 2010,
       editor = {{Goswami}, Aruna and {Reddy}, B. Eswar},
       series = {Astrophysics and Space Science Proceedings},
       volume = {16},
        month = jan,
        pages = {379},
          doi = {10.1007/978-3-642-10352-0_8},
archivePrefix = {arXiv},
       eprint = {1010.2746},
 primaryClass = {astro-ph.SR},
       adsurl = {https://ui.adsabs.harvard.edu/abs/2010ASSP...16..379L}
}

@ARTICLE{zalesky:2022,
       author = {{Zalesky}, Joseph A. and {Saboi}, Kezman and {Line}, Michael R. and {Zhang}, Zhoujian and {Schneider}, Adam C. and {Liu}, Michael C. and {Best}, William M.~J. and {Marley}, Mark S.},
        title = "{A Uniform Retrieval Analysis of Ultra-cool Dwarfs. IV. A Statistical Census from 50 Late-T Dwarfs}",
      journal = {\apj},
         year = 2022,
        month = sep,
       volume = {936},
       number = {1},
          eid = {44},
        pages = {44},
          doi = {10.3847/1538-4357/ac786c},
archivePrefix = {arXiv},
       eprint = {2206.01199},
 primaryClass = {astro-ph.SR},
       adsurl = {https://ui.adsabs.harvard.edu/abs/2022ApJ...936...44Z}
}

@ARTICLE{balmer:2025,
       author = {{Balmer}, William O. and {Franson}, Kyle and {Chomez}, Antoine and {Pueyo}, Laurent and {Stolker}, Tomas and {Lacour}, Sylvestre and {Nowak}, Mathias and {Nasedkin}, Evert and {Bonse}, Markus J. and {Thorngren}, Daniel and {Palma-Bifani}, Paulina and {Molli{\`e}re}, Paul and {Wang}, Jason J. and {Zhang}, Zhoujian and {Chavez}, Amanda and {Kammerer}, Jens and {Blunt}, Sarah and {Bowler}, Brendan P. and {Bonnefoy}, Mickael and {Brandner}, Wolfgang and {Charnay}, Benjamin and {Chauvin}, Gael and {Henning}, Th. and {Lagrange}, A. -M. and {Pourr{\'e}}, Nicolas and {Rickman}, Emily and {De Rosa}, Robert and {Vigan}, Arthur and {Winterhalder}, Thomas},
        title = "{VLTI/GRAVITY Observations of AF Lep b: Preference for Circular Orbits, Cloudy Atmospheres, and a Moderately Enhanced Metallicity}",
      journal = {\aj},
         year = 2025,
        month = jan,
       volume = {169},
       number = {1},
          eid = {30},
        pages = {30},
          doi = {10.3847/1538-3881/ad9265},
archivePrefix = {arXiv},
       eprint = {2411.05917},
 primaryClass = {astro-ph.EP},
       adsurl = {https://ui.adsabs.harvard.edu/abs/2025AJ....169...30B}
}

@ARTICLE{nasedkin:2024b,
       author = {{Nasedkin}, E. and {Molli{\`e}re}, P. and {Lacour}, S. and {Nowak}, M. and {Kreidberg}, L. and {Stolker}, T. and {Wang}, J.~J. and {Balmer}, W.~O. and {Kammerer}, J. and {Shangguan}, J. and {Abuter}, R. and {Amorim}, A. and {Asensio-Torres}, R. and {Benisty}, M. and {Berger}, J. -P. and {Beust}, H. and {Blunt}, S. and {Boccaletti}, A. and {Bonnefoy}, M. and {Bonnet}, H. and {Bordoni}, M.~S. and {Bourdarot}, G. and {Brandner}, W. and {Cantalloube}, F. and {Caselli}, P. and {Charnay}, B. and {Chauvin}, G. and {Chavez}, A. and {Choquet}, E. and {Christiaens}, V. and {Cl{\'e}net}, Y. and {Coud{\'e} Du Foresto}, V. and {Cridland}, A. and {Davies}, R. and {Dembet}, R. and {Dexter}, J. and {Drescher}, A. and {Duvert}, G. and {Eckart}, A. and {Eisenhauer}, F. and {F{\"o}rster Schreiber}, N.~M. and {Garcia}, P. and {Garcia Lopez}, R. and {Gendron}, E. and {Genzel}, R. and {Gillessen}, S. and {Girard}, J.~H. and {Grant}, S. and {Haubois}, X. and {Hei{\ss}el}, G. and {Henning}, Th. and {Hinkley}, S. and {Hippler}, S. and {Houll{\'e}}, M. and {Hubert}, Z. and {Jocou}, L. and {Keppler}, M. and {Kervella}, P. and {Kurtovic}, N.~T. and {Lagrange}, A. -M. and {Lapeyr{\`e}re}, V. and {Le Bouquin}, J. -B. and {Lutz}, D. and {Maire}, A. -L. and {Mang}, F. and {Marleau}, G. -D. and {M{\'e}rand}, A. and {Monnier}, J.~D. and {Mordasini}, C. and {Ott}, T. and {Otten}, G.~P.~P.~L. and {Paladini}, C. and {Paumard}, T. and {Perraut}, K. and {Perrin}, G. and {Pfuhl}, O. and {Pourr{\'e}}, N. and {Pueyo}, L. and {Ribeiro}, D.~C. and {Rickman}, E. and {Ruffio}, J.~B. and {Rustamkulov}, Z. and {Shimizu}, T. and {Sing}, D. and {Stadler}, J. and {Straub}, O. and {Straubmeier}, C. and {Sturm}, E. and {Tacconi}, L.~J. and {van Dishoeck}, E.~F. and {Vigan}, A. and {Vincent}, F. and {von Fellenberg}, S.~D. and {Widmann}, F. and {Winterhalder}, T.~O. and {Woillez}, J. and {Yazici}, {\c{S}}. and {Gravity Collaboration}},
        title = "{Four-of-a-kind? Comprehensive atmospheric characterisation of the HR 8799 planets with VLTI/GRAVITY}",
      journal = {\aap},
         year = 2024,
        month = jul,
       volume = {687},
          eid = {A298},
        pages = {A298},
          doi = {10.1051/0004-6361/202449328},
archivePrefix = {arXiv},
       eprint = {2404.03776},
 primaryClass = {astro-ph.EP},
       adsurl = {https://ui.adsabs.harvard.edu/abs/2024A&A...687A.298N}
}

@ARTICLE{molliere:2025,
       author = {{Molli{\`e}re}, P. and {K{\"u}hnle}, H. and {Matthews}, E.~C. and {Henning}, Th. and {Min}, M. and {Patapis}, P. and {Lagage}, P. -O. and {Waters}, L.~B.~F.~M. and {G{\"u}del}, M. and {J{\"a}ger}, Cornelia and {Zhang}, Z. and {Decin}, L. and {Biller}, B.~A. and {Absil}, O. and {Argyriou}, I. and {Barrado}, D. and {Cossou}, C. and {Glasse}, A. and {Olofsson}, G. and {Pye}, J.~P. and {Rouan}, D. and {Samland}, M. and {Scheithauer}, S. and {Tremblin}, P. and {Whiteford}, N. and {van Dishoeck}, E.~F. and {{\"O}stlin}, G. and {Ray}, T.},
        title = "{Evidence for SiO cloud nucleation in the rogue planet PSO J318}",
      journal = {arXiv e-prints},
         year = 2025,
        month = jul,
          eid = {arXiv:2507.18691},
        pages = {arXiv:2507.18691},
          doi = {10.48550/arXiv.2507.18691},
archivePrefix = {arXiv},
       eprint = {2507.18691},
 primaryClass = {astro-ph.EP},
       adsurl = {https://ui.adsabs.harvard.edu/abs/2025arXiv250718691M}
}

@ARTICLE{tsuji:2002,
       author = {{Tsuji}, Takashi},
        title = "{Dust in the Photospheric Environment: Unified Cloudy Models of M, L, and T Dwarfs}",
      journal = {\apj},
         year = 2002,
        month = aug,
       volume = {575},
       number = {1},
        pages = {264-290},
          doi = {10.1086/341262},
archivePrefix = {arXiv},
       eprint = {astro-ph/0204401},
 primaryClass = {astro-ph},
       adsurl = {https://ui.adsabs.harvard.edu/abs/2002ApJ...575..264T}
}

@ARTICLE{tremblin:2015,
       author = {{Tremblin}, P. and {Amundsen}, D.~S. and {Mourier}, P. and {Baraffe}, I. and {Chabrier}, G. and {Drummond}, B. and {Homeier}, D. and {Venot}, O.},
        title = "{Fingering Convection and Cloudless Models for Cool Brown Dwarf Atmospheres}",
      journal = {\apjl},
         year = 2015,
        month = may,
       volume = {804},
       number = {1},
          eid = {L17},
        pages = {L17},
          doi = {10.1088/2041-8205/804/1/L17},
archivePrefix = {arXiv},
       eprint = {1504.03334},
 primaryClass = {astro-ph.SR},
       adsurl = {https://ui.adsabs.harvard.edu/abs/2015ApJ...804L..17T}
}

@ARTICLE{leconte:2018,
       author = {{Leconte}, J{\'e}r{\'e}my},
        title = "{Why Compositional Convection Cannot Explain Substellar Objects{\textquoteright} Sharp Spectral-type Transitions}",
      journal = {\apjl},
         year = 2018,
        month = feb,
       volume = {853},
       number = {2},
          eid = {L30},
        pages = {L30},
          doi = {10.3847/2041-8213/aaaa61},
archivePrefix = {arXiv},
       eprint = {1801.08142},
 primaryClass = {astro-ph.SR},
       adsurl = {https://ui.adsabs.harvard.edu/abs/2018ApJ...853L..30L}
}

@ARTICLE{yurchenko:2020,
       author = {{Yurchenko}, S.~N. and {Mellor}, Thomas M. and {Freedman}, Richard S. and {Tennyson}, J.},
        title = "{ExoMol line lists - XXXIX. Ro-vibrational molecular line list for CO$_{2}$}",
      journal = {\mnras},
         year = 2020,
        month = aug,
       volume = {496},
       number = {4},
        pages = {5282-5291},
          doi = {10.1093/mnras/staa1874},
archivePrefix = {arXiv},
       eprint = {2007.02122},
 primaryClass = {astro-ph.EP},
       adsurl = {https://ui.adsabs.harvard.edu/abs/2020MNRAS.496.5282Y}
}

@ARTICLE{coles:2019,
       author = {{Coles}, Phillip A. and {Yurchenko}, Sergei N. and {Tennyson}, Jonathan},
        title = "{ExoMol molecular line lists - XXXV. A rotation-vibration line list for hot ammonia}",
      journal = {\mnras},
         year = 2019,
        month = dec,
       volume = {490},
       number = {4},
        pages = {4638-4647},
          doi = {10.1093/mnras/stz2778},
archivePrefix = {arXiv},
       eprint = {1911.10369},
 primaryClass = {astro-ph.SR},
       adsurl = {https://ui.adsabs.harvard.edu/abs/2019MNRAS.490.4638C}
}

@ARTICLE{azzam:2016,
       author = {{Azzam}, Ala'a. A.~A. and {Tennyson}, Jonathan and {Yurchenko}, Sergei N. and {Naumenko}, Olga V.},
        title = "{ExoMol molecular line lists - XVI. The rotation-vibration spectrum of hot H$_{2}$S}",
      journal = {\mnras},
         year = 2016,
        month = aug,
       volume = {460},
       number = {4},
        pages = {4063-4074},
          doi = {10.1093/mnras/stw1133},
archivePrefix = {arXiv},
       eprint = {1607.00499},
 primaryClass = {astro-ph.EP},
       adsurl = {https://ui.adsabs.harvard.edu/abs/2016MNRAS.460.4063A}
}

@ARTICLE{piskunov:1995,
       author = {{Piskunov}, N.~E. and {Kupka}, F. and {Ryabchikova}, T.~A. and {Weiss}, W.~W. and {Jeffery}, C.~S.},
        title = "{VALD: The Vienna Atomic Line Data Base.}",
      journal = {\aaps},
         year = 1995,
        month = sep,
       volume = {112},
        pages = {525},
       adsurl = {https://ui.adsabs.harvard.edu/abs/1995A&AS..112..525P}
}

@ARTICLE{barton:2013,
       author = {{Barton}, Emma J. and {Yurchenko}, Sergei N. and {Tennyson}, Jonathan},
        title = "{ExoMol line lists - II. The ro-vibrational spectrum of SiO}",
      journal = {\mnras},
         year = 2013,
        month = sep,
       volume = {434},
       number = {2},
        pages = {1469-1475},
          doi = {10.1093/mnras/stt1105},
archivePrefix = {arXiv},
       eprint = {1307.2300},
 primaryClass = {astro-ph.SR},
       adsurl = {https://ui.adsabs.harvard.edu/abs/2013MNRAS.434.1469B}
}

@ARTICLE{moran:2024,
       author = {{Moran}, Sarah E. and {Marley}, Mark S. and {Crossley}, Samuel D.},
        title = "{Neglected Silicon Dioxide Polymorphs as Clouds in Substellar Atmospheres}",
      journal = {\apjl},
         year = 2024,
        month = sep,
       volume = {973},
       number = {1},
          eid = {L3},
        pages = {L3},
          doi = {10.3847/2041-8213/ad72e7},
archivePrefix = {arXiv},
       eprint = {2408.00698},
 primaryClass = {astro-ph.EP},
       adsurl = {https://ui.adsabs.harvard.edu/abs/2024ApJ...973L...3M}
}

@ARTICLE{hansen:1971,
       author = {{Hansen}, James E.},
        title = "{Multiple Scattering of Polarized Light in Planetary Atmospheres Part II. Sunlight Reflected by Terrestrial Water Clouds.}",
      journal = {Journal of the Atmospheric Sciences},
         year = 1971,
        month = nov,
       volume = {28},
       number = {8},
        pages = {1400-1426},
          doi = {10.1175/1520-0469(1971)028<1400:MSOPLI>2.0.CO;2},
       adsurl = {https://ui.adsabs.harvard.edu/abs/1971JAtS...28.1400H}
}

@ARTICLE{suarez:2023,
       author = {{Su{\'a}rez}, Genaro and {Vos}, Johanna M. and {Metchev}, Stanimir and {Faherty}, Jacqueline K. and {Cruz}, Kelle},
        title = "{Ultracool Dwarfs Observed with the Spitzer Infrared Spectrograph: Equatorial Latitudes in L Dwarf Atmospheres Are Cloudier}",
      journal = {\apjl},
         year = 2023,
        month = sep,
       volume = {954},
       number = {1},
          eid = {L6},
        pages = {L6},
          doi = {10.3847/2041-8213/acec4b},
archivePrefix = {arXiv},
       eprint = {2308.02093},
 primaryClass = {astro-ph.EP},
       adsurl = {https://ui.adsabs.harvard.edu/abs/2023ApJ...954L...6S}
}

@ARTICLE{rowland:2023,
       author = {{Rowland}, Melanie J. and {Morley}, Caroline V. and {Line}, Michael R.},
        title = "{Toward Robust Atmospheric Retrieval on Cloudy L Dwarfs: the Impact of Thermal and Abundance Profile Assumptions}",
      journal = {\apj},
         year = 2023,
        month = apr,
       volume = {947},
       number = {1},
          eid = {6},
        pages = {6},
          doi = {10.3847/1538-4357/acbb07},
archivePrefix = {arXiv},
       eprint = {2301.08258},
 primaryClass = {astro-ph.SR},
       adsurl = {https://ui.adsabs.harvard.edu/abs/2023ApJ...947....6R}
}

@ARTICLE{davis:2025,
       author = {{Davis}, C. Evan and {Fortney}, Jonathan J. and {Iyer}, Aishwarya and {Mukherjee}, Sagnick and {Morley}, Caroline V. and {Marley}, Mark S. and {Line}, Michael and {Muirhead}, Philip S.},
        title = "{The Sonora Substellar Atmosphere Models. VI. Red Diamondback: Extending Diamondback with SPHINX for Brown Dwarf Early Evolution}",
      journal = {\apj},
         year = 2025,
        month = dec,
       volume = {994},
       number = {2},
          eid = {198},
        pages = {198},
          doi = {10.3847/1538-4357/ae1015},
archivePrefix = {arXiv},
       eprint = {2510.08694},
 primaryClass = {astro-ph.SR},
       adsurl = {https://ui.adsabs.harvard.edu/abs/2025ApJ...994..198D}
}

@ARTICLE{feuillet:2019,
       author = {{Feuillet}, Diane K. and {Frankel}, Neige and {Lind}, Karin and {Frinchaboy}, Peter M. and {Garc{\'\i}a-Hern{\'a}ndez}, D.~A. and {Lane}, Richard R. and {Nitschelm}, Christian and {Roman-Lopes}, Alexandre},
        title = "{Spatial variations in the Milky Way disc metallicity-age relation}",
      journal = {\mnras},
         year = 2019,
        month = oct,
       volume = {489},
       number = {2},
        pages = {1742-1752},
          doi = {10.1093/mnras/stz2221},
archivePrefix = {arXiv},
       eprint = {1908.02772},
 primaryClass = {astro-ph.GA},
       adsurl = {https://ui.adsabs.harvard.edu/abs/2019MNRAS.489.1742F}
}

@ARTICLE{faherty:2024,
       author = {{Faherty}, Jacqueline K. and {Burningham}, Ben and {Gagn{\'e}}, Jonathan and {Su{\'a}rez}, Genaro and {Vos}, Johanna M. and {Alejandro Merchan}, Sherelyn and {Morley}, Caroline V. and {Rowland}, Melanie and {Lacy}, Brianna and {Kiman}, Rocio and {Caselden}, Dan and {Kirkpatrick}, J. Davy and {Meisner}, Aaron and {Schneider}, Adam C. and {Kuchner}, Marc Jason and {Bardalez Gagliuffi}, Daniella Carolina and {Beichman}, Charles and {Eisenhardt}, Peter and {Gelino}, Christopher R. and {Gharib-Nezhad}, Ehsan and {Gonzales}, Eileen and {Marocco}, Federico and {Rothermich}, Austin James and {Whiteford}, Niall},
        title = "{Methane emission from a cool brown dwarf}",
      journal = {\nat},
         year = 2024,
        month = apr,
       volume = {628},
       number = {8008},
        pages = {511-514},
          doi = {10.1038/s41586-024-07190-w},
archivePrefix = {arXiv},
       eprint = {2404.10977},
 primaryClass = {astro-ph.SR},
       adsurl = {https://ui.adsabs.harvard.edu/abs/2024Natur.628..511F}
}

@ARTICLE{faherty:2025,
       author = {{Faherty}, Jacqueline K. and {Meisner}, Aaron M. and {Burningham}, Ben and {Visscher}, Channon and {Line}, Michael and {Su{\'a}rez}, Genaro and {Gagn{\'e}}, Jonathan and {Alejandro Merchan}, Sherelyn and {Rothermich}, Austin James and {Burgasser}, Adam J. and {Schneider}, Adam C. and {Caselden}, Dan and {Kirkpatrick}, J. Davy and {Kuchner}, Marc Jason and {Bardalez Gagliuffi}, Daniella Carolina and {Eisenhardt}, Peter and {Gelino}, Christopher R. and {Gonzales}, Eileen C. and {Marocco}, Federico and {Leggett}, Sandy and {Lodieu}, Nicolas and {Casewell}, Sarah L. and {Tremblin}, Pascal and {Cushing}, Michael and {Zapatero Osorio}, Maria Rosa and {B{\'e}jar}, V{\'\i}ctor J.~S. and {Gauza}, Bartosz and {Wright}, Edward and {Phillips}, Mark W. and {Zhang}, Jun-Yan and {Martin}, Eduardo L.},
        title = "{Silicate precursor silane detected in cold low-metallicity brown dwarf}",
      journal = {\nat},
         year = 2025,
        month = sep,
       volume = {645},
       number = {8079},
        pages = {62-66},
          doi = {10.1038/s41586-025-09369-1},
archivePrefix = {arXiv},
       eprint = {2509.18635},
 primaryClass = {astro-ph.EP},
       adsurl = {https://ui.adsabs.harvard.edu/abs/2025Natur.645...62F}
}

@ARTICLE{nasedkin:2025,
       author = {{Nasedkin}, E. and {Schrader}, M. and {Vos}, J.~M. and {Biller}, B. and {Burningham}, B. and {Cowan}, N.~B. and {Faherty}, J.~K. and {Gonzales}, E. and {Lam}, M.~B. and {McCarthy}, A.~M. and {Muirhead}, P.~S. and {O'Toole}, C. and {Plummer}, M.~K. and {Su{\'a}rez}, G. and {Tan}, X. and {Visscher}, C. and {Whiteford}, N. and {Zhou}, Y.},
        title = "{The JWST weather report: Retrieving temperature variations, auroral heating, and static cloud coverage on SIMP-0136}",
      journal = {\aap},
         year = 2025,
        month = oct,
       volume = {702},
          eid = {A1},
        pages = {A1},
          doi = {10.1051/0004-6361/202555370},
archivePrefix = {arXiv},
       eprint = {2507.07772},
 primaryClass = {astro-ph.EP},
       adsurl = {https://ui.adsabs.harvard.edu/abs/2025A&A...702A...1N}
}

@ARTICLE{oliverosgomez:2025,
       author = {{Oliveros-Gomez}, Natalia and {Manjavacas}, Elena and {Karalidi}, Theodora and {Phillippe}, Myrla and {Campos Estrada}, Beatriz and {Biller}, Beth and {Vos}, Johanna M. and {Faherty}, Jacqueline and {Chen}, Xueqing and {Dupuy}, Trent J. and {Henning}, Thomas and {McCarthy}, Allison M. and {Muirhead}, Philip S. and {Lee}, Elspeth K.~H. and {Tremblin}, Pascal and {Ramirez}, Jasmine and {Suarez}, Genaro and {Sutlieff}, Ben J. and {Tan}, Xianyu and {Crouzet}, Nicolas},
        title = "{The JWST weather report from the nearest brown dwarfs III: Heterogeneous clouds and Thermochemical instabilities as possible drivers of WISE 1049AB's spectroscopic variability}",
      journal = {arXiv e-prints},
         year = 2025,
        month = nov,
          eid = {arXiv:2511.15667},
        pages = {arXiv:2511.15667},
          doi = {10.48550/arXiv.2511.15667},
archivePrefix = {arXiv},
       eprint = {2511.15667},
 primaryClass = {astro-ph.SR},
       adsurl = {https://ui.adsabs.harvard.edu/abs/2025arXiv251115667O}
}

@ARTICLE{biller:2024,
       author = {{Biller}, Beth A. and {Vos}, Johanna M. and {Zhou}, Yifan and {McCarthy}, Allison M. and {Tan}, Xianyu and {Crossfield}, Ian J.~M. and {Whiteford}, Niall and {Suarez}, Genaro and {Faherty}, Jacqueline and {Manjavacas}, Elena and {Chen}, Xueqing and {Liu}, Pengyu and {Sutlieff}, Ben J. and {Limbach}, Mary Anne and {Molliere}, Paul and {Dupuy}, Trent J. and {Oliveros-Gomez}, Natalia and {Muirhead}, Philip S. and {Henning}, Thomas and {Mace}, Gregory and {Crouzet}, Nicolas and {Karalidi}, Theodora and {Morley}, Caroline V. and {Tremblin}, Pascal and {Kataria}, Tiffany},
        title = "{The JWST weather report from the nearest brown dwarfs I: multiperiod JWST NIRSpec + MIRI monitoring of the benchmark binary brown dwarf WISE 1049AB}",
      journal = {\mnras},
         year = 2024,
        month = aug,
       volume = {532},
       number = {2},
        pages = {2207-2233},
          doi = {10.1093/mnras/stae1602},
archivePrefix = {arXiv},
       eprint = {2407.09194},
 primaryClass = {astro-ph.SR},
       adsurl = {https://ui.adsabs.harvard.edu/abs/2024MNRAS.532.2207B}
}

@misc{hansen2019pycma,
  author       = {Nikolaus Hansen and Youhei Akimoto and Petr Baudis},
  title        = {{CMA-ES/pycma} on {G}ithub},
  howpublished = {Zenodo, DOI:10.5281/zenodo.2559634},
  month        = feb,
  year         = 2019,
  doi          = {10.5281/zenodo.2559634},
  url          = {https://doi.org/10.5281/zenodo.2559634},
}

@article{hansen:2001,
    author = {Hansen, Nikolaus and Ostermeier, Andreas},
    title = {Completely Derandomized Self-Adaptation in Evolution Strategies},
    journal = {Evolutionary Computation},
    volume = {9},
    number = {2},
    pages = {159-195},
    year = {2001},
    month = {06},
    issn = {1063-6560},
    doi = {10.1162/106365601750190398},
    url = {https://doi.org/10.1162/106365601750190398},
    eprint = {https://direct.mit.edu/evco/article-pdf/9/2/159/1493523/106365601750190398.pdf},
}

@ARTICLE{rossow:1978,
       author = {{Rossow}, W.~B.},
        title = "{Cloud Microphysics: Analysis of the Clouds of Earth, Venus, Mars, and Jupiter}",
      journal = {\icarus},
         year = 1978,
        month = oct,
       volume = {36},
       number = {1},
        pages = {1-50},
          doi = {10.1016/0019-1035(78)90072-6},
       adsurl = {https://ui.adsabs.harvard.edu/abs/1978Icar...36....1R}
}

@ARTICLE{powell:2018,
       author = {{Powell}, Diana and {Zhang}, Xi and {Gao}, Peter and {Parmentier}, Vivien},
        title = "{Formation of Silicate and Titanium Clouds on Hot Jupiters}",
      journal = {\apj},
         year = 2018,
        month = jun,
       volume = {860},
       number = {1},
          eid = {18},
        pages = {18},
          doi = {10.3847/1538-4357/aac215},
archivePrefix = {arXiv},
       eprint = {1805.01468},
 primaryClass = {astro-ph.EP},
       adsurl = {https://ui.adsabs.harvard.edu/abs/2018ApJ...860...18P}
}

@ARTICLE{charnay:2018,
       author = {{Charnay}, B. and {B{\'e}zard}, B. and {Baudino}, J.-L. and {Bonnefoy}, M. and {Boccaletti}, A. and {Galicher}, R.},
        title = "{A Self-consistent Cloud Model for Brown Dwarfs and Young Giant Exoplanets: Comparison with Photometric and Spectroscopic Observations}",
      journal = {\apj},
         year = 2018,
        month = feb,
       volume = {854},
       number = {2},
          eid = {172},
        pages = {172},
          doi = {10.3847/1538-4357/aaac7d},
archivePrefix = {arXiv},
       eprint = {1711.11483},
 primaryClass = {astro-ph.EP},
       adsurl = {https://ui.adsabs.harvard.edu/abs/2018ApJ...854..172C}
}

@ARTICLE{henning:1996,
       author = {{Henning}, T. and {Stognienko}, R.},
        title = "{Dust opacities for protoplanetary accretion disks: influence of dust aggregates.}",
      journal = {\aap},
         year = 1996,
        month = jul,
       volume = {311},
        pages = {291-303},
       adsurl = {https://ui.adsabs.harvard.edu/abs/1996A&A...311..291H}
}

@ARTICLE{kitzmann:2018b,
       author = {{Kitzmann}, Daniel and {Heng}, Kevin},
        title = "{Optical properties of potential condensates in exoplanetary atmospheres}",
      journal = {\mnras},
         year = 2018,
        month = mar,
       volume = {475},
       number = {1},
        pages = {94-107},
          doi = {10.1093/mnras/stx3141},
archivePrefix = {arXiv},
       eprint = {1710.04946},
 primaryClass = {astro-ph.EP},
       adsurl = {https://ui.adsabs.harvard.edu/abs/2018MNRAS.475...94K}
}

@book{palik1985handbook,
  title     = {Handbook of Optical Constants of Solids},
  editor    = {Edward D. Palik},
  volume    = {1},
  year      = {1985},
  publisher = {Academic Press},
  address   = {Orlando, FL},
  isbn      = {0-12-544420-6}
}

@ARTICLE{henning:1997,
       author = {{Henning}, T. and {Mutschke}, H.},
        title = "{Low-temperature infrared properties of cosmic dust analogues.}",
      journal = {\aap},
         year = 1997,
        month = nov,
       volume = {327},
        pages = {743-754},
       adsurl = {https://ui.adsabs.harvard.edu/abs/1997A&A...327..743H}
}

@ARTICLE{zeidler:2013,
       author = {{Zeidler}, S. and {Posch}, Th. and {Mutschke}, H.},
        title = "{Optical constants of refractory oxides at high temperatures. Mid-infrared properties of corundum, spinel, and {\ensuremath{\alpha}}-quartz, potential carriers of the 13 {\ensuremath{\mu}}m feature}",
      journal = {\aap},
         year = 2013,
        month = may,
       volume = {553},
          eid = {A81},
        pages = {A81},
          doi = {10.1051/0004-6361/201220459},
       adsurl = {https://ui.adsabs.harvard.edu/abs/2013A&A...553A..81Z}
}

@ARTICLE{jaeger:2003,
       author = {{J{\"a}ger}, C. and {Il'in}, V.~B. and {Henning}, Th. and {Mutschke}, H. and {Fabian}, D. and {Semenov}, D. and {Voshchinnikov}, N.},
        title = "{A Database of Optical Constants of Cosmic Dust Analogs}",
      journal = {\jqsrt},
         year = 2003,
        month = jan,
       volume = {79-80},
        pages = {765},
          doi = {10.1016/S0022-4073(02)00301-1},
       adsurl = {https://ui.adsabs.harvard.edu/abs/2003JQSRT..79..765J}
}

@ARTICLE{jaeger:1998,
       author = {{Jaeger}, C. and {Molster}, F.~J. and {Dorschner}, J. and {Henning}, Th. and {Mutschke}, H. and {Waters}, L.~B.~F.~M.},
        title = "{Steps toward interstellar silicate mineralogy. IV. The crystalline revolution}",
      journal = {\aap},
         year = 1998,
        month = nov,
       volume = {339},
        pages = {904-916},
       adsurl = {https://ui.adsabs.harvard.edu/abs/1998A&A...339..904J}
}

@ARTICLE{scott:1996,
       author = {{Scott}, A. and {Duley}, W.~W.},
        title = "{Ultraviolet and Infrared Refractive Indices of Amorphous Silicates}",
      journal = {\apjs},
         year = 1996,
        month = aug,
       volume = {105},
        pages = {401},
          doi = {10.1086/192321},
       adsurl = {https://ui.adsabs.harvard.edu/abs/1996ApJS..105..401S}
}

@ARTICLE{servoin:1973,
       author = {{Servoin}, J.~L. and {Piriou}, B.},
        title = "{Infrared Reflectivity and Raman Scattering of Mg2SiO4 Single Crystal}",
      journal = {Physica Status Solidi B Basic Research},
         year = 1973,
        month = feb,
       volume = {55},
       number = {2},
        pages = {677-686},
          doi = {10.1002/pssb.2220550224},
       adsurl = {https://ui.adsabs.harvard.edu/abs/1973PSSBR..55..677S}
}

@ARTICLE{koike:1995,
       author = {{Koike}, Chiyoe and {Kaito}, Chihiro and {Yamamoto}, Tetsuo and {Shibai}, Hiroshi and {Kimura}, Seiji and {Suto}, Hiroshi},
        title = "{Extinction spectra of corundum in the wavelengths from UV to FIR.}",
      journal = {\icarus},
         year = 1995,
        month = mar,
       volume = {114},
       number = {1},
        pages = {203-214},
          doi = {10.1006/icar.1995.1055},
       adsurl = {https://ui.adsabs.harvard.edu/abs/1995Icar..114..203K}
}

@ARTICLE{jaeger:1994,
       author = {{Jaeger}, C. and {Mutschke}, H. and {Begemann}, B. and {Dorschner}, J. and {Henning}, Th.},
        title = "{Steps toward interstellar silicate mineralogy. I. Laboratory results of a silicate glass of mean cosmic composition.}",
      journal = {\aap},
         year = 1994,
        month = dec,
       volume = {292},
        pages = {641-655},
       adsurl = {https://ui.adsabs.harvard.edu/abs/1994A&A...292..641J}
}

@ARTICLE{dorschner:1995,
       author = {{Dorschner}, J. and {Begemann}, B. and {Henning}, T. and {Jaeger}, C. and {Mutschke}, H.},
        title = "{Steps toward interstellar silicate mineralogy. II. Study of Mg-Fe-silicate glasses of variable composition.}",
      journal = {\aap},
         year = 1995,
        month = aug,
       volume = {300},
        pages = {503},
       adsurl = {https://ui.adsabs.harvard.edu/abs/1995A&A...300..503D}
}

@ARTICLE{sitarz:2000,
       author = {{Sitarz}, M. and {Handke}, M. and {Mozgawa}, W.},
        title = "{Identification of silicooxygen rings in SiO $_{2}$ based on IR spectra}",
      journal = {Spectrochimica Acta Part A: Molecular Spectroscopy},
         year = 2000,
        month = aug,
       volume = {56},
       number = {9},
        pages = {1819-1823},
          doi = {10.1016/S1386-1425(00)00241-9},
       adsurl = {https://ui.adsabs.harvard.edu/abs/2000AcSpA..56.1819S}
}

@ARTICLE{koike:2013,
       author = {{Koike}, C. and {Noguchi}, R. and {Chihara}, H. and {Suto}, H. and {Ohtaka}, O. and {Imai}, Y. and {Matsumoto}, T. and {Tsuchiyama}, A.},
        title = "{Infrared Spectra of Silica Polymorphs and the Conditions of Their Formation}",
      journal = {\apj},
         year = 2013,
        month = nov,
       volume = {778},
       number = {1},
          eid = {60},
        pages = {60},
          doi = {10.1088/0004-637X/778/1/60},
       adsurl = {https://ui.adsabs.harvard.edu/abs/2013ApJ...778...60K}
}

@ARTICLE{gandhi:2023b,
       author = {{Gandhi}, Siddharth and {de Regt}, Sam and {Snellen}, Ignas and {Zhang}, Yapeng and {Rugers}, Benson and {van Leur}, Niels and {Bosschaart}, Quincy},
        title = "{JWST Measurements of $^{13}$C, $^{18}$O, and $^{17}$O in the Atmosphere of Super-Jupiter VHS 1256 b}",
      journal = {\apjl},
         year = 2023,
        month = nov,
       volume = {957},
       number = {2},
          eid = {L36},
        pages = {L36},
          doi = {10.3847/2041-8213/ad07e2},
archivePrefix = {arXiv},
       eprint = {2311.05349},
 primaryClass = {astro-ph.EP},
       adsurl = {https://ui.adsabs.harvard.edu/abs/2023ApJ...957L..36G}
}

@ARTICLE{xuan:2026,
       author = {{Xuan}, Jerry W. and {Ruffio}, Jean-Baptiste and {Chachan}, Yayaati and {Ohno}, Kazumasa and {Kesseli}, Aurora and {Murray-Clay}, Ruth and {Lee}, Eve J. and {Moses}, Julianne I. and {Balmer}, William O. and {Baburaj}, Aneesh and {Blake}, Geoffrey A. and {Johnstone}, Doug and {Zhang}, Yapeng and {Knutson}, Heather A. and {Mawet}, Dimitri and {Beichman}, Charles and {Hodapp}, Klaus and {Perrin}, Marshall D. and {Konopacky}, Quinn and {Meyer}, Michael and {Bryden}, Geoffrey and {Greene}, Thomas P. and {Leisenring}, Jarron and {Ygouf}, Marie and {Benneke}, Bj{\"o}rn and {Inglis}, Julie and {Wallack}, Nicole L.},
        title = "{The Compositions of the HR 8799 Planets Reflect Accretion of Both Solids and Metal-enriched Gas}",
      journal = {\apj},
         year = 2026,
        month = mar,
       volume = {1000},
       number = {1},
          eid = {27},
        pages = {27},
          doi = {10.3847/1538-4357/ae448f},
archivePrefix = {arXiv},
       eprint = {2602.09422},
 primaryClass = {astro-ph.EP},
       adsurl = {https://ui.adsabs.harvard.edu/abs/2026ApJ..1000...27X}
}

@ARTICLE{calamari:2024,
       author = {{Calamari}, Emily and {Faherty}, Jacqueline K. and {Visscher}, Channon and {Gemma}, Marina E. and {Burningham}, Ben and {Rothermich}, Austin},
        title = "{Predicting Cloud Conditions in Substellar Mass Objects Using Ultracool Dwarf Companions}",
      journal = {\apj},
         year = 2024,
        month = mar,
       volume = {963},
       number = {1},
          eid = {67},
        pages = {67},
          doi = {10.3847/1538-4357/ad1f6d},
archivePrefix = {arXiv},
       eprint = {2401.11038},
 primaryClass = {astro-ph.SR},
       adsurl = {https://ui.adsabs.harvard.edu/abs/2024ApJ...963...67C}
}

@ARTICLE{balmer:2023,
       author = {{Balmer}, William O. and {Pueyo}, Laurent and {Stolker}, Tomas and {Reggiani}, Henrique and {Maire}, A.-L. and {Lacour}, S. and {Molli{\`e}re}, P. and {Nowak}, M. and {Sing}, D. and {Pourr{\'e}}, N. and {Blunt}, S. and {Wang}, J.~J. and {Rickman}, E. and {Kammerer}, J. and {Henning}, Th. and {Ward-Duong}, K. and {Abuter}, R. and {Amorim}, A. and {Asensio-Torres}, R. and {Benisty}, M. and {Berger}, J.-P. and {Beust}, H. and {Boccaletti}, A. and {Bohn}, A. and {Bonnefoy}, M. and {Bonnet}, H. and {Bourdarot}, G. and {Brandner}, W. and {Cantalloube}, F. and {Caselli}, P. and {Charnay}, B. and {Chauvin}, G. and {Chavez}, A. and {Choquet}, E. and {Christiaens}, V. and {Cl{\'e}net}, Y. and {Coud{\'e} Du Foresto}, V. and {Cridland}, A. and {Dembet}, R. and {Dexter}, J. and {Drescher}, A. and {Duvert}, G. and {Eckart}, A. and {Eisenhauer}, F. and {Gao}, F. and {Garcia}, P. and {Garcia Lopez}, R. and {Gendron}, E. and {Genzel}, R. and {Gillessen}, S. and {Girard}, J.~H. and {Haubois}, X. and {Hei{\ss}el}, G. and {Hinkley}, S. and {Hippler}, S. and {Horrobin}, M. and {Houll{\'e}}, M. and {Hubert}, Z. and {Jocou}, L. and {Keppler}, M. and {Kervella}, P. and {Kreidberg}, L. and {Lagrange}, A.-M. and {Lapeyr{\`e}re}, V. and {Le Bouquin}, J.-B. and {L{\'e}na}, P. and {Lutz}, D. and {Monnier}, J.~D. and {Mouillet}, D. and {Nasedkin}, E. and {Ott}, T. and {Otten}, G.~P.~P.~L. and {Paladini}, C. and {Paumard}, T. and {Perraut}, K. and {Perrin}, G. and {Pfuhl}, O. and {Rameau}, J. and {Rodet}, L. and {Rousset}, G. and {Rustamkulov}, Z. and {Shangguan}, J. and {Shimizu}, T. and {Stadler}, J. and {Straub}, O. and {Straubmeier}, C. and {Sturm}, E. and {Tacconi}, L.~J. and {van Dishoeck}, E.~F. and {Vigan}, A. and {Vincent}, F. and {von Fellenberg}, S.~D. and {Widmann}, F. and {Wieprecht}, E. and {Wiezorrek}, E. and {Winterhalder}, T. and {Woillez}, J. and {Yazici}, S. and {Young}, A. and {Gravity Collaboration}},
        title = "{VLTI/GRAVITY Observations and Characterization of the Brown Dwarf Companion HD 72946 B}",
      journal = {\apj},
         year = 2023,
        month = oct,
       volume = {956},
       number = {2},
          eid = {99},
        pages = {99},
          doi = {10.3847/1538-4357/acf761},
archivePrefix = {arXiv},
       eprint = {2309.04403},
 primaryClass = {astro-ph.SR},
       adsurl = {https://ui.adsabs.harvard.edu/abs/2023ApJ...956...99B}
}

@ARTICLE{wang:2022,
       author = {{Wang}, Ji and {Kolecki}, Jared R. and {Ruffio}, Jean-Baptiste and {Wang}, Jason J. and {Mawet}, Dimitri and {Baker}, Ashley and {Bartos}, Randall and {Blake}, Geoffrey A. and {Bond}, Charlotte Z. and {Calvin}, Benjamin and {Cetre}, Sylvain and {Delorme}, Jacques-Robert and {Doppmann}, Greg and {Echeverri}, Daniel and {Finnerty}, Luke and {Fitzgerald}, Michael P. and {Jovanovic}, Nemanja and {Liu}, Michael C. and {Lopez}, Ronald and {Morris}, Evan and {Pai Asnodkar}, Anusha and {Pezzato}, Jacklyn and {Ragland}, Sam and {Roy}, Arpita and {Ruane}, Garreth and {Sappey}, Ben and {Schofield}, Tobias and {Skemer}, Andrew and {Venenciano}, Taylor and {Kent Wallace}, J. and {Wallack}, Nicole L. and {Wizinowich}, Peter and {Xuan}, Jerry W.},
        title = "{Retrieving the C and O Abundances of HR 7672 AB: A Solar-type Primary Star with a Benchmark Brown Dwarf}",
      journal = {\aj},
         year = 2022,
        month = apr,
       volume = {163},
       number = {4},
          eid = {189},
        pages = {189},
          doi = {10.3847/1538-3881/ac56e2},
archivePrefix = {arXiv},
       eprint = {2202.02477},
 primaryClass = {astro-ph.EP},
       adsurl = {https://ui.adsabs.harvard.edu/abs/2022AJ....163..189W}
}

@ARTICLE{luna:2021,
       author = {{Luna}, Jessica L. and {Morley}, Caroline V.},
        title = "{Empirically Determining Substellar Cloud Compositions in the Era of the James Webb Space Telescope}",
      journal = {\apj},
         year = 2021,
        month = oct,
       volume = {920},
       number = {2},
          eid = {146},
        pages = {146},
          doi = {10.3847/1538-4357/ac1865},
archivePrefix = {arXiv},
       eprint = {2108.03161},
 primaryClass = {astro-ph.SR},
       adsurl = {https://ui.adsabs.harvard.edu/abs/2021ApJ...920..146L}
}

@ARTICLE{toppani:2006,
       author = {{Toppani}, Alice and {Libourel}, Guy and {Robert}, Fran{\c{c}}ois and {Ghanbaja}, Jaafar},
        title = "{Laboratory condensation of refractory dust in protosolar and circumstellar conditions}",
      journal = {\gca},
         year = 2006,
        month = oct,
       volume = {70},
       number = {19},
        pages = {5035-5060},
          doi = {10.1016/j.gca.2006.05.020},
       adsurl = {https://ui.adsabs.harvard.edu/abs/2006GeCoA..70.5035T}
}

@ARTICLE{fabian:2000,
       author = {{Fabian}, D. and {J{\"a}ger}, C. and {Henning}, Th. and {Dorschner}, J. and {Mutschke}, H.},
        title = "{Steps toward interstellar silicate mineralogy. V. Thermal Evolution of Amorphous Magnesium Silicates and Silica}",
      journal = {\aap},
         year = 2000,
        month = dec,
       volume = {364},
        pages = {282-292},
       adsurl = {https://ui.adsabs.harvard.edu/abs/2000A&A...364..282F}
}

@INCOLLECTION{ebel:2006,
       author = {{Ebel}, D.~S.},
        title = "{Condensation of Rocky Material in Astrophysical Environments}",
    booktitle = {Meteorites and the Early Solar System II},
         year = 2006,
       editor = {{Lauretta}, Dante S. and {McSween}, Harry Y.},
        pages = {253},
       adsurl = {https://ui.adsabs.harvard.edu/abs/2006mess.book..253E}
}

@MISC{jwstprop:2288,
       author = {{Lothringer}, Joshua D. and {Valenti}, Jeff A. and {Gao}, Peter and {Marley}, Mark S. and {Moran}, Sarah E. and {Sing}, David K. and {Wakeford}, Hannah},
        title = "{Formation and impact of silicate clouds on L dwarfs}",
 howpublished = {JWST Proposal. Cycle 1, ID. \#2288},
         year = 2021,
        month = mar,
        pages = {2288},
       adsurl = {https://ui.adsabs.harvard.edu/abs/2021jwst.prop.2288L}
}

@ARTICLE{jakobsen:2022,
       author = {{Jakobsen}, P. and {Ferruit}, P. and {Alves de Oliveira}, C. and {Arribas}, S. and {Bagnasco}, G. and {Barho}, R. and {Beck}, T.~L. and {Birkmann}, S. and {B{\"o}ker}, T. and {Bunker}, A.~J. and {Charlot}, S. and {de Jong}, P. and {de Marchi}, G. and {Ehrenwinkler}, R. and {Falcolini}, M. and {Fels}, R. and {Franx}, M. and {Franz}, D. and {Funke}, M. and {Giardino}, G. and {Gnata}, X. and {Holota}, W. and {Honnen}, K. and {Jensen}, P.~L. and {Jentsch}, M. and {Johnson}, T. and {Jollet}, D. and {Karl}, H. and {Kling}, G. and {K{\"o}hler}, J. and {Kolm}, M.-G. and {Kumari}, N. and {Lander}, M.~E. and {Lemke}, R. and {L{\'o}pez-Caniego}, M. and {L{\"u}tzgendorf}, N. and {Maiolino}, R. and {Manjavacas}, E. and {Marston}, A. and {Maschmann}, M. and {Maurer}, R. and {Messerschmidt}, B. and {Moseley}, S.~H. and {Mosner}, P. and {Mott}, D.~B. and {Muzerolle}, J. and {Pirzkal}, N. and {Pittet}, J.-F. and {Plitzke}, A. and {Posselt}, W. and {Rapp}, B. and {Rauscher}, B.~J. and {Rawle}, T. and {Rix}, H.-W. and {R{\"o}del}, A. and {Rumler}, P. and {Sabbi}, E. and {Salvignol}, J.-C. and {Schmid}, T. and {Sirianni}, M. and {Smith}, C. and {Strada}, P. and {te Plate}, M. and {Valenti}, J. and {Wettemann}, T. and {Wiehe}, T. and {Wiesmayer}, M. and {Willott}, C.~J. and {Wright}, R. and {Zeidler}, P. and {Zincke}, C.},
        title = "{The Near-Infrared Spectrograph (NIRSpec) on the James Webb Space Telescope. I. Overview of the instrument and its capabilities}",
      journal = {\aap},
         year = 2022,
        month = may,
       volume = {661},
          eid = {A80},
        pages = {A80},
          doi = {10.1051/0004-6361/202142663},
archivePrefix = {arXiv},
       eprint = {2202.03305},
 primaryClass = {astro-ph.IM},
       adsurl = {https://ui.adsabs.harvard.edu/abs/2022A&A...661A..80J}
}

@ARTICLE{wells:2015,
       author = {{Wells}, Martyn and {Pel}, J.-W. and {Glasse}, Alistair and {Wright}, G.~S. and {Aitink-Kroes}, Gabby and {Azzollini}, Ruym{\'a}n and {Beard}, Steven and {Brandl}, B.~R. and {Gallie}, Angus and {Geers}, V.~C. and {Glauser}, A.~M. and {Hastings}, Peter and {Henning}, Th. and {Jager}, Rieks and {Justtanont}, K. and {Kruizinga}, Bob and {Lahuis}, Fred and {Lee}, David and {Martinez-Delgado}, I. and {Mart{\'\i}nez-Galarza}, J.~R. and {Meijers}, M. and {Morrison}, Jane E. and {M{\"u}ller}, Friedrich and {Nakos}, Thodori and {O'Sullivan}, Brian and {Oudenhuysen}, Ad and {Parr-Burman}, P. and {Pauwels}, Evert and {Rohloff}, R.-R. and {Schmalzl}, Eva and {Sykes}, Jon and {Thelen}, M.~P. and {van Dishoeck}, E.~F. and {Vandenbussche}, Bart and {Venema}, Lars B. and {Visser}, Huib and {Waters}, L.~B.~F.~M. and {Wright}, David},
        title = "{The Mid-Infrared Instrument for the James Webb Space Telescope, VI: The Medium Resolution Spectrometer}",
      journal = {\pasp},
         year = 2015,
        month = jul,
       volume = {127},
       number = {953},
        pages = {646},
          doi = {10.1086/682281},
archivePrefix = {arXiv},
       eprint = {1508.03070},
 primaryClass = {astro-ph.IM},
       adsurl = {https://ui.adsabs.harvard.edu/abs/2015PASP..127..646W}
}

@INPROCEEDINGS{bushouse:2020,
       author = {{Bushouse}, H.},
        title = "{The JWST Science Calibration Pipeline}",
    booktitle = {Astronomical Data Analysis Software and Systems XXIX},
         year = 2020,
       editor = {{Pizzo}, R. and {Deul}, E.~R. and {Mol}, J.~D. and {de Plaa}, J. and {Verkouter}, H.},
       series = {Astronomical Society of the Pacific Conference Series},
       volume = {527},
        month = jan,
        pages = {583},
       adsurl = {https://ui.adsabs.harvard.edu/abs/2020ASPC..527..583B}
}

@software{jwstpipe1160,
       author = {{Bushouse}, Howard and {Eisenhamer}, Jonathan and {Dencheva}, Nadia and {Davies}, James and {Greenfield}, Perry and {Morrison}, Jane and {Hodge}, Phil and {Simon}, Bernie and {Grumm}, David and {Droettboom}, Michael and {Slavich}, Edward and {Sosey}, Megan and {Pauly}, Tyler and {Miller}, Todd and {Jedrzejewski}, Robert and {Hack}, Warren and {Davis}, David and {Crawford}, Steven and {Law}, David and {Gordon}, Karl and {Regan}, Michael and {Cara}, Mihai and {MacDonald}, Ken and {Bradley}, Larry and {Shanahan}, Clare and {Jamieson}, William and {Teodoro}, Mairan and {Williams}, Thomas and {Pena-Guerrero}, Maria and {Graham}, Brett and {Molter}, Edward and {Brandt}, Timothy and {Hayes}, Christian and {Cooper}, Rachel and {Clarke}, Melanie},
        title = "{JWST Calibration Pipeline}",
         year = 2024,
        month = nov,
          eid = {10.5281/zenodo.14153298},
          doi = {10.5281/zenodo.14153298},
      version = {1.16.1},
    publisher = {Zenodo},
       adsurl = {https://ui.adsabs.harvard.edu/abs/2024zndo..14153298B}
}

@ARTICLE{barber:2026,
       author = {{Barber}, Sam and {Metchev}, Stanimir and {Su{\'a}rez}, Genaro and {Lambier}, Samantha and {Miles-P{\'a}ez}, Paulo A. and {Artigau}, {\'E}tienne and {Biller}, Beth and {Burgasser}, Adam J. and {Faherty}, Jacqueline and {Mace}, Gregory and {Morley}, Caroline V. and {Vos}, Johanna M.},
        title = "{Silicate clouds, SiO, and molecular trends across the L─T sequence: a JWST/MIRI LRS spectral library of brown dwarfs}",
      journal = {\mnras},
         year = 2026,
        month = aug,
       volume = {550},
       number = {4},
          eid = {stag1387},
        pages = {stag1387},
          doi = {10.1093/mnras/stag1387},
archivePrefix = {arXiv},
       eprint = {2607.15521},
 primaryClass = {astro-ph.SR},
       adsurl = {https://ui.adsabs.harvard.edu/abs/2026MNRAS.550g1387B}
}

@dataset{morley_2024_12735103,
author = {Morley, Caroline and
Mukherjee, Sagnick and
Marley, Mark and
Fortney, Jonathan and
Visscher, Channon and
Lupu, Roxana and
Gharib-Nezhad, Ehsan and
Thorngren, Daniel and
Freedman, Richard and
Batalha, Natasha},
title = {The Sonora Substellar Atmosphere Models:
Diamondback
},
year = 2024,
publisher = {Zenodo},
doi = {10.5281/zenodo.12735103},
url = {https://doi.org/10.5281/zenodo.12735103},
}

@misc{Allard2012_SVO,
  author       = {Allard, France and Homeier, Derek and Freytag, Bernd},
  title        = {{BT-Settl} Model Atmospheres},
  year         = {2012},
  howpublished = {SVO Theoretical Spectra Server},
  url          = {https://svo2.cab.inta-csic.es/theory/newov2/},
  note         = {Accessed: 2026}
}

@misc{exoremYGP2021,
  author       = {Charnay, Benjamin},
  year = 2021,
  title        = {{Exo-REM YGP Model Grids}},
  howpublished = {LEISA, Observatoire de Paris},
  url = {https://lesia.obspm.fr/exorem/YGP_grids/old_grids_2021/},
  note         = {Accessed 2026 October 29}
}
\bibliographystyle{aasjournal}

\end{document}